\documentclass{article}   

\usepackage{graphicx}
  \usepackage{chicago}
  \usepackage[a4paper]{geometry}
\newcommand{\vx}{{\bf x}}
\newcommand{\vd}{{\bf d}}

\newcommand{\vs}{{\bf s}}

\newcommand{\vq}{{\bf q}}

\newcommand{\vu}{{\bf u}}

\newcommand{\vr}{{\bf r}}
\newcommand{\vp}{{\bf p}}

\newcommand{\vw}{{\bf w}}
\newcommand{\vk}{{\bf k}}
\newcommand{\ii}{{\rm i}}
\newcommand{\dd}{{\rm d}}
\newcommand{\dr}{\partial}

\newcommand{\rhob}{\overline{\rho}}

\newcommand{\mg}{\big<}
\newcommand{\md}{\big>}

\newcommand{\mC}{\mathcal{C}}

\newcommand{\mM}{\mathcal{M}}
\newcommand{\mP}{\mathcal{P}}

\newcommand{\mH}{\mathcal{H}}
\newcommand{\mW}{\mathcal{W}}
\newcommand{\mF}{\mathcal{F}}

\newcommand{\mS}{\mathcal{S}}

\newcommand{\fun}[2]{\lower3.6pt\vbox{\baselineskip0pt\lineskip.9pt
        \ialign{$\mathsurround=0pt#1\hfill##\hfil$\crcr#2\crcr\sim\crcr}}}

\newcommand{\cum}[1]{{\langle\rho^{#1}\rangle_c}}   % one-point density cumulants
\newcommand{\mom}[1]{{\langle\rho^{#1}\rangle}}   % one-point density moments

\newcommand{\de}{\delta}
\newcommand{\Dirac}{\delta_{\rm Dirac}}

\newcommand{\sym}{{\hbox{sym.}}}

\newcommand{\Nb}{\overline{N}}
\newcommand{\Ppoisson}{P_{\rm Poisson}}
\newcommand{\mgP}{\mathcal{D}}

\newcommand{\ci}{{\cis}}
\newcommand{\bb}{I}
\newcommand{\bc}{J}
\newcommand{\deltam}{\delta_{\rm m}}
\newcommand{\rtheta}{{\Theta}}
\newcommand{\fw}{w}

\newcommand{\cis}{{<}}
\newcommand{\iis}{{\nabla}}
\newcommand{\plus}{{+}}
\newcommand{\minus}{{-}}

\newcommand{\htheta}{\hat{\theta}}
\newcommand{\etaf}{{\eta_{f}}}
\newcommand{\etai}{{\eta_{i}}}

\newcommand{\reg}{{\rm reg}}

\newcommand{\lin}{{\rm lin.}}
\newcommand{\zel}{{\rm Zel.}}

\newcommand{\dpartial}{{\delta}}

\newcommand{\oneloop}{{\rm 1-loop}}
\newcommand{\twoloops}{{\rm 2-loop}}
\newcommand{\hs}{\hat{s}}
\newcommand{\hrho}{\hat{\rho}}
\newcommand{\Dr}{{\Delta R}}

\newcommand\binom[2]
{\left(\begin{array}{c}
#1\\
#2
\end{array}\right)}

\newcommand{\threeloops}{{\rm 3-loop}}
\newcommand{\ploops}{{\rm p-loop}}

\newcommand{\Mpc}{{\rm Mpc}}

\usepackage{color}
\definecolor{Red}{rgb}{0.65,0.08,0.05}
\definecolor{Blue}{rgb}{0.05,0.08,0.65}
\definecolor{Purple}{RGB}{143,51,143}

\title{\bf The evolution of the large-scale structure of the universe: beyond the linear regime\footnote{Lectures given to the Les Houches Summer School ÒPost-Planck Cosmology,Ó 8 July - 2 August 2013}}
\author{ Francis Bernardeau\\
\\
 \small{Institut de Physique Th\'eorique, CEA, IPhT, F-91191 Gif-sur-Yvette,}\\
 \small{CNRS, URA 2306, F-91191 Gif-sur-Yvette, France}}
\date{}

\begin{document}

\maketitle

\begin{abstract}
These lecture notes introduce analytical tools, methods and results describing the growth of cosmological structure beyond the linear regime.
The presentation is focused on the single flow regime of the Vlasov-Poisson equation describing 
the development of gravitational instabilities in a pressureless fluid.\\
These notes  include the introduction of diagrammatic representations of the standard perturbation theory with 
applications to the calculation of the so-called loop contributions to the power spectra. A large part of these notes 
is devoted to the exploration of the convergence
properties of these terms from the contribution of both the long-wave modes and the short-wave modes. It is shown that 
the former can be addressed with the help of the eikonal approximation while the latter effect is partially screened out. The resulting performances of the two-loop corrections  of the power spectra are then presented. \\
Finally other avenues that use different methods are explored. In particular it is shown how joint density and profile probability distribution
functions can be constructed out of the multiple-variable cumulant generating functions computed at tree order.
%\end{quote}
\end{abstract}

\tableofcontents
%\maintext

%&&&&&&&&&&&&&&&&&&&&&&&&&&&&&&&&&
\section{Introduction}
%&&&&&&&&&&&&&&&&&&&&&&&&&&&&&&&&&

Observations made by the Planck mission and their interpretation represent undoubtedly the crown on our understanding 
of the development of gravitational instabilities throughout the cosmological ages \shortcite{2013arXiv1303.5062P}. It is nothing but the outcome of 
a scientific program that started about 30 years ago when the Cold Dark Matter (CDM) type cosmological
models were first formulated \shortcite{1982ApJ...263L...1P,1984Natur.311..517B,1984ApJ...285L..45B}. The Planck observations that reach exquisite precision 
can indeed be explained with the help of only a few ingredients, an energy and matter content of the universe, 
a basic inflationary scenario. And within this scenario, the quality of robustness of the  results is utterly impressive 
(see contribution of F.R. Bouchet, this volume, for more details).  This is also the best that could be achieved out of the  physics of linear 
perturbations in cosmology (see contribution of Matias Zaldarriaga, this volume), from the physics of the early universe, 
including an inflationary stage, the identification and evolution of the metric perturbations in  
general relativity  to the development of perturbations are they re-enter within the Hubble radius including the physics of 
recombination captured by a Boltzmann equation.

The next frontier in our ability to confront observations and theory relies on the observational of large-scale
structure of the universe in a regime which will necessarily incorporate non-linear effects, even though it could be at a mild level.
This is the case in current and  the next generation of surveys that should be able to exploit a large fraction
of the observational universe, (see contribution of Will Percival, this volume). The analysis of these
observations definitely requires tools and methods that go beyond the linear theory, and incidentally beyond the exploitation of
statistical properties of a Gaussian or nearly Gaussian field. 

The traditional approach to study the development of large-scale structure is to make use of 
numerical simulations (see contribution of Romain Teyssier). But although they provide very valuable insights into 
the late time development of gravitational instabilities they do not easily incorporate large dynamical range and 
are not necessary able to explore large variety of cosmological parameters or models.  Obtaining results from first principles remains therefore 
extremely precious. The analytical investigations we will present here will be of interest 
for scales that range in between the linear physics and the fully nonlinear evolution, say between 
100 and a few Mpc, and allows to apprehend the early departure from linear growth.

The purpose of these lectures is to provide specific tools and techniques  and also to chart  the state of the art for calculations from first principles.  This is a domain that witnessed important development over the last few years with the introduction of novel methods. Standard perturbation theory (PT) calculations were described in \shortciteN{2002PhR...367....1B} but since then  a lot of efforts have been devoted to the development of alternative analytical methods that  try to improve upon standard perturbation theory calculations. The first significant progress in this line of calculations is the Renormalized Perturbation Theory (RPT) proposition  \shortcite{2006PhRvD..73f3519C} followed among other propositions by the closure theory \shortcite{2008ApJ...674..617T} and the time flow equations approach \shortcite{2008JCAP...10..036P}. Latest developments 
present results up to three-loop order (as in \shortciteNP{2012arXiv1211.1571B,2013arXiv1309.3308B}) with effective proposition 
such as MPTbreeze \shortcite{2012MNRAS.427.2537C} and RegPT \shortcite{2012PhRvD..86j3528T} that incorporate 2-loop order calculations and are accompanied by publicly released codes.  Concurrently, better understanding of the mathematical structure of the equation have been obtained, in particular the role of the extended galilean invariance \shortcite{2013arXiv1304.1546B,2013JCAP...05..031P,2013NuPhB.873..514K,2013arXiv1309.3557C} or how to compute 
the impact of long wave modes through the eikonal approximations \shortcite{2012PhRvD..85f3509B}. Finally, approaches that aim at circumventing  the intrinsic 
limitation of first principle calculations, namely the single flow approximation, have been put forward. They are hybrid approaches, based on effective field modeling 
\shortcite{2012JHEP...09..082C,2012JCAP...01..019P}, that could fruitfully complement the standard approaches. The methods aforementioned aim at computing corrective terms to the basic observable such as power spectra. Many other avenues have been explored that aim at computing quantities that are complementary to poly-spectra
observations and predictions. Some of these methods and techniques will also be described.

The plan of the lectures is the following. In Section \ref{sec:Vlasov}, the single flow Vlasov-Poisson system is derived. 
In Section \ref{sec:linear} the Green function of this system is presented, including the case of the multiple fluids and for 
an arbitrary background. With the help of Section \ref{sec:Stats} that presents the basic tools
 of statistical analysis of  classical random fields, we can move into the description of the nonlinear system in Section \ref{sec:nonlinear} including its  diagrammatic representation. The following sections, \ref{sec:IR} to \ref{sec:couplings}, present the structure of mode couplings for the one-loop and two-loop corrections to the power spectrum. These sections contain a presentation of the eikonal approximation and of the $\Gamma-$expansion theorem on which schemes such as the RPT and RegPT are 
based on and a presentation of the converging properties of the diagrams. The remaining sections of the notes present alternative approaches. In Section 
\ref{sec:otherPT} alternative PT methods are presented
 with a particular emphasis on Lagrangian PT. In Section \ref{sec:otherobs} examples of other observables are considered. Specific tools, based on  the cumulant 
generating function, namely the derivation of the inverse Laplace transform and of the Edgeworth expansion,
 are given. Finally in Sections \ref{sec:PDFs} and \ref{sec:profiles}  
 we will see how one can take advantage of the spherical collapse solution  to derive 
  the cumulant generating function of the density in concentric cells. It is used to derive respectively the probability distribution functions of the density and its profile. 
Some perspectives are presented in Section \ref{sec:perspectives}.

%\tableofcontents

%&&&&&&&&&&&&&&&&&&&&&&&&&&&&&&&&&
\section{The single flow Vlasov-Poisson equation}
\label{sec:Vlasov}
%&&&&&&&&&&&&&&&&&&&&&&&&&&&&&&&&&

We owe to G. Lema\^{i}tre in papers dating back \citeyear{1931MNRAS..91..490L} and \citeyear{1934PNAS...20...12L} the idea that the large-scale structure of the universe grew because of gravitational 
instabilities out of primordial small inhomogeneities. In the early thirties, however, basic knowledge on the thermal history of the universe were missing (not to mention the 
existence of dark matter) and this is only recently that this work program could be completed. The description of the growth of gravitational instabilities in the local 
universe is described in detail in classical textbooks \cite{1980lssu.book.....P} and more specifically in \shortciteN{2002PhR...367....1B} regarding various basic 
aspects of perturbation theories. 
  
The object of this section is to describe the growth of perturbation in the local universe where the Newton dynamical laws are assumed to apply. To be more precise 
it is assumed that the distances of interest are small compared to the curvature radius of the universe so that, as long as the local gravitational potentials are small, 
general relativity effects are effectively negligible.

\subsection{The Vlasov equation}

So let us start assuming that the universe is full of dust like particles whose only interaction is gravitational. 
For simplicity it is assumed that their masses are all the same, $m$.

The first step of the calculation is to introduce the phase space density function, 
$
f(\vx,\vp)\,\dd^3\vx\,\dd^3\vp,
$
which is the number of particles per volume element $\dd^3\vx\,\dd^3\vp$
where the position $\vx$ of the particles is expressed in comoving coordinates and 
the particle conjugate momentum $\vp$ reads
\begin{equation} 
\vp={\vu\, m\,a},
\end{equation}
where $a$ is the expansion factor, 
$\vu$ is the peculiar velocity, i.e. the difference of the physical velocity of the Hubble expansion.
Then the conservation of the particles together with the Liouville theorem when the the two-body interactions can be 
neglected\footnote{Otherwise we would have to  use the Boltzmann equation.}, 
implies that the total time derivative of $f$ vanishes so that
\begin{eqnarray}
{\dd f\over \dd t}= {\dr\over\dr t}f(\vx,\vp,t)+\frac{\dd \vx}{\dd t}{\dr\over\dr\vx}f(\vx,\vp,t)+
\frac{\dd\vp}{\dd t}{\dr\over\dr\vp}f(\vx,\vp,t)=0.\label{Vlasov}
\end{eqnarray}
This is the Vlasov equation. The approach we are following here is actually very general and can be applied in a variety of contexts including the early stages of the 
dynamics. We restrict ourselves to the sub-Hubble scales that will make 
our derivation simple. The time variation of the position can be expressed in terms of $\vp$ and one gets
\begin{equation}
\frac{\dd\vx}{\dd t}={\vp\over m\,a^2}.
\end{equation}
The time variation of the momentum in general can be obtained from the geodesic equation. Assuming 
the metric perturbations are small and for scales much below the Hubble scale we have
\begin{equation}
\frac{\dd\vp}{\dd t}=-m\nabla_{\vx}\Phi(\vx,t)
\end{equation}
where $\Phi$ is the potential. We recall that in the context of metric perturbation in an expanding universe the potential 
$\Phi(\vx)$ is sourced by the density contrast (of all species). In our context we simply have
\begin{equation}
\Delta \Phi(\vx)={4 \pi G m\over a}\left(\int f(\vx,\vp,t)
\dd^3\vp- \overline{n}\right)\label{Poisson1}
\end{equation}
where $\overline{n}$ is the spatial average of  $\int\!f(\vx,\vp,t)\dd^3\vp$ and we then have
\begin{equation}
 {\dr\over\dr t}f(\vx,\vp,t)+{\vp\over m\,a^2}{\dr\over\dr\vx}f(\vx,\vp,t)-m\nabla_{\vx}\Phi(\vx,t){\dr\over\dr\vp}f(\vx,\vp,t)=0.\label{Vlasov1}\\
\end{equation}
The system  (\ref{Vlasov1}, \ref{Poisson1}) forms the \emph{Vlasov-Poisson} equation. This is precisely the set of equations the $N$-body simulations attempt to 
solve.

We can now derive the basic conservation equations we are going to use from the first 2 moments of the Vlasov equation.
Let us define the density field per volume $\dd^{3}\vr$ as
\begin{equation}
\rho(\vx,t)=\frac{m}{a^{3}}\int\dd^{3}\vp\,f(\vx,\vp).
\end{equation}
It can be decomposed in an homogeneous form and an
inhomogeneous form, 
\begin{equation}
\rho(\vx,t)=\rhob(t)(1+\delta(\vx,t)).
\end{equation}
Note that $\rhob(t)$ the spatial averaged of $\rho(\vx,t)$ should behave like $a(t)^{-3}$ for non relativistic species.
One should then define the higher order moment of the phase space distribution: the mean velocity flow
is defined as (for each component),
\begin{equation}
u_{i}(\vx,t)=\frac{1}{\int\dd^{3}\vp\,f(\vx,\vp,t)} \int\dd^{3}\vp\,\frac{p_{i}}{ma}f(\vx,\vp,t),
\end{equation}
and the second moment defines the velocity dispersion $\sigma_{ij}(\vx,t)$,
\begin{equation}
u_{i}(\vx,t)u_{j}(\vx,t)+\sigma_{ij}(\vx,t)=\frac{1}{\int\dd^{3}\vp\,f(\vx,\vp,t)} \int\dd^{3}\vp\,\frac{p_{i}}{ma}\,\frac{p_{j}}{ma}\,f(\vx,\vp,t).
\end{equation}
The first two moments of the Vlasov equation give then the conservation and Euler equations, respectively
\begin{equation}
{\partial\delta(\vx,t)\over\partial t}+{1\over
a}\left[(1+\delta(\vx,t))\vu_i(\vx,t)\right]_{,i}=0\label{Continuity}
\end{equation}
and
\begin{equation} 
{\dr\vu_i(\vx,t)\over\dr t}+{\dot{a}\over a}\vu_i(\vx,t)+{1\over
a}\vu_j(\vx,t)\,\vu_{i}(\vx,t)_{,j}=-{1\over a}\Phi(\vx,t)_{,i}-{\left(\rho(\vx,t)\,\sigma_{ij}(\vx,t)\right)_{,j}\over \rho(\vx,t)\,a}
\label{Euler}.
\end{equation}
The first term of the right hand side of eqn (\ref{Euler}) is the gravitational force, the second is due to the pressure force which
in general can be anisotropic. There are subsequent equations that can be written for the whole hierarchy of the velocity moments
and depending of the physical situation, the hierarchy can be truncated if microphysics  dictates a relation between the pressure 
tensor and the local density (this is the case for perfect fluid), if for some reasons the higher order moments become negligible
as it is the case in the early phase of gravitational dynamics. 

In the context we are interested in, the velocity actually vanishes until the formation of the first caustics.

\subsection{Single flow approximation}

\begin{figure}
 \includegraphics[width=.9\textwidth]{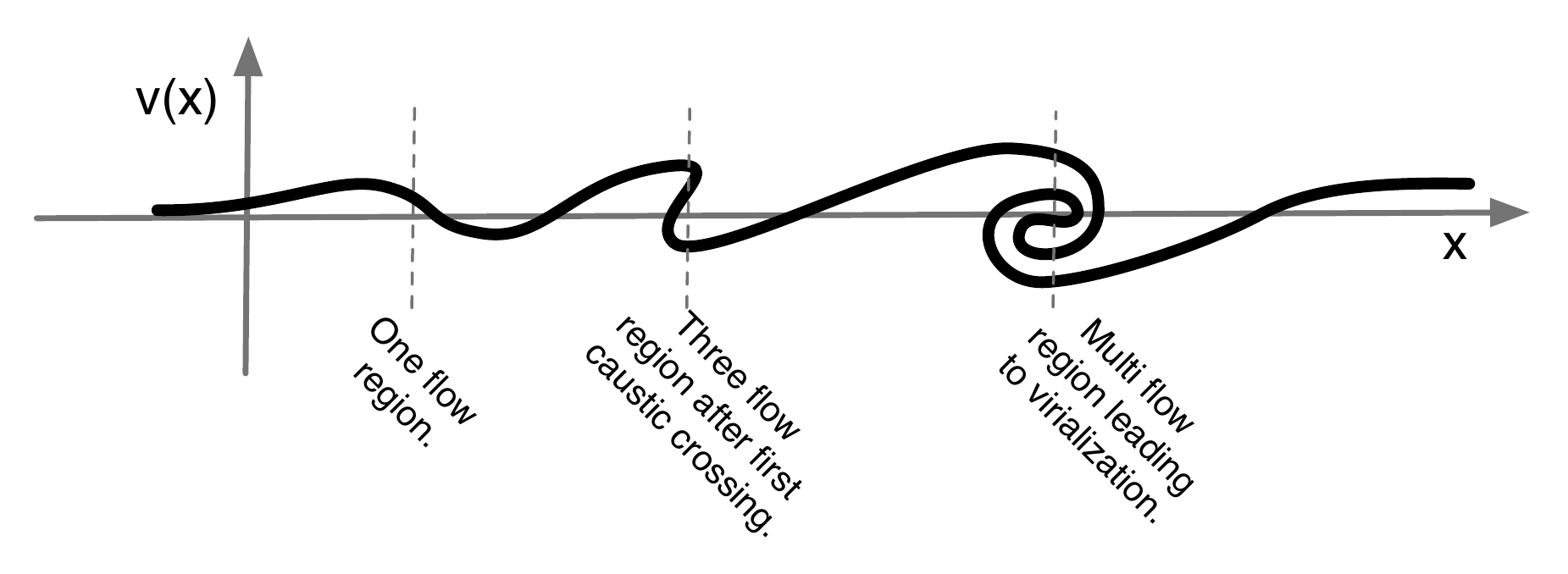}
\caption{Schematic description of phase space after the first shell crossings and emergence of multi-flow regions. The figure is for 1D dynamics. From left to right, 
one can see regions with growing number of flows after dark matter caustic crossings.}
\label{flows}
\end{figure}

The early stages of the gravitational instabilities are indeed characterized, assuming the matter is non-relativistic,  by a negligible velocity dispersion
when it is compared to the velocity flows, i.e. much smaller than the velocity gradients induced by the density fluctuations of the scales of interest. This is the 
\emph{single flow approximation}. It simply states that one can assume
\begin{equation}
f(\vx,\vp,t)={a^3\,\rho(\vx,t)\over
m}\,\delta^{(3)}[\vp-m\,a\,\vu(\vx,t)],
\end{equation}
to a good approximation. This approximation will naturally break at the time of shell crossings when different flows -- pulled toward one-another by gravity -- cross. A 
sketch of what the phase space looks like is shown on Fig. \ref{flows}. The multi-flow regions will eventually lead to the formation of astrophysical objects through a 
complicated phase of relaxation (virialization, see \shortciteN{1987gady.book.....B} for some hints on how that could take place). After shell crossings very little analytical results are known and one should
rely on $N$-body codes. Within this approximation we simply have $\sigma_{ij}(\vx,t)=0$.

The Vlasov-Poisson equation in the single flow regime is the system that will be studied throughout these lecture notes, from linear
to non nonlinear regime. 

\subsection{The curl modes}

In the single flow regime, one can note that the source term of the Euler equation is potential, implying that it cannot generate any curl 
mode in the velocity field. More precisely, one can decompose any three-dimensional field in a gradient part and 
a curl part
\begin{equation}
\vu_{i}(\vx)=\psi(\vx)_{,i}+\vw_{i}(\vx)
\end{equation}
where  $\vw_{i,i}=0$. Defining the local vorticity as 
\begin{equation}
\omega_{k}(\vx)=\epsilon^{ijk}\vu_{i}(\vx)_{,j}
\end{equation}
where $\epsilon^{ijk}$ is the totally anti-symmetric Levi-Civita tensor one can easily show that
\begin{equation}
\omega_{k}(\vx)=\epsilon^{ijk}\vw_{i}(\vx)_{,j}
\end{equation}
and that applying the operator $\epsilon^{ijk}\nabla_{j}$ to the Euler equation one gets (see \shortciteNP{2002PhR...367....1B})
\begin{equation}
\frac{\partial}{\partial t}\omega_{k}+\frac{\dot a}{a}\omega_{k}
-\epsilon^{ijk}\epsilon^{lmi}\left(\vu_{l}\,\omega_{m}\right)_{,j}=0.
\end{equation}
This equation actually expresses the fact that the vorticity is conserved throughout the expansion. In the linear regime that is when the last term of this equation is 
dropped it simply means that the vorticity scales like $1/a$. In the subsequent stage of the dynamics the vorticity can only grow in contracting regions but it is still 
somehow conserved, it cannot be created out of potential modes only. 
That will be the case until shell crossing where the anisotropic velocity dispersion can then induce vorticity. This has been explicitly demonstrated 
in various studies~\shortcite{1999A&A...343..663P,2007A&A...476...31V,2009PhRvD..80d3504P}. As a consequence, in the following, curl modes in the vector flied will always be neglected.

%%%%%%%%%%%%%%%%%%%%%%%%%%%%%%%%%%%%%%%%%
\section{The linear theory}
\label{sec:linear}

We now proceed to explore the linear regime of the Vlasov-Poisson system. One objective 
is to make contact with earlier stages of the gravitational dynamics and the second is to introduce the notion of Green function we will
use in the following.

\subsection{The linear modes}

The linearization of the motion equation is obtained when one assumes that the terms $\left[\delta(\vx,t))\vu_i(\vx,t)\right]_{,i}$
and $\vu_j(\vx,t)\,\vu_{i}(\vx,t)_{,j}$ in respectively the continuity and the Euler equation vanish. This is obtained when both the density 
contrast and the velocity gradients in units of $H$ are negligible.
The linearized system is obtained in terms of the velocity divergence
\begin{equation}
\theta(\vx,t)=\frac{1}{aH}u_{i,i}
\end{equation}
so that the system now reads
\begin{eqnarray}
\frac{\partial}{\partial t}\delta(\vx,t)+H\theta(\vx,t)&=&0\\
\frac{\partial}{\partial t}\theta(\vx,t)+2H\theta+\frac{\dot H}{H}\theta(\vx,t)&=&-\frac{3}{2}H\Omega_{m}(t)\delta(\vx,t)
\end{eqnarray}
after taking the divergence of the Euler equation. We have introduced here the Hubble parameter $H=\dot{a}/a$
and used the Friedman equation $H^{2}=8\pi/3\ G\rho_{c}(t)$ together with the definition of $\Omega_{m}=\rhob(t)/\rho_{c}(t)$.

The resolution of this system is now simple. It can be obtained after eliminating the velocity divergence and 
one gets a second order dynamical equation,
\begin{equation}
\label{DustEvol}
{\dr^2\over\dr\,t^2}\delta(\vx,t)+2\, H{\dr\over\dr\,t}\delta(\vx,t)-\frac{3}{2}\,H^{2}\,\Omega_{m}\,\delta(\vx,t)=0,
\end{equation}
for the density contrast. It is to be noted that the spatial coordinates are here just labels: there is no operator acting
of the physical coordinates. This is quite a unique feature in the growth of instabilities in a pressureless fluid. 
That implies in particular that the linear growth rate of the fluctuations will be independent on scale. 
The time dependence of the linear solution is given by the two solutions of
\begin{equation}
\ddot{D}+2\,H\,\dot{D}-{3\over 2}\,H^2\,\Omega_{m}\,D=0,\label{eqDevol}
\end{equation}
one of which is decaying and the other is growing with time.
For an Einstein de-Sitter (EdS) background (a universe with no curvature and with a critical matter density) the solutions read
\begin{equation}
D^{\rm EdS}_+(t)\propto t^{2/3},\ \ D^{\rm EdS}_-(t)\propto 1/t,
\end{equation}
that is $D^{\rm EdS}_+(t)$ is proportional to the expansion factor. 
This result gives the time scale of the growth of structure. This is what permits a direct comparison between the 
amplitude of the metric perturbations at recombination and the density perturbation in the local universe.  
Note that it implies that the potential, for the corresponding mode, is constant (see the Poisson equation). 

In the following we will later see how these results can be extended to other background evolution.

\subsection{The Green functions}

The previous results show that the linear density field can be written in general
\begin{equation}
\delta(\vx,t)=\delta_{+}(\vx)D_{+}(t)+\delta_{-}(\vx)D_{-}(t)
\end{equation}
and 
\begin{equation}
\theta(\vx,t)=-\frac{\dd}{\dd \log a}D_{+}\ \delta_{+}(\vx)-\frac{\dd}{\dd \log a}D_{-}\delta_{-}(\vx).
\end{equation}
The actual growing and decaying modes can then be obtained by inverting this system. For instance for an Einstein de Sitter background one gets
\begin{eqnarray}
\delta_{+}(\vx)D_{+}(t)&=&\frac{D_{+}(t)}{D_{+}(t_{0})}\left[\frac{3}{5}\delta(\vx,t_{0})-\frac{2}{5}\theta(\vx,t_{0})\right],\\
\delta_{-}(\vx)D_{-}(t)&=&\frac{D_{-}(t)}{D_{-}(t_{0})}\left[\frac{2}{5}\delta(\vx,t_{0})+\frac{2}{5}\theta(\vx,t_{0})\right],
\end{eqnarray}
and similar results for the velocity divergence.
Following \shortciteN{1998MNRAS.299.1097S}, this result can be encapsulated in a simple form after one introduces the 
 doublet  $\Psi_a(\vk,\tau)$,
\begin{equation}
\Psi_a(\vk,\tau) \equiv \Big( \delta(\vx,t),\ -\theta(\vx,t) \Big),
\label{2vectorEdS}
\end{equation}
where $a$ is an index whose value is either 1 (for the density component) or $2$ (for the velocity component).
The linear growth solution can now be written
\begin{equation}
\Psi_{a}(\vx,t)=g_{a}^{\ b}(t,t_{0})\Psi_{b}(\vx,t_{0})
\end{equation}
where $g_{a}^{b}$ is the Green function of the system. It is usually written with the following time variable,
\begin{equation}
\eta=\log D_{+}
\end{equation}
(not to be mistaken with the conformal time). For an Einstein de Sitter universe, we have explicitly,
\begin{equation}
g_{a}^{\ b}(\eta,\eta_{0})=\frac{e^{\eta-\eta_{0}}}{5}\left[
\begin{array}{cc}
3&2\\
3&2
\end{array}
\right]
+
\frac{e^{-\frac{3}{2}(\eta-\eta_{0})}}{5}\left[
\begin{array}{cc}
2&-2\\
-3&3
\end{array}
\right].\label{expressiongab}
\end{equation}
We will see in the following that, provided the doublet $\Psi_{a}$ is properly defined, this form
remains practically unchanged for any background.

\subsection{The general background case}
 
For a general background, it is fruitful to extent the definition of the doublet to,
\begin{equation}
\Psi_a(\vx,\eta) \equiv \Big( \delta(\vx,\eta),\ -\frac{1}{f_{+}}\theta(\vx,\eta) \Big),
\label{2vector}
\end{equation}
where 
\begin{equation}
f_{+}=\frac{\dd \log D_{+}}{\dd \log a}
\end{equation}
Defining $\htheta= -\theta(\vx,\eta)/{f_{+}}$ and for the time variable $\eta$, 
the linearized motion equations indeed read
\begin{eqnarray}
\frac{\partial}{\partial\eta}\delta(\vx,\eta)-\htheta(\vx,\eta)&=&0\\
\frac{\partial}{\partial\eta}\htheta(\vx,\eta)
\left(\frac{3}{2}\frac{\Omega_{m}}{f_{+}^{2}}-1\right)\htheta-\frac{3}{2}\frac{\Omega_{m}}{f_{+}^{2}}\delta(\vx,\eta)&=&0,
\end{eqnarray}
which can be rewritten as
\begin{equation}
\frac{\partial}{\partial\eta}\Psi_{a}(\vx,\eta)+\Omega_{a}^{\ b}(\eta)\Psi_{b}(\vx,\eta)=0,
\end{equation}
with
\begin{equation}
\Omega_{a}^{\ b}(\eta)=\left(
\begin{array}{ccc}
0&\ &-1\\
-\frac{3}{2}\frac{\Omega_{m}}{f_{+}^{2}}&\ &\frac{3}{2}\frac{\Omega_{m}}{f_{+}^{2}}-1
\end{array}
\right).
\label{Omegaabdef}
\end{equation}
In general, the formal solution of this system can be written in terms of a Green function $g_{a}^{\ b}(\eta,\eta_{0})$. The latter satisfies 
the differential equation
\begin{equation}
\frac{\partial}{\partial\eta}g_{a}^{\ b}(\eta,\eta_{0})+\Omega_{a}^{\ c}(\eta)\,g_{c}^{\ b}(\eta,\eta_{0})=0
\end{equation}
with the condition
\begin{equation}
g_{a}^{\ b}(\eta_{0},\eta_{0})=\delta_{a}^{\ b},
\label{gabbound}
\end{equation}
where $\delta_{a}^{\ b}$ is the identity matrix. It is to be noted that the Green
function can formally be written in terms of the peculiar solutions of the systems. For instance if one considers the growing and decaying
solution $u_{a}^{(+)}(\eta)$ and $u_{a}^{(-)}(\eta)$,  the Green function can be written
\begin{equation}
g_{a}^{\ b}(\eta,\eta_{0})=\sum_{\alpha=-,+}u_{a}^{(\alpha)}(\eta)c^{b}_{(\alpha)}(\eta_{0})
\label{genexpgab}
\end{equation}
where the constants $c^{b}_{(\alpha)}(\eta_{0})$ are set such that (\ref{gabbound}) is satisfied.

With the definition (\ref{2vector}) of $\Psi_{a}$, the growing and decaying modes are, to a (surprisingly) good approximation, given by
\begin{equation}
u_{a}^{(+)}=e^{\eta}\left(
\begin{array}{c}1\\1\end{array}
\right),\ \ u_{a}^{(-)}=e^{-\frac{3}{2}\eta}\left(
\begin{array}{c}1\\-3/2\end{array}
\right).
\end{equation}
This is due to the fact that $\Omega_{m}/f_{+}^{2}\approx 1$ in most regimes and models we consider.

As a result, in practice, we will always use the form (\ref{expressiongab}) for the Green function.

\subsection{The two-fluid case}

\begin{figure}
\centering
 \includegraphics[width=7cm]{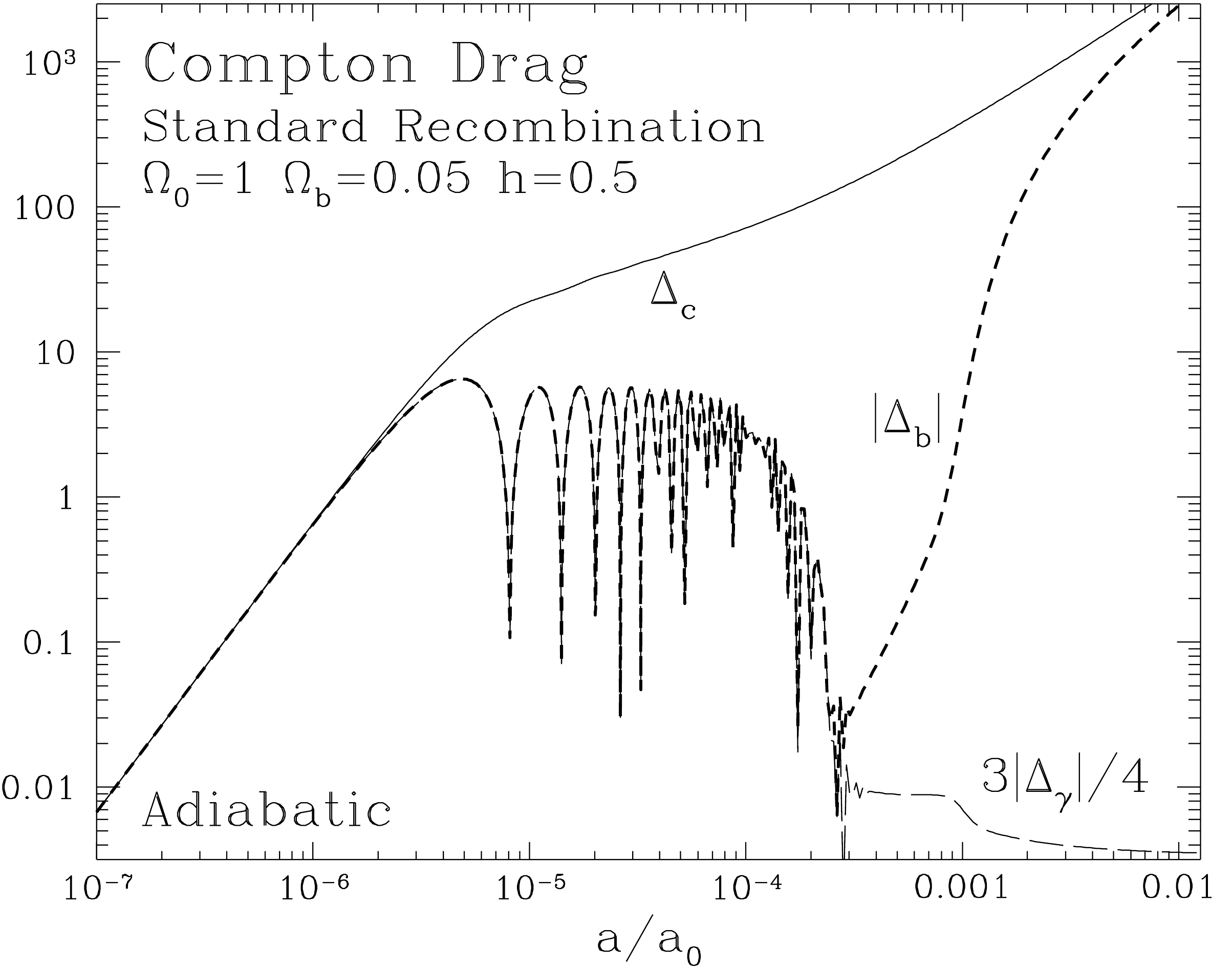}
\caption{Cosmic evolution of the generalized density contrast $\Delta_{i}$ for radiation, 
$\Delta_{\gamma}$, dark matter, $\Delta_c$, and baryons, $\Delta_b$. 
One can see that instabilities start earlier for the dark matter component than for the baryon component. After decoupling 
the two-fluid follows the Vlasov-Poisson system and the linear solution is the superposition of the adiabatic growing 
mode and iso-density modes. From \protect \shortciteN{1995astro.ph..8126H}.}
\label{drag}
\end{figure}

In this paragraph I review the formalism for PT calculations in the presence of multiple pressureless fluids.
The equations modeling multiple pressureless fluids and the resulting Green functions were first presented in \shortciteN{2010PhRvD..81b3524S}. 

We assume that the Universe is filled with pressureless fluids with only gravitational interactions.
For $N$ fluids, we  denote each fluid by a subscript $\bb$ ($\bb=1,\ldots, N$).
Then, for each fluid the continuity equation reads
\begin{equation}
\frac{\partial}{\partial t} \delta_{\bb}+\frac1a \left((1+\delta_{\bb})u_{\bb}^{i}\right)_{,i}=0\;,
\end{equation}
while the Euler equation reads
\begin{equation}
\frac{\partial}{\partial t}u_{\bb}^{i}+H u_{\bb}^{i}+\frac{1}{a}u_{\bb}^{j}u_{\bb,j}^{i}=-\frac{1}{a}\phi_{,i}\;.
\end{equation}
It is to be noted that the source term of the Poisson equation, $\Delta \phi=4\pi\,G\,a^{2}\rhob\,\deltam$,
is obviously the total mass density contrast defined by,
\begin{equation}
 \rhob\,\deltam=\sum_{I}\rho_{I}\delta_{I}.
\end{equation}
Note that this system gives exactly the same motion equations for the total mass density contrast as the single flow case. It
then admits the same linear solutions.

The linear behavior of the multi flow
equations is conveniently described by the introduction of the multiplet $\Psi_a$ ($a=1,\ldots, 2N$) 
\shortcite{2001NYASA.927...13S,2010PhRvD..81b3524S,2012PhRvD..85f3509B}, 
\begin{equation}
\Psi_{a}=\left(
\delta_{1},
\htheta_{1},
\delta_{2},
\htheta_{2},
\dots
\right)^T \;, \label{Psi_def_2}
\end{equation}
with $
\htheta_{\bb} \equiv - {\theta_{\bb}}/{f_{+}(t)}$. The growth rate $f_{+}$ is defined as the logarithmic change of the 
growth factor with the expansion, $f_{+} \equiv \dd \ln D_+/ \dd a$, with $\dd \eta \equiv \dd \ln D_{+}$ where 
$D_{+}$ is the (unchanged) growing linear growth rate of the total density contrast.
As before, the equations of motion can be recapped in the form,
\begin{equation}
\frac{\partial}{\partial \eta}\Psi_{a}(\vx,\eta)+\Omega_{a}^{\ b}(\eta)\Psi_{b}(\vx,\eta)=0,
\end{equation}
where the non-vanishing matrix elements of  $\Omega_{a}^{\ b}$ are given by
\begin{eqnarray}
\Omega_{(2\bb-1)}^{\ (2\bb)}=-1\;, \ \ 
\Omega_{(2\bb)}^{\ (2\bb)}=\frac{3}{2}\frac{\Omega_{\rm m}}{f_{+}^{2}}-1\;, \ \ 
\Omega_{(2\bb)}^{\ (2\bc-1)}=-\frac{3}{2}\frac{\Omega_{\rm m}}{f_{+}^{2}}\fw_{q} \;
\end{eqnarray}
and where we denote by $\fw_\bb \equiv \Omega_\bb /\Omega_{\rm m}$ the relative fraction of the fluid $\bb$.

This system has adiabatic solution, for which the fluids have all the same density contrast and velocity divergence, and
\emph{iso-density } modes. The latter are obtained under the constraint that the total density contrast vanishes, i.e.
\begin{equation}
\label{constraint_iso}
\deltam =\sum_{\bb}\ \fw_{\bb}\,\delta_{\bb}=0 \;.
\end{equation}
When we consider only 2 fluids, there are 2 such modes, one decaying and one constant in time. In the following we denote by ``$\plus$'' and ``$\minus$''  the 
growing and decaying adiabatic modes, respectively, and by ``$\iis$'' and ``$\cis$'' the decaying and constant iso-density modes.
Since under the constraint (\ref{constraint_iso}) the evolution equations decouple, the time dependence of these modes can be easily inferred. One solution is given 
by
\begin{eqnarray}
\rtheta_{\bb}^{(\iis)}(\eta)&\propto& \exp\left[-\int^{\eta}\dd\eta'\; \left(\frac{3}{2}\frac{\Omega_{\rm m}}{f_{+}^{2}}-1\right)\right] \;, \\
\delta_{\bb}^{(\iis)}(\eta)&=&\int^\eta \dd \eta'\; \rtheta_{\bb}^{(\iis)}(\eta') \;,
\end{eqnarray}
with
\begin{equation}
\sum_\bb \fw_\bb \rtheta_{\bb}^{(\iis)}=0 \;, \label{Theta_sum_i}
\end{equation}
which automatically ensures (\ref{constraint_iso}).
Note that, because  $\Omega_{\rm m}/f_{\plus}^2$ departs only weakly 
from the value taken in an EdS cosmology, i.e.~$\Omega_{\rm m}/f_{+}^2=1$, the iso-density modes are expected to depart very weakly from
\begin{equation}
\rtheta_{\bb}^{(\iis)}(\eta)\propto \exp(-\eta/2)\; ,\qquad \delta_{\bb}^{(\iis)}(\eta)=-2\rtheta_{\bb}^{(\iis)}(\eta) \;.\label{i_mode}
\end{equation}
A second set of isodensity modes is given by
\begin{equation}
\rtheta_{\bb}^{(\ci)}(\eta)=0\; ,\qquad \delta_{\bb}^{(\ci)}(\eta)=\hbox{Constant}\;, \label{ci_mode}
\end{equation}
under the condition that eqn~(\ref{constraint_iso}) is satisfied.

To be specific, let us concentrate now on the case of two fluids and assume an EdS background.
In this case, the growing and decaying solutions are then proportional, respectively, to
\begin{eqnarray}
u_{a}^{(\plus)} & =&\left(1,1,1,1\right)^T \;,\\
u_{a}^{(\minus)} & =&\left(1,-3/2,1,-3/2\right)^T \;.
\end{eqnarray}
Moreover, the isodensity modes are proportional to
\begin{eqnarray}
u_{a}^{(\iis)} & =&\left(-2 \fw_2 ,\fw_2,2\fw_1,-\fw_1\right)^T \;, \\
u_{a}^{(\ci)} & =&\left(\fw_2 ,0,-\fw_1,0\right)^T \;.
\end{eqnarray}

If necessary we are then in the position to write down the linear propagator $g_{a}^{\ b}(\eta,\eta_0)$, 
from the general form of eqn~(\ref{genexpgab}).

On Fig. \ref{drag} (taken from \citeNP{1995astro.ph..8126H}) one can see that just after recombination the two-fluid system, CDM and baryons, evolves  in a way that 
involves both adiabatic and iso-density modes. The reason is that, at small scales, the baryons 
remain tightly coupled to the photons so that after recombination, their density contrasts and velocity modes are both damped.

%%%%%%%%%%%%%%%%%%%%%%%%%%%%%%%%%%%%%%%%%%%%%%%%%%
\section{Modes and statistics}
\label{sec:Stats}

\subsection{The origin of stochasticity}

\begin{figure}
\centering
 \includegraphics[width=9cm]{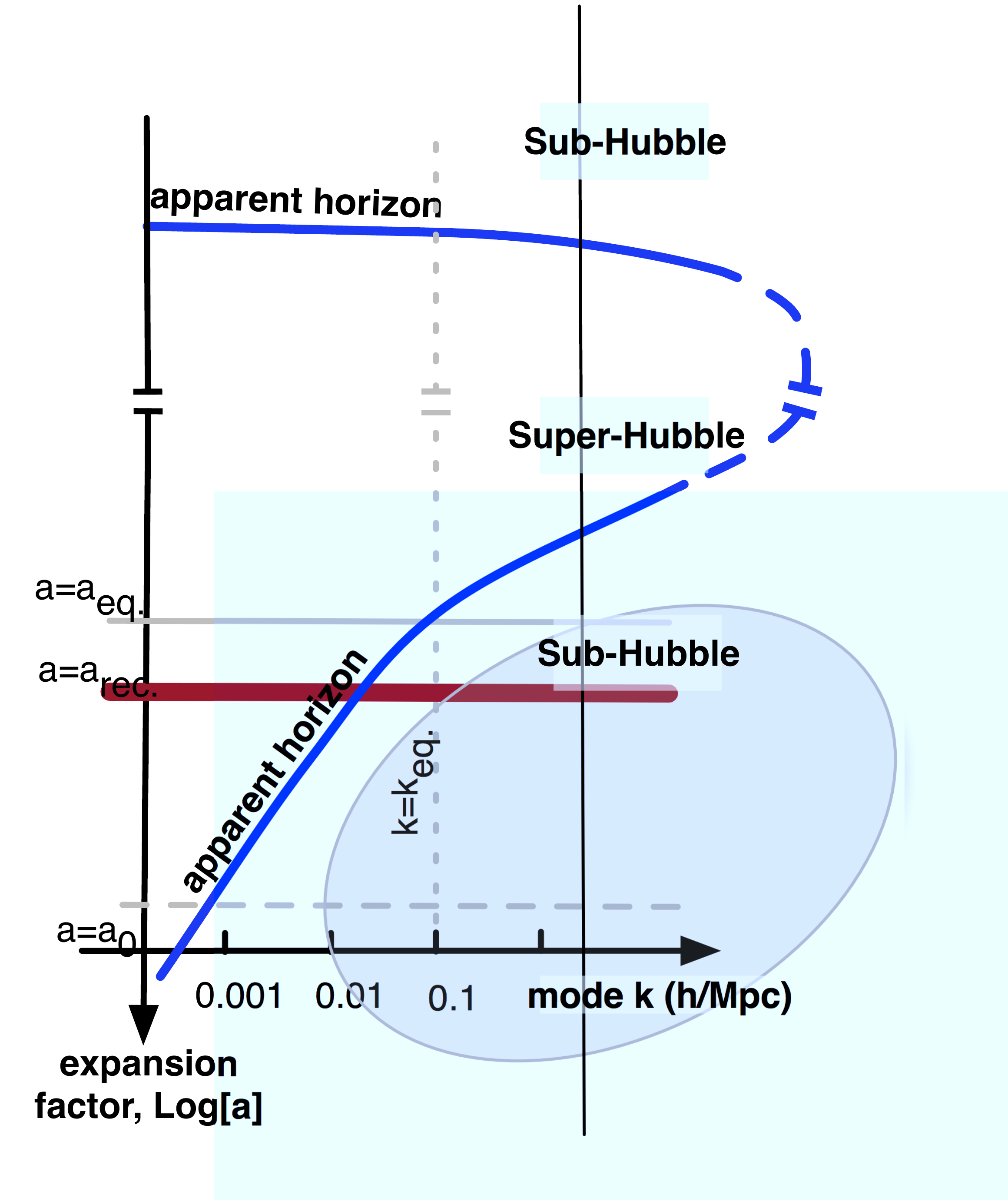}
\caption{Development  of linear metric perturbation across scale and time. This plot shows
the different stages of perturbation growth as a function of the wave modes. For short wave-length (high k),
(apparent) horizon crossing takes place before equality (time at which the radiation density and the matter density are
equal). The growth of structure in the CDM component in then stalled from horizon crossing to equality, and resumes
afterwards. Baryons participate in the growth of structure only after recombination. 
The motion equations we wrote are valid in the gray zone, at sub-Hubble scales and for pressureless fluids. 
In case of single field inflation, primordial fluid fluctuations are adiabatic: all fluid fluctuations that we see today
then originate from a single scalar field degree of freedom. The nonlinear growth of perturbation will take place
at low redshift and for large wave modes (bottom right of the figure).
}
\label{InflationRegime}
\end{figure}

In models of inflation the stochastic properties of the fields
originate from quantum fluctuations of a scalar field, the inflaton. It
is beyond the scope of this review to describe inflationary models in
any detail. We instead refer the reader to standard reviews for a
complete discussion~\cite{2005hep.th....3203L,2000cils.book.....L,1999PhR...314....1L}.  It is worth however
recalling that in such models (at least for the simplest single-field
models within the slow-roll approximation) all fluctuations originate
from scalar adiabatic perturbations.  During the inflationary phase
the energy density of the universe is dominated by the density stored
in the inflaton field. This field has quantum fluctuations that can be
decomposed in Fourier modes using the creation and annihilation
operators $a^{\dag}_{\vk}$ and $a_{\vk}$ for a wave mode $\vk$,
\begin{equation}
\de\varphi=
\int\dd^3 \vk\left[a_{\vk}\,\psi_k(t)\,\exp(\ii\vk.\vx)+
a^{\dag}_{\vk}\,\psi_k^*(t)\,\exp(-\ii\vk.\vx)\right].
\end{equation}
The operators obey the standard commutation relation,
\begin{equation}
[a_{\vk},a^{\dag}_{-\vk'}]=\de_D(\vk+\vk'),
\label{fb:comm}
\end{equation}
and the mode functions $\psi_k(t)$ are obtained from the Klein-Gordon
equation for $\varphi$ in an expanding Universe. We give here its
expression for a de-Sitter metric (i.e. when the spatial sections are
flat and $H$ is constant),

\begin{equation}
\psi_k(t)={H\over (2\,k)^{1/2}\,k}\, \left(\ii+{k\over
\,a\,H}\right)\,\exp\left[{\ii\,k\over a\,H}\right],
\label{fb:psik}
\end{equation}
where $a$ and $H$ are respectively the expansion factor and the
Hubble constant that are determined by the overall content of the
Universe through the Friedmann equations.

When the modes exit the Hubble radius, $k/(aH)\ll1$, one can see from
eqn~(\ref{fb:psik}) that the dominant mode reads, 
\begin{equation}
\varphi_{\vk}\approx {iH\over \sqrt{2}
k^{3/2}}\,\left(a_{\vk}+a_{-\vk}^{\dag}\right),\ \
\de\varphi=\int\dd^3\vk\,\varphi_{\vk}\,e^{\ii\,\vk.\vx}.
\label{fb:phikinfl}
\end{equation}
Therefore these modes are all proportional to
$a_{\vk}+a_{-\vk}^{\dag}$.  One important consequence of this is that
the quantum nature of the fluctuations has
disappeared~\shortcite{1985PhRvD..32.1899G,1998IJMPD...7..455K}: any combinations of
$\varphi_{\vk}$ commute with each other.  The field $\varphi$ can then
be seen as a classic stochastic field where  \emph{ensemble averages
identify with vacuum expectation values}, 
\begin{equation}
\langle
... \rangle\equiv\langle 0\vert...\vert0\rangle.  
\end{equation}

After the inflationary phase the modes re-enter the Hubble
radius. They leave imprints of their energy fluctuations in the
gravitational potential, the statistical properties of which can
therefore be deduced from eqns (\ref{fb:comm}, \ref{fb:phikinfl}).
All subsequent stochasticity that appears in the cosmic fields can
thus be expressed in terms of the random variable $\varphi_{\vk}$.
The linear theory calculation precisely tells us how each mode, in each fluid component, grows 
across time, i.e. it provides us with the so-called transfer functions, $T_{a}(\vk,\eta,\eta_{0})$, defined as
\begin{equation}
\delta_{a}(\vk,\eta)=T_{a}(\vk,\eta,\eta_{0})\de\varphi(\vk,\eta_{0})
\label{transferdef}
\end{equation}
where $\eta_{0}$ is a time which corresponds to an arbitrarily early time.

\subsection{Statistical homogeneity and isotropy}

In the following the density contrast will be decomposed in Fourier modes that, for a flat universe, are defined such as
\begin{equation}
\de(\vx)=\int\frac{\dd^3\vk}{(2\pi)^{3/2}}\,\de(\vk)\,\exp(\ii\vk\cdot \vx)
\label{dek}
\end{equation}
or equivalently
\begin{equation}
\de(\vk)=\int\frac{\dd^3\vx}{(2\pi)^{3/2}}\,\de(\vx)\,\exp(-\ii\vk\cdot \vx).
\label{dek2}
\end{equation}
The observable quantities of interest are actually the statistical properties of such a field, whether it is represented in real space or in Fourier space. The 
\emph{Cosmological Principle}, e.g. that the assumption that the Universe is statically isotropic and homogeneous, implies that real space correlators are 
homogeneous and isotropic which for instance implies that $\mg\de(\vx)\de(\vx+\vr)\md$ is a function of the separation $r$ only. This defines the two-point correlation 
function,
\begin{equation}
\xi(r)=\mg\de(\vx)\de(\vx+\vr)\md.
\label{eq:xidef}
\end{equation}
In Fourier space, the two point correlator of the Fourier modes then takes the form,
\begin{eqnarray}
\mg\de(\vk)\de(\vk')\md&=&\int \frac{\dd^3\vx}{(2\pi)^{3/2}}
\frac{\dd^3\vr}{(2\pi)^{3/2}}\,\xi(r)\,
\exp[-\ii(\vk+\vk')\cdot \vx-\ii\vk'\cdot\vr]\nonumber\\
&=&\Dirac(\vk+\vk')\int {\dd^3\vr}\,
\xi(r)\, \exp(\ii \vk\cdot \vr)
\nonumber\\
&\equiv&\Dirac(\vk+\vk')\,P(k),
\label{pk}
\end{eqnarray}
where $P(k)$ is the \emph{power spectrum} of the density field, e.g. the cross-correlation matrix is symmetric in Fourier space. 

All these relations apply to the observed fields. However in case of the primordial fluctuations, the field $\de\varphi(\vk)$ corresponds to a
free field from a quantum mechanical point of view. That makes it eventually a Gaussian classical field. As such it obeys the Wick
theorem. The latter tells us that higher order correlators can then be entirely constructed from the power spectrum (from pair associations) 
through the relations,
\begin{eqnarray}
\mg\de\varphi(\vk_1)\dots\de\varphi(\vk_{2p+1})\md&=&0\label{fb:Wick1}\\
\mg\de\varphi(\vk_1)\dots\de\varphi(\vk_{2p})\md&=&\!\!\!
\sum_{\rm pair\  associations}\ \ \prod_{p\ \rm pairs\ (i,j)}\!\!
\mg\de\varphi(\vk_i)\de\varphi(\vk_j)\md. \label{fb:Wick2}
\end{eqnarray}
These relations apply as well to any linear combinations of the primordial field, and therefore to any field computed in the linear regime.

\subsection{Moments and cumulants}

In the nonlinear regime however, fields also exhibit higher order 
non-trivial correlation functions that cannot be reconstructed from the two-point order
correlators. They are defined as the \emph{connected} part (denoted with
subscript $c$) of the joint ensemble average of fields in an
arbitrarily number of locations. Formally, for the density field, it reads,
\begin{eqnarray}
\langle \de(\vx_1),\dots,\de(\vx_N)\rangle_c
&=&\langle\de(\vx_1),\dots,\de(\vx_N)\rangle-\nonumber\\
&&\ \ \ -
\sum_{\mS\in {\mP}\left(\{\vx_1,\dots,\vx_n\}\right)}
\prod_{s_i\in\mS}\ 
\langle\de(\vx_{s_i(1)}),\dots, \de(\vx_{s_i(\#s_i)})\rangle,
\label{fb:xindef}
\end{eqnarray}
where the sum is made over the proper partitions (any partition except
the set itself) of $\{\vx_1,\dots,\vx_N\}$ and $s_i$ is thus a subset
of $\{\vx_1,\dots,\vx_N\}$ contained in partition $\mS$.  When the
average of $\de(\vx)$ is defined as zero, only partitions that contain
no singlets contribute.

\begin{figure}
\centering 
\includegraphics[width=.6\textwidth]{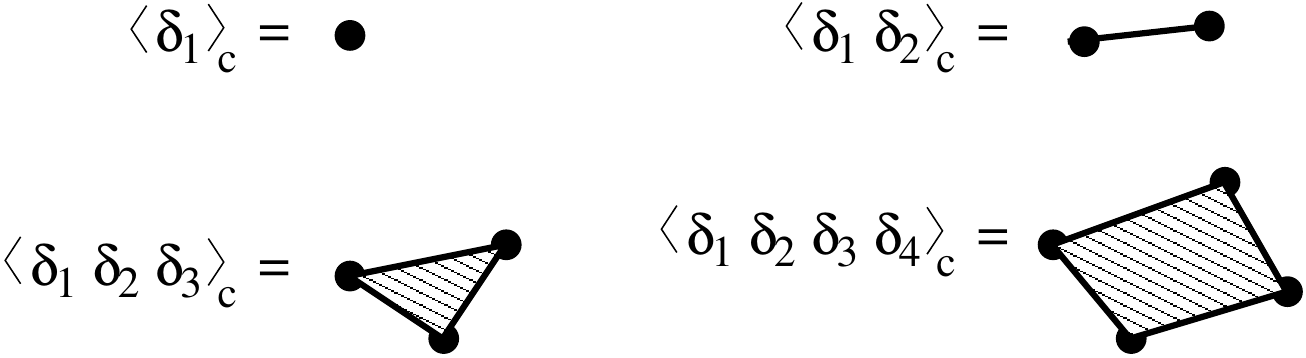}
%\vspace{4 cm}
%\special{hscale=80 vscale=80 voffset=0 hoffset=30 psfile=fig1.eps}
\caption{Representation of the connected part of the moments.}
\label{fb:conn}
\end{figure}

\begin{figure}
\centering 
\includegraphics[width=.8\textwidth]{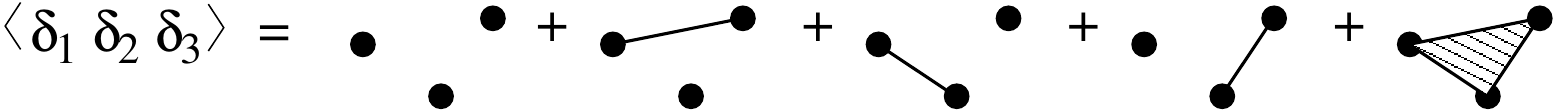}
\caption{Writing of the three-point moment in terms of connected parts.}
\label{fb:3pts}
\end{figure}

The decomposition in connected and non-connected parts can be easily
visualized. It means that any ensemble average can be decomposed in
a product of connected parts. They are defined for instance in
Fig.~\ref{fb:conn}. The tree-point moment is ``written'' in
Fig.~\ref{fb:3pts}.
Because of
homogeneity of space $\langle\de(\vk_1)\dots\de(\vk_N)\rangle_c$ is
always proportional to $\de_D(\vk_1+\dots+\vk_N)$. Then we
can define $P_N(\vk_1,\dots,\vk_N)$ with
\begin{equation}
\langle\de(\vk_1)\dots\de(\vk_N)\rangle_c=
\de_D(\vk_1+\dots+\vk_N)\,P_N(\vk_1,\dots,\vk_N).
\label{fb:Pndef}
\end{equation}
One case of particular interest is for $n=3$, the bispectrum, which is usually denoted by
$B(\vk_1,\vk_2,\vk_3)$. Note that it depends on 2 wave modes only, and it depends on 3 independent variables
characterizing the triangle formed by the 3 wave modes (for instance 2 lengths and 1 angle). 

\subsection{Moment and cumulant generating functions}

It is convenient to define a function from which all moments can be
generated, namely the {\em moment generating function}. It can be defined\footnote{It is to be noted however that the existence of moments -- which itself is not guaranteed for any stochastic process --
does not ensure the existence of their generating function as the series defined in (\ref{fb:mMdef}) can have a vanishing 
converging radius. Such a case is encountered for a lognormal distribution for instance and it implies that 
the moments of such a stochastic process do not uniquely define the probability distribution function,
see \shortciteN{Stoyanov87} for instance for details.} for any number of
random variables. Here we give its definition for the local density field.
It is defined by
\begin{equation}
\mM(t)\equiv\sum_{p=0}^{\infty}{\mom{p}\over p!}t^p=\langle \exp(t\de)\rangle.
\label{fb:mMdef}
\end{equation}
The moments can obviously obtained by subsequent derivatives of this
function at the origin $t=0$. A cumulant generating function can
similarly be defined by
\begin{equation}
\mC(t)\equiv\sum_{p=2}^{\infty}{\cum{p}\over p!}t^p.
\end{equation}
A fundamental result is that the cumulant generating function is given
by the logarithm of the moment generation function (see e.g. appendix
D in~\shortciteN{BDFN92} for a proof)
\begin{equation}
\mM(t)=\exp[\mC(t)].
\label{momtocum}
\end{equation}
In case of a Gaussian probability distribution function, this is straightforward to check since
$\langle \exp(t\de)\rangle=\exp(\sigma^2t^2/2)$.

%%%%%%%%%%%%%%%%%%%%%%%%%%%%%%%%%%%%%%%%%%%%%%
\section{The nonlinear equations}
\label{sec:nonlinear}
\subsection{A field representation of the nonlinear motion equations}

We now move to a full representation of the equations of motion, including the nonlinear terms that have
been neglected so far. For that it is more convenient to go into the Fourier modes. More specifically we have,
\begin{eqnarray}
\left[\delta(\vx)u_{i}(\vx)\right]_{,i}&\!\!=\!\!&
\int\frac{\dd^{3}\vk_{1}}{(2\pi)^{3/2}}\frac{\dd^{3}\vk_{2}}{(2\pi)^{3/2}}\,
\delta(\vk_{1})\theta(\vk_{2})\frac{\vk_{2}.(\vk_{1}+\vk_{2})}{k_{2}^{2}}\,\exp(\ii(\vk_{1}+\vk_{2}).\vx),\\
\left[u_{j}(\vx)u_{i,j}(\vx)\right]_{,i}&\!\!=\!\!&
\int\frac{\dd^{3}\vk_{1}}{(2\pi)^{3/2}}\frac{\dd^{3}\vk_{2}}{(2\pi)^{3/2}}\,
\theta(\vk_{1})\theta(\vk_{2})\,\frac{\vk_{1}\cdot\vk_{2}\,\vk_{1}\cdot(\vk_{1}+\vk_{2})}{k_{1}^{2}k_{2}^{2}}
\,\exp(\ii(\vk_{1}+\vk_{2}).\vx).
\end{eqnarray}
Plugging these expressions and taking the Fourier transform of these terms, one eventually gets,
\shortcite{2002PhR...367....1B} 
\begin{eqnarray}
\frac{\partial}{\partial \eta} \Psi_a(\vk,\eta) + \Omega_{a}^{\ b}(\eta) \Psi_b(\vk,\eta) &=& 
\gamma_{a}^{\ bc}(\vk_1,\vk_2) \ \Psi_b(\vk_1,\eta) \ \Psi_c(\vk_2,\eta),
\label{FullEoM}
\end{eqnarray}
where $ \Omega_{a}^{\ b}(\eta) $ is defined in eqn (\ref{Omegaabdef}) and where (and that will be the case
henceforth) we use the convention that  repeated
Fourier arguments are integrated over and the Einstein convention on repeated indices, and 
where the \emph{symmetrized vertex} matrix $\gamma_{a}^{\,bc}$ describes the non linear 
interactions between different Fourier modes. Its components are given by
\begin{eqnarray}
\gamma_{2}^{\ 22}(\vk_1,\vk_2)&=&\Dirac(\vk-\vk_1-\vk_2) \ {|\vk_1+\vk_2|^2 (\vk_1
\cdot\vk_2 )\over{2 k_1^2 k_2^2}}, \nonumber \\
\gamma_{1}^{\ 21}(\vk_1,\vk_2)&=&\Dirac(\vk-\vk_1-\vk_2) \  {(\vk_1+\vk_2) \cdot
\vk_1\over{2 k_1^2}},
\label{vertexdefinition}
\end{eqnarray}
$\gamma_{a}^{\ bc}(\vk_1,\vk_2)=\gamma_{a}^{\ cb}(\vk_2,\vk_1)$, and $\gamma=0$ 
otherwise, where $\Dirac$ denotes the Dirac distribution function. The matrix  $\gamma_{a}^{\ bc}$ is independent 
on time (and on the background evolution) and encodes all the
non-linear couplings of the system.

One can then take advantage of the knowledge of the Green function of this system to write a formal
solution of eqn (\ref{FullEoM}) \cite{1998MNRAS.299.1097S,2001NYASA.927...13S,2006PhRvD..73f3519C}, as
\begin{eqnarray}
\Psi_a(\vk,\eta) &=& g_{a}^{\ b}(\eta) \ \Psi_b(\vk,\eta_{0}) + \nonumber\\
&+&\int_{\eta_{0}}^{\eta}  {\dd \eta'} \ g_{a}^{\ b}(\eta,\eta') \ 
\gamma_{b}^{\ cd}(\vk_1,\vk_2) \Psi_c(\vk_1,\eta') \Psi_d(\vk_2,\eta'),
\label{eomi}
\end{eqnarray}
where  $\Psi_a(\vk,\eta_0)$ denotes the initial conditions. 

In the following calculations we will be using the value of the $\Omega_{a}^{\ b}$ matrix to be that of the Einstein de Sitter background
that is effectively assuming that $D_{-}$ scales like $D_{+}^{-3/2}$. This is known to be a very good approximation even in the context of a $\Lambda-$CDM universe 
(see for instance \shortciteN{2012arXiv1207.1465C} for an explicit investigation of the consequences of this approximation).

\subsection{Diagrammatic representations}

\begin{figure}[ht] %  figure placement: here, top, bottom, or page
   \centering
 \includegraphics[width=7.5cm]{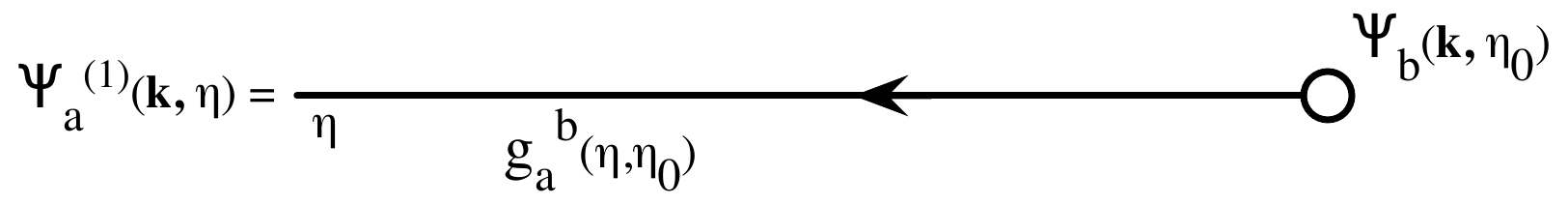} 
   \caption{Diagrammatic representation of the linear propagator. $\Psi_{b}$ represents the initial conditions and $g_{a}^{\ b}$ is the time dependent propagator. This 
diagram value is the linear solution of the motion equation.}
   \label{PsiLinear}
\end{figure} 

\begin{figure}[ht] %  figure placement: here, top, bottom, or page
   \centering
 \includegraphics[width=7cm]{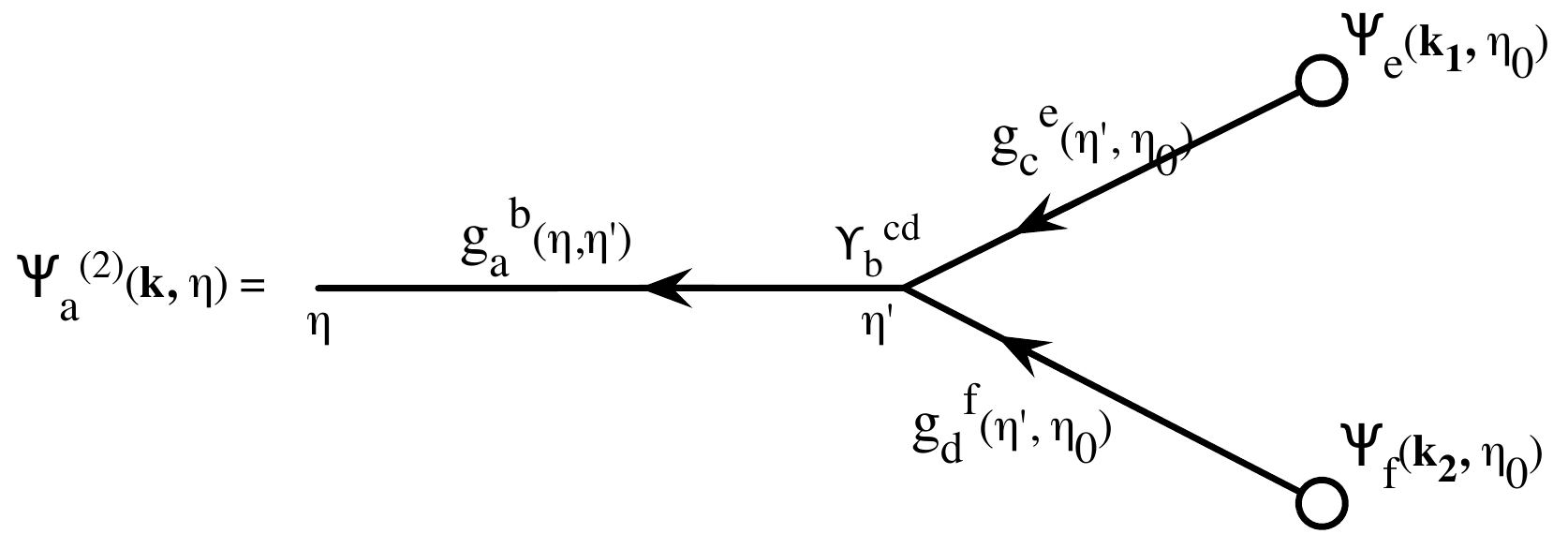} 
   \caption{Diagrammatic representation of the fields at second order. This diagram value is given by eqn (\ref{eomi}) when one replaces $\Psi_{c}$ and $\Psi_{d}$ in 
the second term of the right hand side by their linear expressions. In the diagram, each time one encounters a vertex, a time integration and a Dirac function in the wave modes is implicitly assumed.}
   \label{Psi2Order}
\end{figure} 

\begin{figure}[ht] %  figure placement: here, top, bottom, or page
   \centering
 \includegraphics[width=7cm]{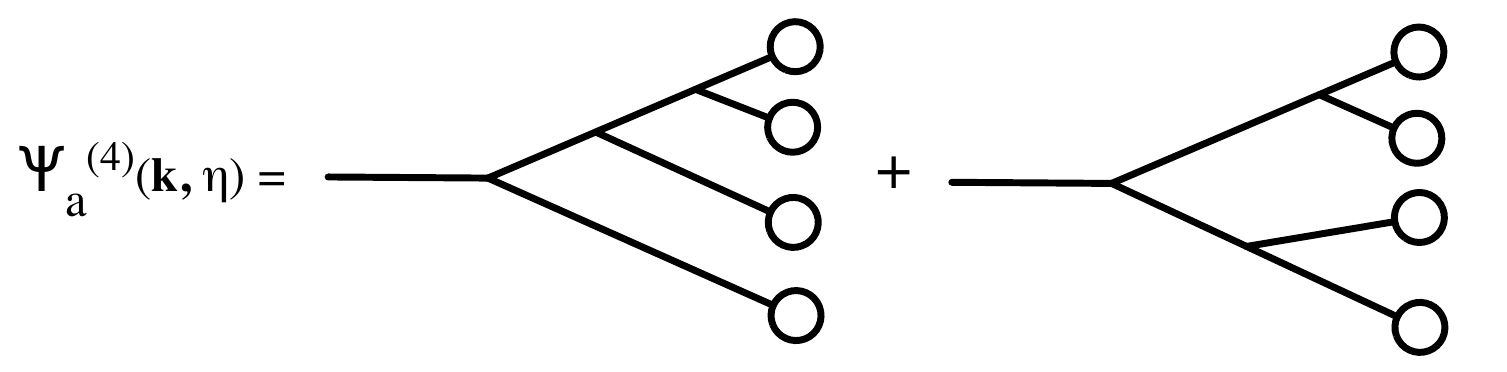} 
   \caption{Diagrammatic representation of the fields at fourth order. Three different diagrams are found to contribute.}
   \label{Psi4Order}
\end{figure} 

One of a nice feature of eqn (\ref{eomi}) is that it admits simple diagrammatic representations in a way very similar to Feynman diagrams.

A detailed description of the procedure to draw diagrams and compute their values can be found in \citeN{2006PhRvD..73f3519C}, we can briefly summarize these 
rules here as follows\footnote{An alternative representation can be found in \shortciteN{2007JCAP...06..026M}.}. In such representations, the open circles represent the initial conditions $\Psi_b(\vk,\eta_{0})$, where $b=1$ ($b=2$) corresponds to the 
density (velocity divergence) field, and the line emerging from it carries a wavenumber $\vk$. Lines are time-oriented (with time direction represented by an arrow) 
and have different indices at both ends, say $a$ and $b$. Each line represents linear evolution described by the propagator $g_{a}^{\,b}(\etaf-\etai)$ from time $\etai
$ to time $\etaf$.  The simplest diagram is presented on Fig. \ref{PsiLinear}. It represents the linear growth solution.

Each nonlinear interaction between modes is represented by a vertex, which due to quadratic nonlinearities in the equations of motion is the convergence point of 
necessarily two incoming lines, with wavenumber say $\vk_{1}$ and $\vk_{2}$, and one outgoing line with wavenumber $\vk=\vk_1+\vk_2$. Each vertex in a 
diagram then represents the matrix $\gamma_{a}^{\,bc}(\vk_{1},\vk_{2})$. It is further understood  that internal indices are summed over and interaction times are 
integrated over the full interval $[0,\etaf]$ as for instance in Fig. \ref{Psi2Order} where we present the diagrammatic expression
of the second order expression of the fields. It is obtained after two  modes computed at linear order, $\vk_{1}$  and $\vk_{2}$ interact together. Note that after the 
interaction the wave mode which is created is not necessarily in the growing modes. As a result, on this diagram
both modes are propagating along the left line.
This construction can obviously be extended to higher order in perturbation theory. On Fig. \ref{Psi4Order} the diagrams contributing to fourth order are shown.

\begin{figure}[ht] %  figure placement: here, top, bottom, or page
   \centering
 \includegraphics[width=6cm]{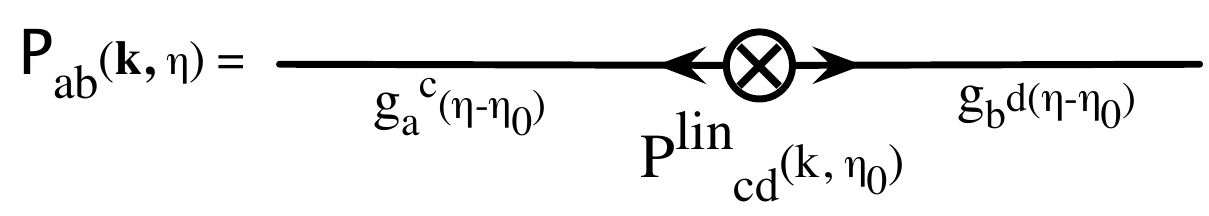} 
   \caption{Diagrammatic representation of the power spectrum at linear order. The symbol $\otimes$ represents the linear power spectrum in the (adiabatic) growing 
mode. }
   \label{LinPowerSpec}
\end{figure}

Relevant statistical quantities are obtained however once ensemble average are taken. Assuming Gaussian initial conditions, 
one then can
apply the Wick theorem to all the factors representing the initial field values 
that appear in diagrams (or product of diagrams) of interest. In practice, at least for
these notes, the diagrams will all be computed assuming the initial conditions correspond effectively to the  adiabatic linear growing mode.
The simplest of such diagram is presented on Fig. \ref{LinPowerSpec}. It corresponds to the ensemble average of 
$\langle 
\Psi_{a}(\vk,\eta)
\Psi_{b}(\vk,\eta)
\rangle$
and it makes intervene the linear power spectrum represented by $\otimes$. 
The previous construction can obviously be extended to any number of fields. The next diagrams will inevitably make intervene loops
(in their diagram representation). One idea we will pursue here is to take advantage of such expansions to explore 
the density spectrum at 1-loop order and 2-loop order, also called at Next-to Leading Order and Next to Next to Leading Order (respectively NLO
and NNLO).

\begin{figure}[ht] %  figure placement: here, top, bottom, or page
   \centering
 \includegraphics[width=10cm]{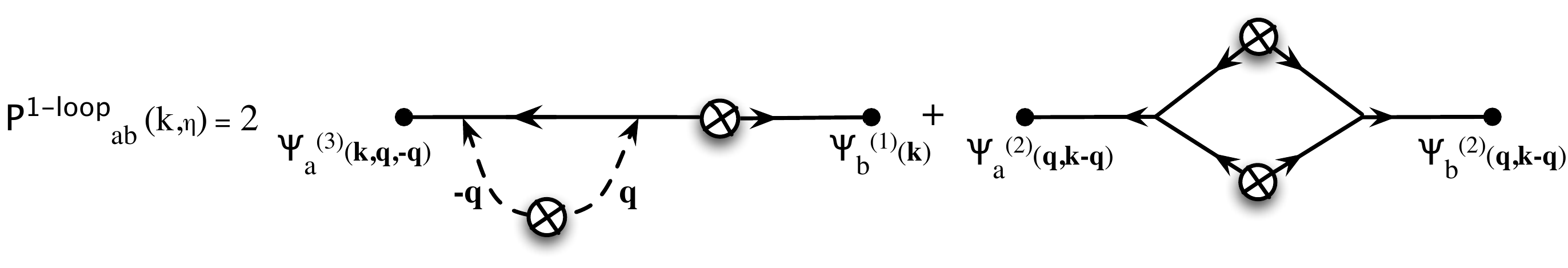} 
   \caption{Diagrammatic representation of the 2 terms contributing to the power spectrum at one-loop order.}
   \label{OneLoopPowerSpec}
\end{figure}

The 2 diagrams contributing to the power spectrum at NLO are shown on Fig. \ref{OneLoopPowerSpec}. The first calculation of such
contributions were done in the 90's \shortcite{1992PhRvD..46..585M,1996ApJ...456...43J,1996ApJS..105...37S}.

\subsection{Scaling of solutions}
\label{fb:subsecscaling}

It is interesting to compute the way subsequent orders in perturbation theory scale with the linear solution. As the vertices and the time
integrations are both dimensionless operations, one can easily show that the $p-$th order expression of the density field is 
of the order of the power $p$ of the linear density field. In other words, there are kernels functions $F_{a}^{(n)}$ such that
\begin{eqnarray}
\Psi^{(p)}_{a}(\vk,\eta)&=&\int\frac{\dd\vk_{1}}{(2\pi)^{3/2}}\dots\frac{\dd\vk_{p}}{(2\pi)^{3/2}}\Dirac{\left(\vk-\vk_{1\dots p}\right)}\nonumber\\
&&\times\mF^{(p)}_{a}(\vk_{1},\dots,\vk_{p};\eta)\delta_{+}(\vk_{1},\eta)\dots\delta_{+}(\vk_{p},\eta)
\end{eqnarray}
where $\delta_{+}(\vk_{i},\eta)$ is the linear growing mode for wave modes $\vk_{i}$,  $\vk_{1\dots p}=\vk_{1}+\dots+\vk_{p}$ and where
$\mF^{(p)}_{a}(\vk_{1},\dots,\vk_{p};\eta)$ are dimensionless functions of the wave modes and are a priori time dependent.
For an Einstein-de Sitter background, the functions $\mF^{(p)}_{a}(\vk_{1},\dots,\vk_{p};\eta)$ are actually time independent
and in general depends only very weakly on time (and henceforth on the cosmological parameters.)
The functions $\mF^{(p)}_{a}$ are usually noted  $F_{p}$ and $G_{p}$ for respectively $a=1$ and $a=2$. For instance it is easy to show that
\begin{equation}
F_{2}(\vk_{1},\vk_{2})=\frac{5}{7}+\frac{1}{2}\frac{\vk_{1}\cdot\vk_{2}}{k_{1}^{2}}+\frac{1}{2}\frac{\vk_{1}\cdot\vk_{2}}{k_{2}^{2}}
+\frac{2}{7}\frac{(\vk_{1}\cdot\vk_{2})^{2}}{k_{1}^{2}\,k_{2}^{2}}
\end{equation}
for an Einstein-de Sitter universe. For an arbitrary background the coefficient $5/7$ and $2/7$ are slightly 
altered but only very weakly \shortcite{1992ApJ...394L...5B}.

It can be noted that this kernel is very general and is actually directly observable. Indeed for Gaussian initial conditions the first non-vanishing
contribution to the bi-spectrum is obtained when one, and only one,  factor is written at second order in the initial field,
\begin{equation}
\langle\delta(\vk_{1})\delta(\vk_{2})\delta(\vk_{3})\rangle_{c}=
\langle\delta^{(1)}(\vk_{1})\delta^{(1)}(\vk_{2})\delta^{(2)}(\vk_{3})\rangle_{c}+\sym
\end{equation}
and it is easy to show that it eventually reads
\begin{equation}
\langle\delta(\vk_{1})\delta(\vk_{2})\delta(\vk_{3})\rangle_{c}=\Dirac(\vk_{1}+\vk_{2}+\vk_{3})
\left[2\,F_{2}(\vk_{1},\vk_{2})P^{\lin}(k_{1})P^{\lin}(k_{2})+\sym
\right]
\end{equation}
where $\sym$ refers to 2 extra terms obtained by circular changes of the indices.
The important consequence of this form is that the bispectrum therefore scales like the square of the power spectrum.
In particular the reduced bispectrum defined as
\begin{equation}
Q(\vk_{1},\vk_{2},\vk_{3})=\frac{B(\vk_{1},\vk_{2},\vk_{3})}{P(k_{1})P(k_{2})+P(k_{2})P(k_{3})+P(k_{3})P(k_{1})}
\end{equation}
is expected to have a time independent amplitude at early time.

More generally, the connected p-point correlators at lowest order in perturbation theory scale like the power $p-1$ of the 
2-point correlators \shortcite{1984ApJ...279..499F,1986ApJ...311....6G}: it comes from the fact that 
in order to connect $p$ points using the Wick theorem one needs at  least $p-1$ lines
connecting a product of $2(p-1)$ fields taken at linear order. For instance the local $p-$order cumulant of the local density 
contrast $\langle\delta^{p}\rangle_{c}$ scales like,
\begin{equation}
\langle\delta^{p}\rangle_{c}\sim \langle \delta^{2}\rangle^{p-1}.
\end{equation}
In the last sections of these notes, more detailed presentation of these relations will be given.

Other consequences of these scaling results concern the p-loop corrections to the power spectrum. Indeed one expects 
to have
\begin{equation}
P^{\rm p-loop}(k)\sim P^{\lin}(k)\ \left[\int \frac{\dd q}{q}\ q^{3} P(q)\right]^{p}.
\end{equation}
At least that would be the case if the linear power spectrum peaked at wave modes about $k$. This is not necessarily the case.
In the following we explore the mode coupling structure, i.e. how modes $q$ are contributing to the corrective terms to
the power spectrum depending on whether they are much smaller (infrared domain) or much larger (ultra-violet domain) than $k$.

\subsection{Time flow equations}
\label{TRG}

Although in this presentation we will focus on the diagrammatic representation of the motion equations, and their integral form, there exists an 
alternative set of differential equations that gives the time dependence of the multi-point spectra. These equations form 
a hierarchy and we will give here the first two.
The evolution equation (\ref{FullEoM}) indeed allows to compute the time derivative of products such as  $\Psi_{a}(\vk,\eta)\Psi_{b}(\vk',\eta)$
or  $\Psi_{a}(\vk_{1},\eta)\Psi_{b}(\vk_{2},\eta)\Psi_{c}(\vk_{3},\eta)$.
After taking their ensemble averages, it leads respectively to the following equations (see \shortciteNP{2008JCAP...10..036P}),
\begin{eqnarray}
\label{TRG:power}
&&\frac{\partial}{\partial\eta}\,P_{ab}({\bf k},\eta) = - \Omega_{a}^{\,c} ({\bf k}\,,\eta)P_{cb}({\bf k},\eta)  - \Omega_{b}^{\,c} ({\bf k},\eta)P_{ac}({\bf k}\,,\eta) \\
&&+\int d^3 q\, \left[ \gamma_{a}^{\,cd}({\bf -q},{\bf q-k})\,B_{bcd}({\bf k},\,{\bf -q},\,{\bf q-k};\eta)
%\right.\nonumber\\
%&&\left. 
+ \gamma_{b}^{\,cd}({\bf -q},{\bf q-k})\,B_{acd}({\bf k},\,{\bf -q},\,{\bf q-k};\eta)\right],\nonumber
\end{eqnarray}
and
\begin{eqnarray}
\label{TRG:bispec}
&&\frac{\partial}{\partial\eta}
B_{abc}({\bf k},\,{\bf -q},\,{\bf q-k};\eta) =  - \Omega_{a}^{\,d} ({\bf k}\,,\eta)B_{dbc}({\bf k},\,{\bf -q},\,{\bf q-k};\eta)\\
&&- \Omega_{b}^{\,d} ({\bf -q}\,,\eta)B_{adc}({\bf k},\,{\bf -q},\,{\bf q-k};\eta)
%\nonumber\\
%&&\qquad\qquad\qquad\qquad\qquad\quad
- \Omega_{c}^{\,d} ({\bf q-k}\,,\eta)B_{abd}({\bf k},\,{\bf -q},\,{\bf q-k};\eta)\nonumber\\
&&+ 2 \left[ \gamma_{a}^{\,de}({\bf -q},{\bf q-k}) P_{db}({\bf q}\,,\eta)P_{ec}({\bf k-q},\eta)
%\right.\nonumber\\
%&&\qquad\qquad\qquad\qquad\quad 
+\gamma_{b}^{\,de}({\bf q-k},{\bf k}) P_{dc}({\bf k-q}\,,\eta)P_{ea}({\bf k},\eta)
\right.\nonumber\\
&&\quad\quad
+\left. \gamma_{c}^{\,de}({\bf k},\,{\bf -q}) P_{da}({\bf k}\,,\eta)P_{eb}({\bf q},\eta)\right],\nonumber
\end{eqnarray}
where in the latter equation the connected parts of the four-point correlators have been dropped. It can be easily checked that 
taking the power spectrum at linear order in the right hand side of eqn (\ref{TRG:bispec}) gives back the standard perturbation theory 
results at one-loop order. Those results can then be used as an alternative scheme to obtain Perturbation Theory results without relying
neither on the explicit form of the Green function nor on diagrammatic expansions. It is then of interest for systems that are richer
than pure dark matter systems (for instance with massive neutrinos as in \shortciteNP{2009JCAP...06..017L}).  

This approach has been advocated though as an alternative approach to standard perturbation theory. Indeed in eqn (\ref{TRG:bispec})
if one uses the non-linear power spectrum then eqns (\ref{TRG:power}) and (\ref{TRG:bispec}) form a closed system of equation which 
provides a NLO calculation of the power spectrum which is distinct from standard Perturbation Theory result. 
In principle such expansions can be pursued to higher order
introducing a system which involves also the tri-spectrum, etc.

%%%%%%%%%%%%%%%%%%%%%%%%%%
\section{The infrared domain and the eikonal approximation}
\label{sec:IR}

On of the reason for exploring the mode coupling structure is that the vertices $\gamma_{a}^{\ bc}(\vk_{1},\vk_{2})$ can be large 
when the ratio $k_{1}/k_{2}$ (or its inverse) gets large. We will see that it corresponds to contributions coming from the infrared (IR) 
domain.

\subsection{The IR behavior for the 1-loop corrections to the power spectrum}

\begin{figure}[ht] %  figure placement: here, top, bottom, or page
   \centering
 \includegraphics[width=10cm]{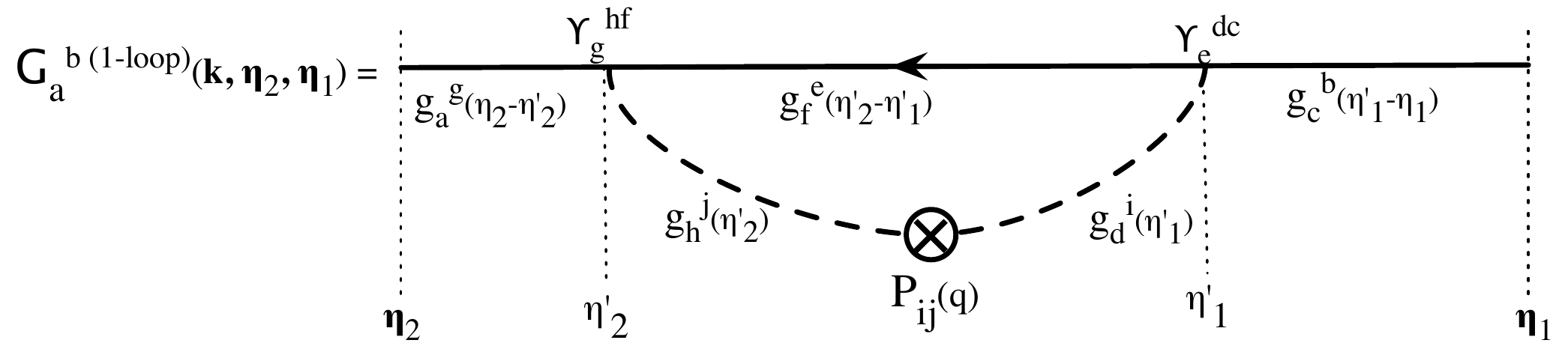} 
   \caption{The one-loop correction diagram to the propagator.}
   \label{Gab1loop}
\end{figure}

Let us first consider the one-loop correction to the power spectrum. We want to compute the amplitude of the 
one-loop correction when the wave mode that circulates in the loop $\vq$ is much smaller than the mode of interest, $\vk$.

Let us consider more specially the left diagram appearing on Fig. \ref{OneLoopPowerSpec} which is partly reproduced in
more detail on Fig. \ref{Gab1loop}. In this approximation the incoming modes from the loop are either $\vq$ or $-\vq$. As
one assumes $q\ll k$, the modes on the horizontal line is always $\vk$. We can then compute the expression of each of
the vertices. The one on the right (at time $\eta'_{1}$) then reads,
\begin{equation}
\gamma_{e}^{\ dc}\approx \frac{1}{2}\frac{\vq.\vk}{q^{2}}\ \delta_{e}^{\,c}\ \delta_{2}^{\,d}
\,\Dirac(\vk-\vk_{1}).
\end{equation}
Anticipating the next paragraph we can call this the eikonal limit of the vertex. We note two key properties: it is diagonal in the
$c$ and $e$ indices (of the main line) and it selects only the second component of the incoming waves.
The value of the other vertex can be similarly evaluated,
\begin{equation}
\gamma_{g}^{\,hf}\approx -\frac{1}{2}\frac{\vq.(\vk+\vq)}{q^{2}}\ \delta_{g}^{\,f}\ \delta_{2}^{\,h}
\,\Dirac(\vk-\vk_{1})\approx
-\frac{1}{2}\frac{\vq.\vk}{q^{2}}\ \delta_{g}^{\,f}\ \delta_{2}^{\,h}\,\Dirac(\vk-\vk_{1}).
\end{equation}
The actual computation of the diagram then requires to
\begin{itemize}
\setlength{\parskip}{-.1cm plus4mm minus2mm}
\item sum over all internal indices;
\item integrate the intermediate time variables, $\eta'_{1}$ and $\eta'_{2}$, over adequate intervals;
\item integrate over the angle between $\vq$ and $\vk$.
\end{itemize}
The first step is now easy to perform and relies on generic properties of Green functions; the algebraic structure along the $\vk$-line indeed 
reads
\begin{eqnarray}
g_{a}^{\ g}(\eta_{2}-\eta'_{2})\ \delta_{g}^{\,f}\ g_{f}^{\ e}(\eta'_{2}-\eta'_{1})\ \delta_{e}^{\,c}\ 
g_{c}^{\ b}(\eta'_{1}-\eta_{1})\hspace{-6cm}&&\nonumber\\
&=&g_{a}^{\ f}(\eta_{2}-\eta'_{2})\ g_{f}^{\ c}(\eta'_{2}-\eta'_{1})\ g_{c}^{\ b}(\eta'_{1}-\eta_{1})\nonumber\\
&=&
g_{a}^{\ b}(\eta_{2}-\eta_{1}).
\end{eqnarray} 
The integration over the relative angle between $\vq$ and $\vk$ can then be done straightforwardly:
\begin{equation}
\int\dd^{3}\vq\frac{(\vq.\vk)^{2}}{q^{4}}=\frac{4\pi}{3}\int q^{2}\dd q \frac{k^{2}}{q^{2}}.
\end{equation}
We can now insert this result into the first diagram of Fig. \ref{OneLoopPowerSpec}. It leads to 
\begin{equation}
P_{ab}^{1-loop, \#1}(k)=P_{ab}^{\lin}(k)\left[1-k^{2}\sigma_{d}^{2}\right]
\end{equation}
with
\begin{equation}
\sigma_{d}^{2}(\eta)=\frac{4\pi}{3}\int \dd q \left(e^{\eta}-e^{\eta_{0}}\right)^{2}\,P_{22}^{\lin}(q)
\end{equation}
if the incoming velocity modes are the in linear growing mode. $\sigma_{d}$ can easily be interpreted as the r.m.s. of the 1D 
displacement field. 
A very similar calculation can be done for the second diagram of Fig. \ref{OneLoopPowerSpec}. Using the same approximation
one gets
\begin{equation}
P_{ab}^{1-loop, \#2}(k)=P_{ab}^{\lin}(k)\left[1+k^{2}\sigma_{d}^{2}\right]
\end{equation}
and the two contributions actually cancel\footnote{The actual calculations should be done with care in particular for 
a correct determination of the symmetry factors.}.
This cancellation was first noted by \citeN{1996ApJ...456...43J} and by \citeN{1996ApJS..105...37S}. In the following we explicitly show how it can be extended to all 
orders in perturbation theory with the help of the so called eikonal approximation (introduced in 
\shortciteNP{2012PhRvD..85f3509B}). 

\subsection{The eikonal approximation}
\label{sec:eikonal}

The eikonal approximation has been used in various contexts. 
It originally comes from the equations of wave propagation: it is a standard approximation
which leads to the laws of geometric optics. It has also been used in the context of Quantum Electro-Dynamics, in a manner very similar to 
the way we are going to use it, to exponentiate the effect of soft photon modes on the propagator of 
electrons~(see for instance \citeNP{Abarbanel:1969ek}).

In the context of gravitational dynamics, it is based on the decomposition of right hand side of eqn (\ref{FullEoM})  into 2 domains, the soft domain where one of the 
mode is much smaller than the other one, and one hard domain where the two interacting wave-modes
are of the same order. Let us assume for simplicity that the soft domain is obtained for $k_{1}\ll k_{2}$, then in the source term of
(\ref{FullEoM}) one should have $\vk=\vk_{2}$. The contribution corresponding to that domain can then be viewed as a corrective term
to the \emph{linear} evolution of the mode $\vk$. In other words the motion equation can then be written,
\begin{eqnarray}
&&\frac{\partial}{\partial \eta} \Psi_a(\vk,\eta) + \Omega_{a}^{\ b}(\eta) \Psi_b(\vk,\eta) -\Xi_{a}^{b}(\vk,\eta)\Psi_{b}(\vk,\eta)\nonumber\\
&&\hspace{5cm}=\left[\gamma_{a}^{\ bc}(\vk_1,\vk_2) \ \Psi_b(\vk_1,\eta) \ \Psi_c(\vk_2,\eta)\right]_{\mH},
\label{EikEoM}
\end{eqnarray}
with
\begin{equation}
\Xi_{a}^{\ b}(\vk,\eta)\equiv  2 \int_{{\mS}}\dd^{3}\vq \; %\left[\egamma_{acb}(\vk,\vq,\vk)+
\gamma_{a}^{\ bc}(\vk,\vq)
%\right]
\Psi_{c}(\vq,\eta)  \;. \label{fb:Xidef}
\end{equation}
The key point to make eqn (\ref{EikEoM}) sensible  is that in eqn~(\ref{fb:Xidef}) the domain of integration is restricted to soft momenta for which $q\ll k$, the soft 
domain. Conversely, on the right-hand side of eqn~(\ref{EikEoM}) the convolution is done excluding the soft domain, i.e.~it is over hard modes or modes of 
comparable size. 

Note that  $\Xi_{a}^{\ b}(\vk,\eta)$ is a random coefficient. It depends on the initial conditions but is (assumed to be) independent on the mode whose
evolution we are interested in. \emph{The equation (\ref{EikEoM}) can then be viewed as the motion equation of cosmological modes 
in a random medium with large-scale modes.} It allows to compute how the long wave modes alter the growth of structure. In that
context the vertex value appearing in this expression corresponds to that used for the computation of the one-loop
correction to the power spectrum in the IR domain in the previous paragraph,
\begin{equation}
\gamma_{a}^{\ bc}(\vk_{1},\vq)\approx \frac{1}{2}\frac{\vq.\vk}{q^{2}}\ \delta_{a}^{\,b}\ \delta_{2}^{\,c}\,
\Dirac(\vk-\vk_{1}).
\end{equation}

We restrict  here  our analysis to the property of a single fluid (or if there are two fluids to the adiabatic\footnote{Non-adiabatic large-scale 
modes can also be considered and lead to non-trivial phenomena such as the damping of the small scale fluctuations due to large relative 
displacement between baryons and dark matter particles as first discussed by \citeN{2010PhRvD..82h3520T} and later derived within the
eikonal approximation scheme \shortcite{2013PhRvD..87d3530B}.} modes). 
It leads to an explicit expression for $\Xi_{a}^{\ b}(\vk,\eta)$,
\begin{equation}
\Xi_{a}^{\ b} (\vk,\eta)  = \Xi(\vk,\eta) \; \delta_{ab} \;, \ \ 
 \Xi(\vk,\eta) \equiv \int_{{\mS}}\dd^{3}\vq \; \frac{\vk \cdot\vq}{q^{2}} \; 
 \htheta (\vq,\eta)  \;. \label{Xi_2}
\end{equation}
Note that only the velocity field $\htheta$ (and not the density field $\delta$) contributes to $\Xi_{a}^{\ b}$. Furthermore, as $\htheta(\vx,\eta)$ is real $\htheta(-\vq)=
\htheta^*(\vq)$ and thus $\Xi$ is  purely imaginary.

We can now explore the behavior of this new dynamical system. In particular the impact of the IR modes are now all encoded
in the $\Xi_{a}^{\ b}(\vk,\eta)$ coefficient and so are all encoded in the linearized equation of motion (obtained when the right hand side
of eq (\ref{EikEoM}) is dropped).  That equation admits new Green functions $\xi_{a}^{\ b}(\vk,\eta,\eta')$ that can be explicitly computed. They
satisfy the equation
\begin{equation}
\left(\frac{\partial}{\partial \eta}-\Xi(\vk,\eta)\right)\xi_{a}^{\ b}(\vk,\eta,\eta')+\Omega_{a}^{\ c}(\eta)\xi_{c}^{\ b}(\vk,\eta,\eta')=0\;, \label{xiEOM}
\end{equation}
In the case of a single fluid, as discussed here, eqn~(\ref{xiEOM}) can be easily solved. Taking into account the boundary condition $\xi_{a}^{\ b}(\vk,\eta,\eta)=
\delta_{a}^{\ b}$, one obtains
\begin{equation}
\xi_{a}^{\ b}(\vk,\eta,\eta_0)=g_{a}^{\ b}(\eta,\eta_0)\exp\left(\int^{\eta}_{\eta_0}\dd\eta'\ \Xi(\vk,\eta')\right)\;. \label{resummed_1}
\end{equation}

The argument of the exponential is the time integral of the velocity projected along the direction $\vk$, i.e.~the \emph{displacement} component along $\vk$, that is
\begin{equation}
\xi_{a}^{\ b}(\vk,\eta,\eta')=g_{a}^{\ b}(\eta,\eta_0)\exp\left(\ii\vk\cdot\vd(\eta,\eta')\right)\;. \label{resummed_2}
\end{equation}
 where $\vd(\eta,\eta')$ is the total displacement induced by the long wave modes between time $\eta'$ and $\eta$.
Note that eqn~(\ref{resummed_1}) is valid irrespectively of the fact that the incoming modes in $\Xi$ are in the growing mode or not.

The consequences of this result are multifold. In particular it explicitly gives the impact of the long-wave modes on the growth
of structure: they are entirely captured by a phase shift in the propagator values  which is proportional to $\vk.\vd(\eta,\eta')$. If one now 
considers any contribution to any equal time multi-point spectrum, it is easy to see that the total phase shifts exactly cancel out such that the
long wave modes have no impact on the equal time correlators. It generalizes to any order the result of the previous paragraph
(see detailed derivation of this property in \shortciteNP{2013PhRvD..87d3530B,2013arXiv1304.1546B}). 

The second consequence is that it is now possible to compute resumed propagators. This is at the heart of the so-called RPT 
and RegPT propositions described in the next section. 

\subsection{The extended Galilean invariance from the equivalence principle}

The previous result is actually closely related to an invariance sometimes called the 
extended Galilean invariance \shortcite{2013NuPhB.873..514K,2013JCAP...05..031P}, which actually 
derives from the equivalence principle as shown in \shortciteN{2013arXiv1309.3557C}, which the 
pressureless Vlasov-Poisson system satisfies\footnote{Interestingly the
same invariance is satisfied by the Navier-Stokes equations with
constant density \shortcite{2000tufl.book.....P}.}. 
It should be clear that this invariance significantly extends that of a mere Galilean invariance as it states that the development of the gravitational instabilities in a given patch of the universe is the same irrespectively of the fact that this patch is accelerated or not.
More explicitly the
motion equations are invariant under the following transformations\footnote{To get a fully valid transformation one should also change
the gravitational potential gradient in such a way that the source term of the Euler equation is changed in $f_{i}\to f_{i}+
\frac{\dd^{2}}{\dd\eta^{2}}s_{i}(\eta)+\frac{1}{2}\frac{\dd}{\dd\eta}s_{i}(\eta)$ which leaves the Poisson equation unchanged.},
\begin{eqnarray}
x_{i}\to x_{i}-s_{i}(\eta),\ \ u_{i}\to u_{i}+\frac{\dd}{\dd \eta}s_{i}(\eta),
\end{eqnarray}
where $\vs(\eta)$ is an arbitrary time dependent  vector. Under this transformation the linear propagator of the theory is precisely
changed into,
\begin{equation}
g_{a}^{\ b}(\eta,\eta')\to g_{a}^{\ b}(\eta,\eta')\exp\left(\ii\vk.[\vs(\eta)-\vs(\eta')]\right),
\end{equation}
which is reminiscent of eqn (\ref{resummed_2}).
In other words the adiabatic long wave modes can entirely be absorbed by an extended  Galilean transformation leaving no imprint on equal time correlators.

Interestingly though, the existence of a transformation under which the equations are invariant leads to so-called Ward identities the simplest 
of which we reproduce here,
\begin{eqnarray}
\lim_{q\to 0}B(\vq,\vk,-\vk-\vq;\eta,\eta',\eta'')&=&\nonumber\\
&&\hspace{-2cm}-\frac{\vq.\vk}{q^{2}}\,P^{\lin}(q;\eta,\eta)\,P^{\lin}(k;\eta',\eta'')\,\left({e^{\eta'-\eta}-e^{\eta''-\eta}}\right)
\label{ward1}
\end{eqnarray}
where $P^{\lin}(k;\eta,\eta')$ is the linear unequal time power spectrum taken between time $\eta$ and $\eta'$
and $B(\vk_{1},\vk_{2},-\vk_{1}-\vk_{2};\eta_{1},\eta_{2},\eta_{3})$ is the unequal time bispectrum, i.e.
\begin{eqnarray}
&&\langle\delta(\vk_{1},\eta_{1})\delta(\vk_{2},\eta_{2})\delta(\vk_{3},\eta_{3})\rangle=\nonumber\\
&&\hspace{2.5cm}\Dirac(\vk_{1}+\vk_{2}+\vk_{3})
B(\vk_{1},\vk_{2},-\vk_{1}-\vk_{2};\eta_{1},\eta_{2},\eta_{3}))
\rangle.
\end{eqnarray}
The identity (\ref{ward1}) involves unequal time correlators and as such is probably very difficult to implement in actual observations. 

\section{The $\Gamma-$ expansion}

%\textbf{mentioner autres ressummation schemes}

\subsection{The general formalism and theorem}

\begin{figure}
\centering
\includegraphics[width=10cm]{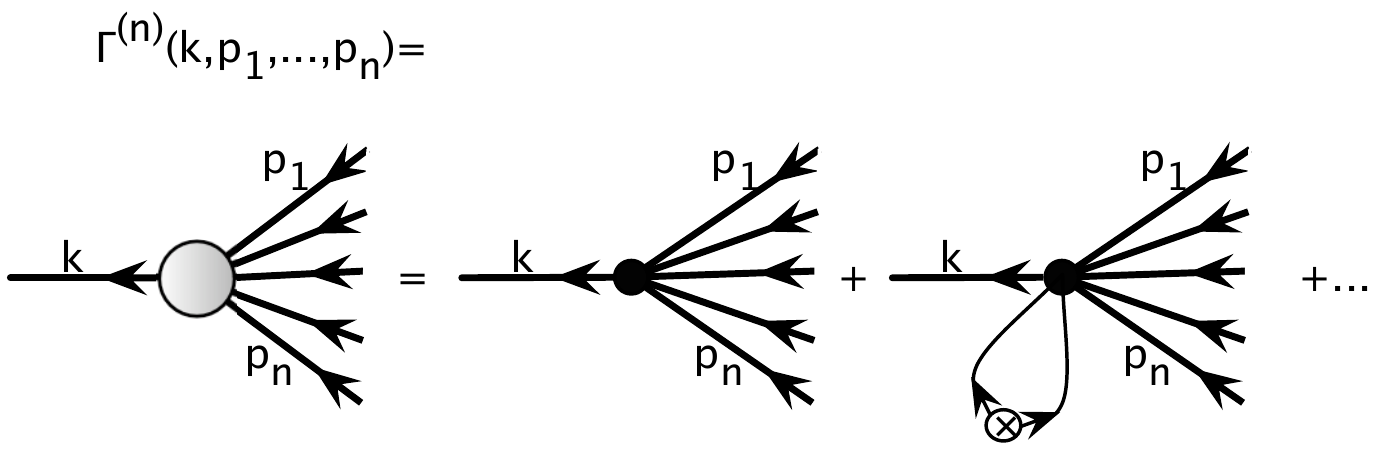} 
%\centerline{\epsfig {figure=Gamma5.eps,width=10cm}}
\caption{Representation of the first two terms of the multi-point
  propagator $\Gamma_{g}^{(n)}$ in a perturbative
  expansion. $\Gamma_{g}^{(n)}$ represents the average value of the
  emerging nonlinear mode $\vk$ given $n$ initial modes in the linear
  regime. Here we show the first two contributions: tree-level and
  one-loop. Note that each object represents a collection of
  (topologically) different diagrams: each black dot represents a set of trees that connect respectively $n+1$ lines for the first term, $n+2$ for the second.}
\label{Gamma5}
\end{figure}

\begin{figure}
\centering
\includegraphics[width=8cm]{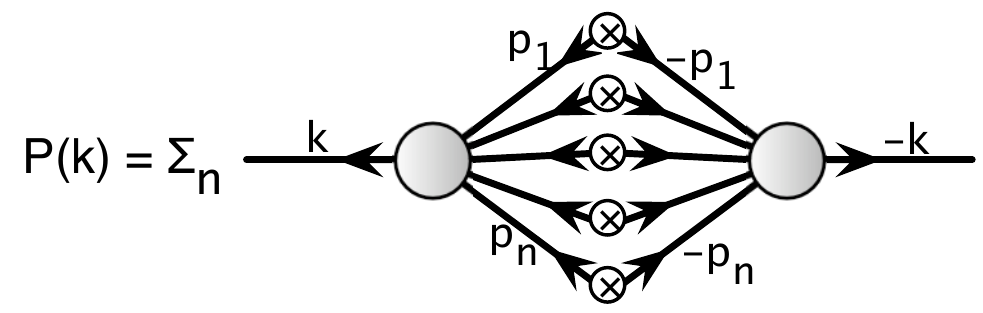} 
%\centerline{\epsfig {figure= BispectreResum.eps,width=10cm}}
\caption{Representation of the resummation rule given by eqn (\ref{PkRecons}). For Gaussian initial conditions, the
power spectrum can be seen as a sum of squares of $\Gamma^{(p)}$ functions.}
\label{SpectreResum}
\end{figure}

\begin{figure}
\centering
\includegraphics[width=8cm]{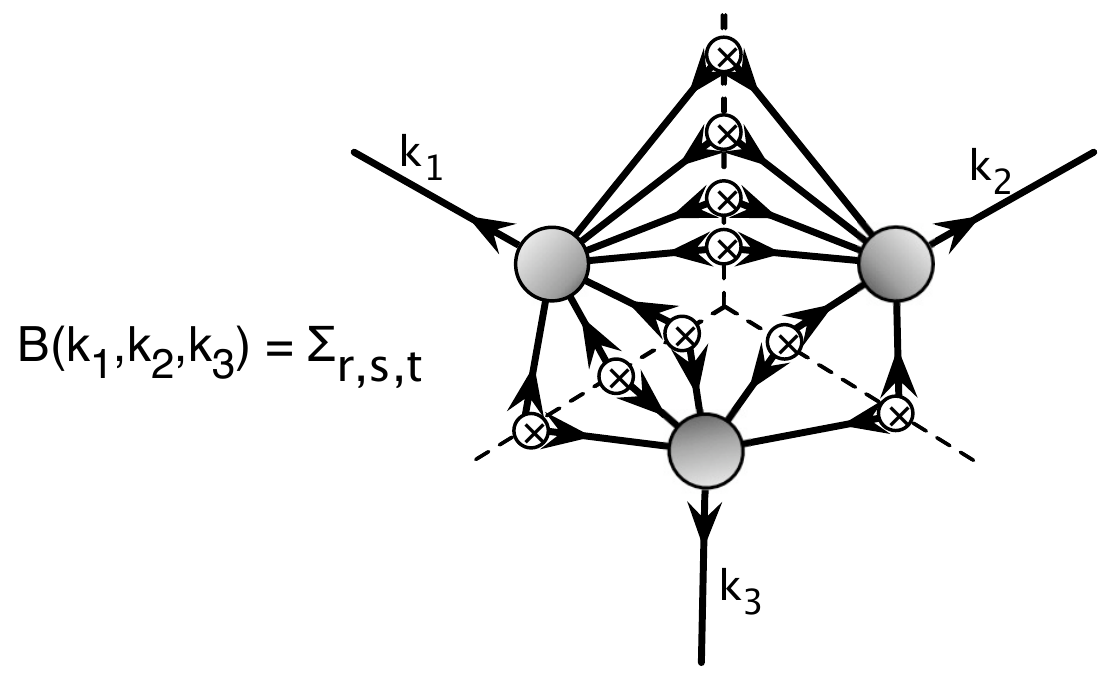} 
%\centerline{\epsfig {figure= BispectreResum.eps,width=10cm}}
\caption{Representation of the resummation rule given by eqn (\ref{BkExpansion}). For Gaussian initial conditions, the
bispectrum can be seen as a sum of product of $\Gamma^{(p)}$ functions.}
\label{BiSpectreResum}
\end{figure}

The so-called $\Gamma-$expansion is a very general result that can be applied to any nonlinear field transformation.
It has been explicitly demonstrated in the context of nonlinear gravitational dynamics 
\shortcite{2012PhRvD..85l3519B} but results regarding biasing schemes 
can also be derived in the very same framework 
\shortcite{2006PhRvD..74j3512M,2013PhRvD..88b3515S}.
So let us consider a field $g(\vx,\eta)$ and its Fourier modes $g(\vk,\eta)$. Let us now assume that $g(\vk,\eta)$ can be 
expanded in terms of an initial Gaussian  random field $\zeta(\vk)$,
\begin{equation}
g(\vk,\eta)=\sum_{n}
\int\frac{\dd^{3}\vk_{1}}{(2\pi)^{3/2}}\dots\frac{\dd^{3}\vk_{n}}{(2\pi)^{3/2}}
\delta_{D}(\vk-\sum_{i}\vk_{i})
g_{n}(\vk_{1},\dots,\vk_{n},\eta)\zeta(\vk_{1})\dots\zeta(\vk_{n}),
\label{gexpansion}
\end{equation}
where the $g_{n}$ functions are time dependent functions. In particular $g_{1}(\eta,\vk)$ is nothing but
the linear solution for that particular fluid. 
One question that one might want to ask is how the growth rate of the fluctuation is actually affected by the
fluctuations.
Multipoint propagators are precisely defined as the ensemble averages (over fluctuations in the medium) of the infinitesimal response
of the system to an initial perturbation.
More precisely we can define the $\Gamma^{(p)}$ functions as,
\begin{equation}
\frac{1}{p!}\left\langle \frac{\dpartial g(\vk,\eta)}{\dpartial\zeta(\vq_{1})\dots \dpartial\zeta(\vq_{p})}\right\rangle
\equiv \Dirac\Big(\vk-\sum_{r=1}^{p}\vq_{r}\Big)\, \Gamma_{g}^{(p)}(\vq_{1},\dots,\vq_{p}).
\end{equation}
To obtained the values of $\Gamma^{(p)}$ out of the $g_{n}(\eta,\vk_{1},\dots,\vk_{n})$ function, in each term starting with $n=p$
in eq (\ref{gexpansion}) one should take out $p$ modes and compute the ensemble average of the remaining terms. A diagrammatic representation 
of this procedure can be found on Fig. \ref{Gamma5}. 
The $\Gamma^{(p)}$ function are thus nonlinear objects on their own. Note a useful property,
\begin{equation}
\Gamma_{g_{1}}^{(p)}(-\vq_{1},\dots,-\vq_{p})=\Gamma_{g_{1}}^{(p)}(\vq_{1},\dots,\vq_{p})
\end{equation}
 from parity symmetry of the field $g$.

The key result explicitly demonstrated in  \shortciteN{2012PhRvD..85l3519B} is then that the power spectrum of the $g$ field, or actually the cross spectra 
of any of two such fields $g_{1}$ and $g_{2}$,  can then be reconstructed out 
of the multipoint propagators  following the formulae,
\begin{eqnarray}
\mg g_{1}(\vk) g_{2}(\vk')\md
&=&
\Dirac\Big(\vk+\vk'\Big)
\sum_{p}{p!}
\int\dd^3\vq_{1}\dots \dd^3\vq_{p}
\nonumber\\
&&\hspace{-2.5cm}\times 
\Dirac(\vk-\vq_{1\dots p}) 
\Gamma_{g_{1}}^{(p)}(\vq_{1},\dots,\vq_{p})
\Gamma_{g_{2}}^{(p)}(\vq_{1},\dots,\vq_{p})
P_{\zeta}(q_{1})\dots P_{\zeta}(q_{p}),
\label{PkRecons}
\end{eqnarray}
when $\zeta$ is Gaussian distributed.  Note that when one considers the spectrum of a given field, the formulae is then 
a sum of square terms making each contribution positive. 
As it is clear from the very beginning this is a very general construction which is valid in particular for the evolved 
density and velocity fields. Note that an extension of this relation exists for non-Gaussian initial conditions \shortcite{2010PhRvD..82h3507B}.

The previous construction for the spectra can actually be extended to any high-order spectra. The formal 
expression for the general term is a bit cumbersome. However, for the bispectrum it is still possible to write its formal expression,
\begin{eqnarray}
\mg g_{1}(\vk_{1}) g_{2}(\vk_{2}) g_{c}(\vk_{3})\md=\sum_{r,s,t} \binom{r+s}{r}\binom{s+t}{s}\binom{t+r}{t} r! s! t! \hspace{-10cm}&&
\nonumber\\
&&
\int
\Pi_{i=1}^{r}\dd^3\vq_{i}P_{\zeta}(q_{i})\ 
\Pi_{j=1}^{s}\dd^3\vq_{j}P_{\zeta}(q_{j})\ 
\Pi_{k=1}^{t}\dd^3\vq_{k}P_{\zeta}(q_{k})\ 
%\ \dd^3\vq'_{1}\dots\dd^3\vq'_{s}
%\ \dd^3\vq''_{1}\dots\dd^3\vq''_{t}\ 
%P_{\zeta}(q_{1})\dots P_{0}(q_{r})\ P_{\zeta}(q'_{1})\dots P_{\zeta}(q'_{s})\ P_{\zeta}(q''_{1})\dots
%P_{\zeta}(q''_{t})
\nonumber\\
&&\,\times
\Dirac(\vk_{1}-\vq_{1\dots r}-\vq'_{1\dots s})\ 
\Dirac(\vk_{2}+\vq'_{1\dots s}-\vq''_{1\dots t})\ 
\Dirac(\vk_{3}+\vq''_{1\dots t}+\vq_{1\dots r})   \nonumber \\
&&\,\times
\Gamma^{(r+s)}_a\left(\vq_{1},\dots,\vq_{r},\vq'_{1},\dots,\vq'_{s}\right) 
\Gamma^{(s+t)}_b\left(-\vq'_{1},\dots,-\vq'_{s},\vq''_{1},\dots,\vq''_{t}\right)\nonumber\\
&&\times\,
\Gamma^{(t+r)}_c\left(-\vq''_{1},\dots,-\vq''_{t},-\vq_{1},\dots,-\vq_{r}\right).
\label{BkExpansion}
\end{eqnarray}
This reconstruction is illustrated on Fig. \ref{BiSpectreResum} where it is shown how the product of three $\Gamma-$functions
can be ``glued'' together for the computation of bispectra. 

\subsection{The case of gravitational dynamics}

We now move to the application of this formalism to the cosmological density and velocity fields.
To be exact one should take into account the index structure of the theory which we overlooked in the previous paragraph.
For instant the $\Gamma^{(1)}$ has now an algebraic structure as,
\begin{equation}
\mg\frac{\dpartial \Psi_{a}(\vk,\etaf)}{\dpartial
\Psi_{b}(\vk',\etai)}\md=\Dirac{\left(\vk-\vk'\right)}\,\Gamma_{a}^{(1)b}(k,\etaf,\etai),
\label{Gabdef}
\end{equation} 
where we also allow ourselves to consider the initial time in eqn~(\ref{gexpansion}) as a free parameter.
Usually this quantity is noted,
\begin{equation}
G_{a}^{\ b}(k,\etaf,\etai)\equiv \Gamma_{a}^{(1)b}(k,\etaf,\etai).
\end{equation} 
It is nothing but the generalization of the linear propagator. The idea of replacing the linear propagator by a 
resumed propagator is at the heart  of the so-called RPT approach 
\shortcite{2006PhRvD..73f3519C,2006PhRvD..73f3520C}.

The expression for $G_{a}^{\ b}$ can be computed order by order in perturbation theory.
Such results can be obtained from a formal expansion of $\Psi_{a}(\vk,\eta)$ with respect to the initial field,
\begin{equation}
\Psi_{a}(\vk,\eta)=\sum_{n=1}^{\infty}\Psi_{a}^{(n)}(\vk,\eta)
\label{PsiExpansion}
\end{equation}
with
\begin{eqnarray}
\Psi_{a}^{(n)}(\vk,\eta)=
\mF_{a}^{(n)b_1 b_2 \ldots b_n}(\vk_{1},\dots,\vk_{n};\eta)
\Phi_{b_1}(\vk_{1})\dots\Phi_{b_n}(\vk_{n})
\label{mFndef}
\end{eqnarray}
where $\mF^{(n)}$ are fully symmetric functions of the wave-vectors.  Note that these functions have
in general a non-trivial time dependence because they also include sub-leading terms in $\eta$. 
Their fastest growing term is of course given by the well known $\{F_n,G_n\}$ kernels in PT (assuming
growing mode initial conditions),
\begin{eqnarray}
\mF^{(n)b_1 b_2 \ldots b_n}_{a}(\vk_{1},\dots,\vk_{n};\eta)u_{b_{1}}\ldots u_{b_{n}}&=&\nonumber\\
&&\hspace{-4cm}\Dirac(\vk-\vk_{1\dots n})\exp(n\eta)\ \left(F_n(\vk_1,..,\vk_n),G_n(\vk_1,..,\vk_n)\right)^{T}
\end{eqnarray}
whose two elements correspond to the density or velocity divergence fields respectively.

\subsection{The RPT formulation}

\begin{figure}
\centering
\includegraphics[width=8cm]{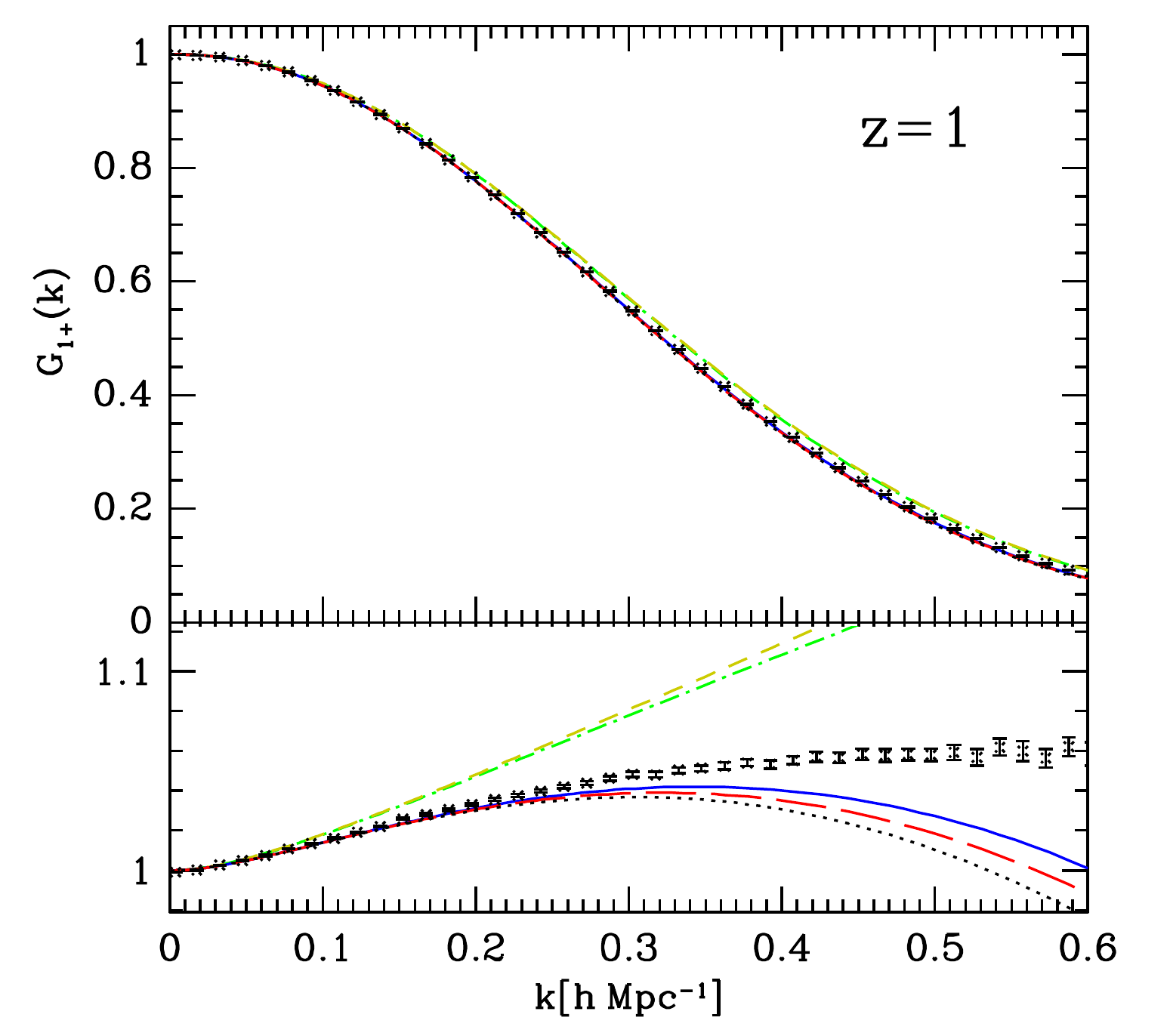} 
%\centerline{\epsfig {figure= BispectreResum.eps,width=10cm}}
\caption{The predicted resumed propagator (lines) compared with numerical simulations (symbols with error bars) at $z=1$. 
The top panel shows the $G_{1+}(k)$  as a function of $k$ and the bottom panel shows $G_{1+}(k) \exp(-k^{2}\sigma_{d}^{2}/2)$.
The dashed and dot-dashed lines are the RPT and RegPT form at 1-loop order. The bottom lines shows the 2-loop predictions.}
\label{GabResum}
\end{figure}

We are now in position to give the explicit form of the RPT proposition described by \citeANP{2006PhRvD..73f3519C} \citeyear{2006PhRvD..73f3519C}.
It is based on the construction of the propagator $G_{a}^{\ b}(k,\eta)$ that encompasses both its full one-loop 
correction and the effect of the long-wave modes computed at all orders.
The explicit form adopted in the original paper is too cumbersome to be re-derived here in detail but a form
that shares its properties can easily be obtained. The impact of the long-wave modes can indeed be all incorporated with the help
of the eikonal approximation where the ``naked' Green function, $g_{a}^{\ b}(\eta,\eta')$, is transformed into the ``dressed'' one,
$\xi_{a}^{\ b}(\eta,\eta')$. Then  armed with the results from Section \ref{sec:eikonal} we can first compute the ensemble average 
of $\xi_{a}^{\ b}(\vk,\eta)$ with respect to the long-wave modes. When those modes are assumed to be in the linear
regime and therefore to be Gaussian distributed, the use of the relation (\ref{momtocum}), leads to,
\begin{equation}
\langle\xi_{a}^{\ b}(\vk,\eta)\rangle=g_{a}^{\ b}(\eta,\eta')\,\exp\left[-\frac{1}{2}k^{2}\sigma_{d}^{2}(\eta,\eta')\right],
\end{equation}
where $\sigma_{d}(\eta,\eta')$ is given by,
\begin{equation}
\sigma_{d}^{2}(\eta,\eta')=\frac{1}{3}\langle\vd^2(\eta,\eta') \rangle=(e^\eta-e^{\eta'})^2\,\int\frac{\dd^{3} \vq}{3q^{2}}\ P^{\lin}(q).
\end{equation} 
This resummation result is the key result  on which the RPT scheme is based.
It is furthermore possible to include the full contribution of the one-loop contribution as given by the diagram of Fig.
\ref{Gab1loop}. Indeed this diagram itself can be computed within the eikonal framework and it leads to
\begin{equation}
G_{a}^{\oneloop\,b}(\vk,\eta,\eta')\to G_{a}^{\oneloop\,b}(\vk,\eta,\eta')\ \exp(\ii\vk.\vd(\eta,\eta'))
\end{equation}
the ensemble average of which over the long-wave modes can also be taken. A possible global form 
 in then the following  \shortcite{2012PhRvD..85l3519B},
\begin{eqnarray}
G_{a}^{\reg\,b}(k,\etaf,\etai)&=&\left[g_{a}^{\ b}(\eta,\eta')+\delta G_{a}^{\oneloop\,b}(k,\eta,\eta')+\frac{1}{2}k^{2}\sigma_{d}^{2}(\eta,\eta') g_{a}^{\,b}(\eta,\eta')\right]
\nonumber\\
&&\times \exp\left(-\frac{k^2\sigma_{d}^{2}(\eta,\eta')}{2}\right),
\label{Gabpredict}
\end{eqnarray}
which has the expected behavior for both the large $k$ and in the low $k$ domains.

On Fig. \ref{GabResum} we show the performance of the RPT type prediction on the behavior of the propagator. The measured 
values of $G_{1+}(k)\equiv G_{1}^{\ b}(k)u_{b}^{(+)}$ are compared with the analytical predictions up to 2 loop order. The RPT
formulation used the 1-loop order predictions only.

 \subsection{The MPTbreeze and RegPT formulation}

\begin{figure}
\centering
\includegraphics[width=12cm]{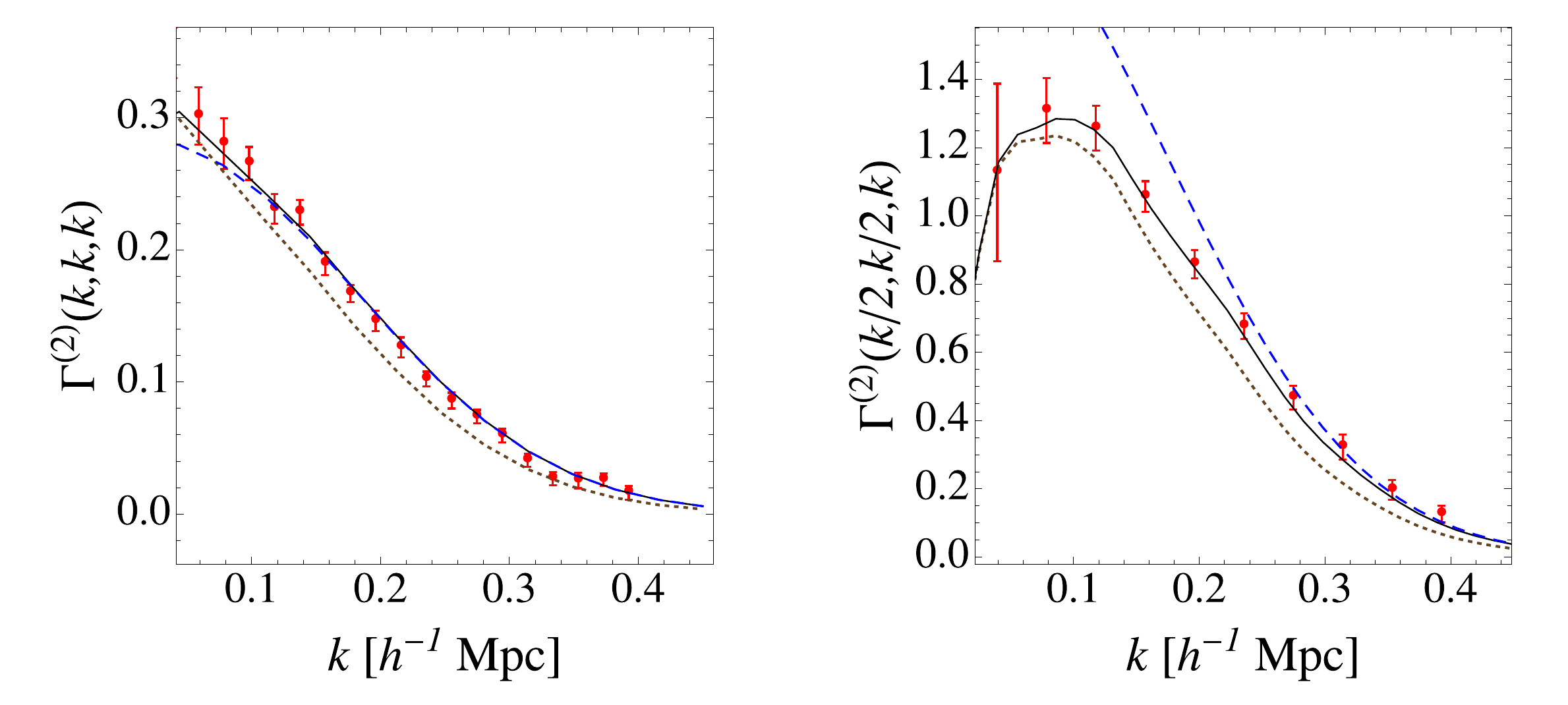} 
%\centerline{\epsfig {figure= BispectreResum.eps,width=10cm}}
\caption{The predicted resumed $\Gamma^{(2)}$ for two different configurations up to one loop order (solid lines) compared with numerical
simulations (symbols with error bars). The dotted lines are the tree oder predictions and the dashed lines the 
one-loop order neglecting the binning effects.}
\label{GamabcResum}
\end{figure}

Other resummation schemes, MPTbreeze \shortcite{2012MNRAS.427.2537C} and RegPT \shortcite{2012PhRvD..86j3528T} proposed afterwards take full 
advantage\footnote{They also have the advantage to be faster to compute.} of the $\Gamma-$expansion presented in the
previous section. It is based on a similar construction to (\ref{Gabpredict}) applied  to the $p-$point propagator. More specifically the following form can be used,
\begin{eqnarray}
\Gamma_{a}^{\reg\,b_{1}\dots b_{p}}(\vk_{1},\dots,\vk_{p},\eta,\eta')&=&\nonumber\\
&&\hspace{-4cm}\left[\Gamma_{a}^{{\rm tree} b_{1}\dots b_{p}}(\vk_{1},\dots,\vk_{p},\eta,\eta')
+\delta \Gamma_{a}^{\oneloop b_{1}\dots b_{p}}(\vk_{1},\dots,\vk_{p},\eta,\eta')
\right.\nonumber\\
&&\hspace{-4cm}\left.+\frac{1}{2}k^{2}\sigma^{2}_{d}(\eta,\eta')  \Gamma_{a}^{{\rm
      tree}\ b_{1}\dots
    b_{p}}(\vk_{1},\dots,\vk_{p},\eta,\eta')\right]\exp\left(-\frac{k^2\sigma^{2}_{d}(\eta,\eta') }{2}\right)
\label{regGammaponeloop}
\end{eqnarray}
where $k=\vert\vk_{1}+\dots+\vk_{p}\vert$. 
By construction this form is such that it has both the correct one-loop correction and the correct large-$k$ behavior. A comparison for the $p=2$
case is presented on Fig. \ref{GamabcResum}. 

It can then be incorporated in a $\Gamma-$expansion scheme. In practice the MPTbreeze and RegPT codes exploit this formalism up to 
2-loop order, that is $G_{a}^{\ b}(k,\eta)$ is computed to to 2-loop order (1-loop order only for MPTbreeze), $\Gamma^{(2)b_{1}b_{2}}_{a}$ to 1-loop
order and $\Gamma^{(3)b_{1}b_{2}b_{3}}_{a}$ at tree order.

Note that all these approaches exhibit similar properties. In particular the predicted power spectra are damped at large $k$ with
an overall factor of the form, $\exp(-k^{2} \sigma_{d}^{2})$. This result is at variance with one consequence of the 
extended Galilean invariance, that is that equal-time spectra should be independent on the long-wave modes that
participate in the values of $ \sigma_{d}$. At two-loop order though the lowest order dependence on $\sigma_{d}$ is  $k^{6}\sigma_{d}^{6}$
(all lower orders in $\sigma_{d}$ cancel out). The remaining dependence that we have in these schemes actually 
signal the validity range of their predictions.

\subsection{The performances of perturbation theory at NNLO}

\begin{figure}
\centering
\includegraphics[height=4.2cm]{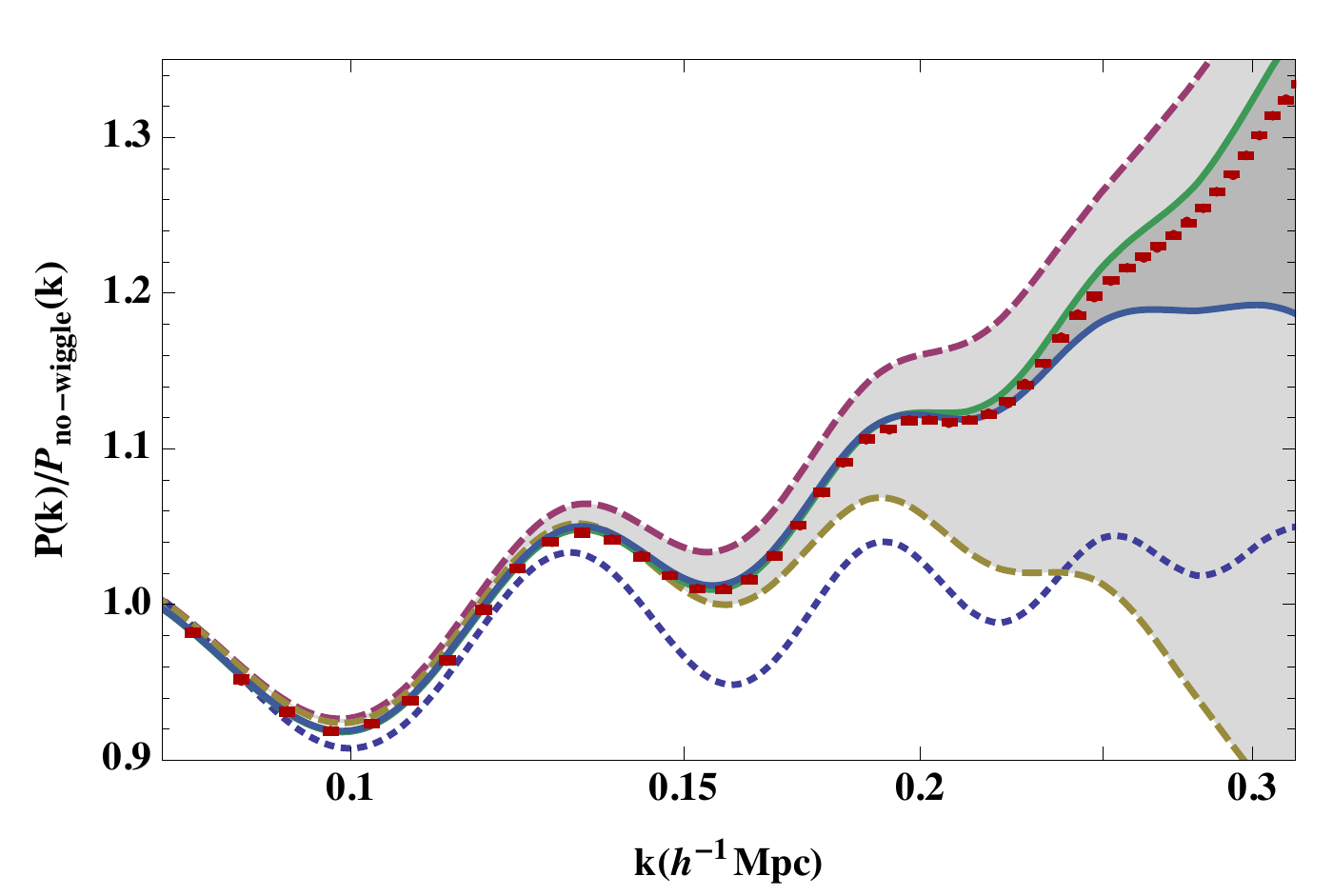} 
\includegraphics[height=4.2cm]{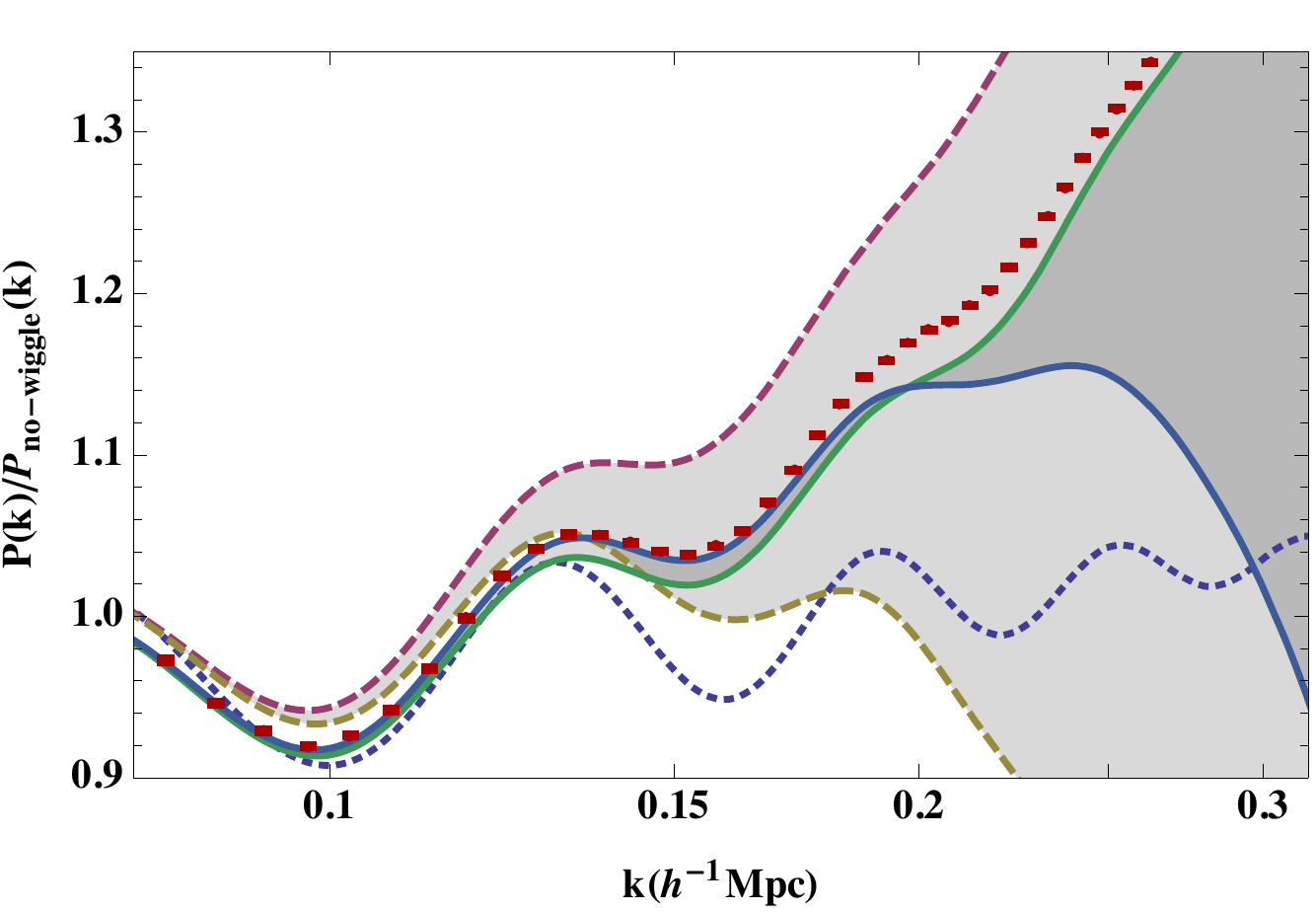} 
\caption{Comparison of PT results with $N-$body results for the power spectrum at $z=1$ (left panel) and $z=0.35$ (right panel). 
The dotted line is the linear prediction; the dashed lines are the standard PT and RegPT NLO predictions
and the solid lines the  NNLO predictions. The grey area show the departure between these predictions at one-loop
order (light grey) and 2-loop order (darker grey).}
\label{ComPk2loop}
\end{figure}

\begin{figure}
\centering
\includegraphics[width=8cm]{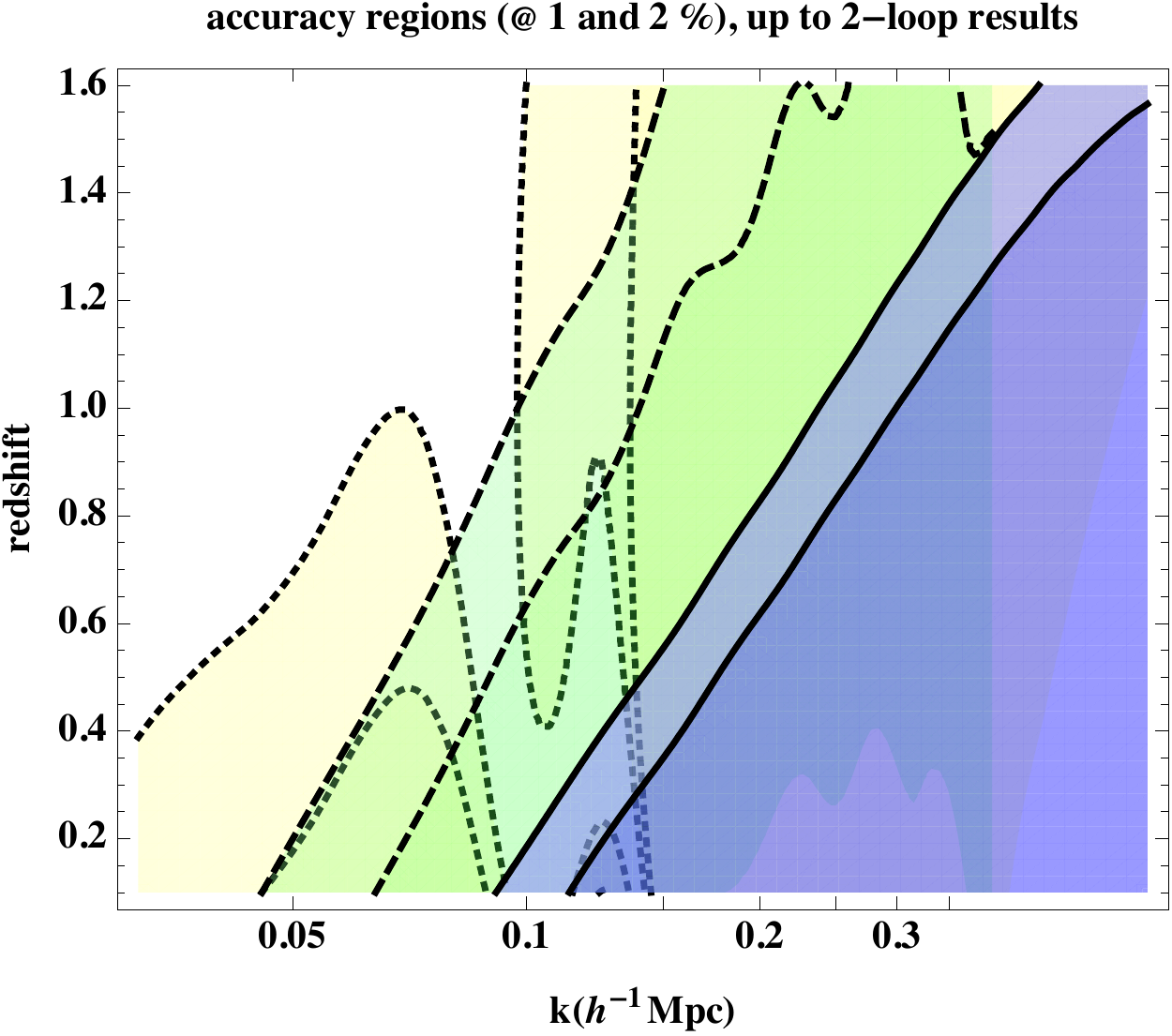} 
\caption{A tentative chart of the accuracy domain of the PT results with contour plots of the 1 and 2\% accuracy 
region for the linear, NLO, and NNLO predictions for the power spectrum. The calculations have been made 
for the WMAP5 cosmological parameters.}
\label{Accuracy2loop}
\end{figure}

To conclude this section we present the performances of the perturbation theory at two-loop order. 
On Fig. \ref{ComPk2loop} we present the performances of the RegPT up to 2-loop order compared to standard PT
and compared to numerical results. The solid lines show the two-loop order predictions for the standard PT and the RegPT schemes.
They agree with another up to wave-modes that are significantly larger than the linear theory validity regime. They also
extent significantly the validity regime of the one-loop results. This is particular striking at redshift 1 and above. 

These findings are summarized on Fig. \ref{Accuracy2loop} which tentatively charts the performances of the linear, one and
two-loop predictions. The dotted lines correspond to the 1\% and 2\% accuracy lines of the linear theory. The contour lines
are obtained from a comparison between the linear and the standard two-loop results. The dashed lines show the same results
for the standard one-loop predictions and the solid lines show an estimate of those lines from a consistency comparison between
different two-loop schemes.

%%%%%%%%%%%%%%%%%%%%%%%%%%%%%%%%%%%%%%%%%%%%%%%%%%%%%
\section{Mode coupling structure}
\label{sec:couplings}

We are now interested more explicitly in the coupling structure of the terms that contribute to the loop corrections to the power 
spectra. In the previous sections we proved that they were safe from infrared divergences. One of the aim of this section is 
to explore in more detail their ultra-violet (UV) convergence properties.

\subsection{Scalings in the long wave mode regime}

Before we detail the converging properties of the loop diagrams, let us first make a simple remark regarding the $\Gamma-$expansion.
One of the nice feature of this approach is that it tends to hierarchize the contribution in $k$ range. Indeed the lowest order contribution
in $k$ from the loop corrections to the propagator (first diagram in Fig. \ref{Gab1loop}) is in $k^{2}$ whereas the lowest contribution from 
the second diagram is in $k^{4}$. This basic property will extend to all orders in the sense that the $p=1$ term in the sum (\ref{PkRecons})
will contain order $k^{2}$ corrections, and $p\le 2$ terms will contain order $k^{4}$ terms. As we will see in the following this behavior is intimately related to the UV behavior of the theory: as the vertex coefficients are homogeneous in wave modes, large powers of $k$ correspond from better converging contributions in the UV domain.

\subsection{Kernels and integrands}

In order to better visualize these properties, 
let us define the functions $K_{ab}(k,q)$ as the kernels describing the linear response of the non-linear power spectrum to the linear
power spectrum in such a way that,
\begin{equation}
\label{kernel:def}
\delta P_{ab}^{\sharp}(k)=\int \frac{\dd q}{q}\ K_{ab}^{\sharp}(k,q)P^{\lin}(q).
\end{equation}
The kernels can actually be considered for any approximate scheme, or any any diagram contributing to the power
spectrum. 
Note that the kernel functions depend themselves a priori on the initial power spectrum: at one-loop order $K_{ab}^{\oneloop}(k,q)$
is just a $\Dirac$ function, it is linear in the linear power spectrum at one loop order, etc.

These functions give, for each order, the impact of a linear mode $q$ on the amplitude of the late time mode $k$ we are interested in. In particular it tells how the 
small-scale modes affect the large-scale modes under consideration.  In the following we will focus our interest in understanding the 
high-$q$ behavior of the kernel functions $K_{ab}(k,q)$ for the propagators. They were considered in detail in
\shortciteN{2012arXiv1211.1571B} and in \shortciteN{2013arXiv1309.3308B} up to three-loop order.

\begin{figure}[ht] %  figure placement: here, top, bottom, or page
   \centering
   \includegraphics[width=8cm]{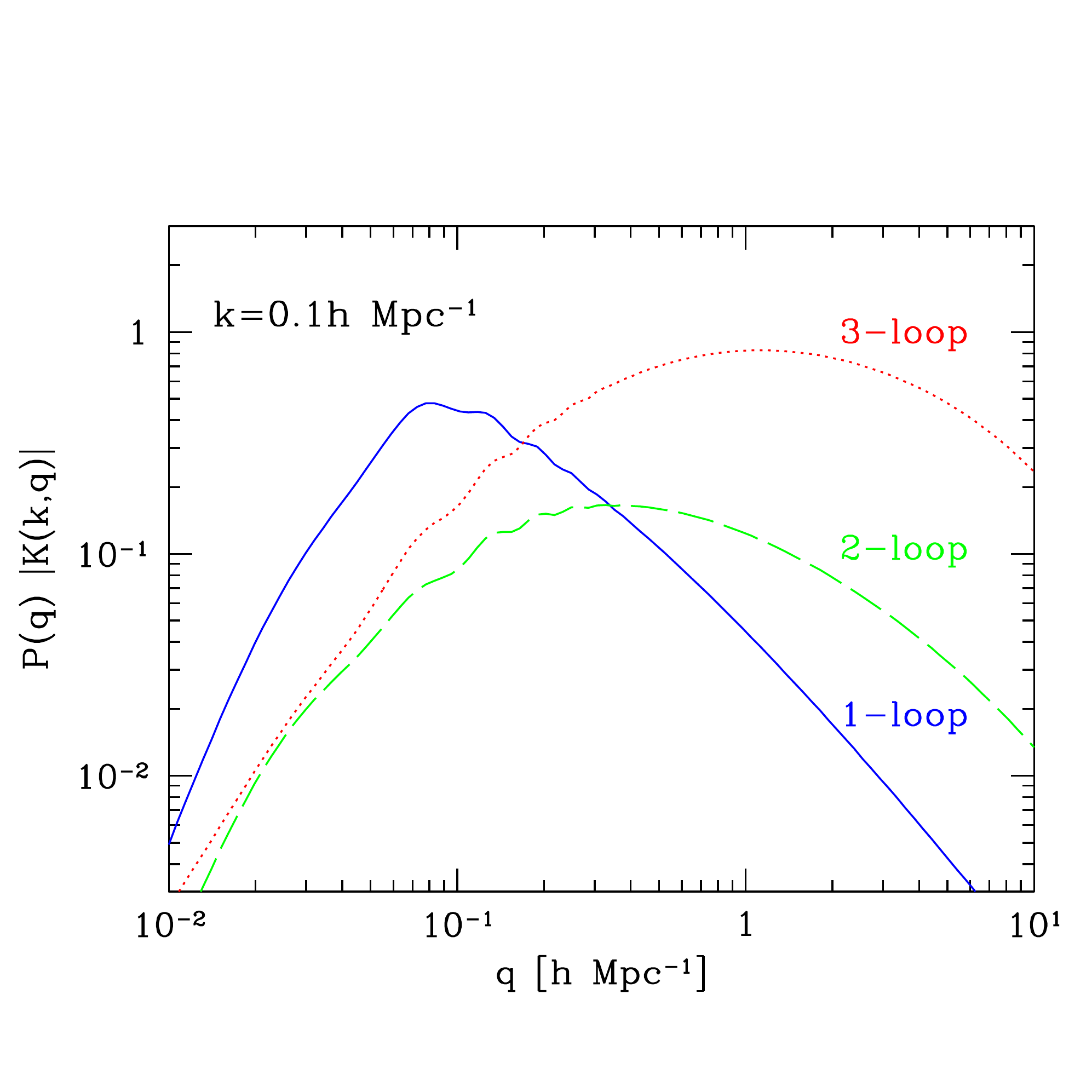}
   \caption{The shape of the kernel functions (actually the integrand in eqn \ref{kernel:def}) for the propagator $P^{\lin}(q)K^{\oneloop}_{1+}(k,q)$ (blue solid line), $P^{\lin}(q)K_{1+}^{\twoloops}(k,q)$ (green dashed line)  for $k=0.1h/\Mpc$ and $P^{\lin}(q)K_{1+}^{\threeloops}(k,q)$ (red dotted line)  as a function of $q$ for $z=0.5$ for the density component and for linear growing modes, from \protect\shortciteN{2012arXiv1211.1571B}. A similar plot can be found in \protect\shortciteN{2013arXiv1309.3308B}  for the power spectrum.}
   \label{Kernels}
\end{figure}

In Fig. \ref{Kernels} we show the shape of the propagator kernel functions at one,  two-loop and three-loop order for $k=0.1 h/\Mpc$
(in the latter case it was obtained from numerical simulations).  The solid line corresponds to the one-loop expression. As can be seen it is rather peaked at $q
\approx k$ and we more explicitly have
\begin{eqnarray}
K_{1+}^{\oneloop}(k,q)P^{\lin}(q)&=&\frac{176 \pi}{315}\,k^{2}\, q\ P^{\lin}(q)\ \ {\rm for}\ q\gg k.
\label{K1plusAsymp}
\end{eqnarray}
Note that convergence of the 1-loop contribution is then ensured for  a spectral index smaller than $-1$. 

At two-loop order, the behavior is qualitatively different. The function peaks rather for $q=0.5h/\Mpc$, \emph{irrespectively} of the value for $k$
(when $k<0.5h/\Mpc$). We note that,
\begin{equation}
K_{1+}^{\twoloops}(k,q)P_{0}(q)\sim k^{2} q^{2} P_{0}(q)\ \ {\rm for}\ q\gg k,\label{K2plusAsymp}
\end{equation}
so that the convergence is now secured for a spectral index smaller than $-2$ only.

\subsection{General convergence properties in the UV demain}

Let us now generalize these results. As mentioned in  \citeN{Scoccimarro:1996se} the total gradient nature of the coupling terms in the
motion equations implies that,
\begin{equation}
\mF_{a}^{(n)}(\vq_{1},\dots,\vq_{n})\sim \frac{\vert \sum_{j}q_{j}\vert^{2}}{q_{i}^{2}}\ \ {\rm when}\ q_{i}\gg \vert \sum_{j}q_{j}\vert,
\label{IRasymp}
\end{equation}
whenever one of the $q_{i}$ is much larger than the sum. This can be seen at an elementary level on the properties of the vertex functions
that  vanish when the sum of the argument goes to 0.
The property (\ref{IRasymp}) has direct consequences on the properties of the loop corrections. The $p-$loop correction to the
propagator takes indeed the form,
\begin{equation}
G^{\ploops}(k,\eta)=(4\pi)^{p}\int q_{1}^{2}\dd q_{1}\,\dots\, q_{p}^{2}\dd q_{p}\ P_{0}(q_{1})\dots P_{0}(q_{p})\,\mF^{\ploops}(k,q_{1},\dots,q_{p})
\end{equation}
with
\begin{equation}
\mF(k,q_{1},\dots,q_{p})=\frac{k^{2}}{\sum_{i }q_{i}^{2}}\alpha_{p}(k/q_{1},\dots, k/q_{p})
\end{equation}
where $\alpha(k/q_{1},\dots,k/q_{p})$ is finite in all regimes. This property determines the converging properties of
the $p-$loop correction to the propagator. Clearly the converging properties of the loop corrections deteriorate. It can easily be shown that formally, for a power law 
spectrum of index $n_{s}$, the convergence is ensured only when,
\begin{equation}
n_{s}<-3+\frac{2}{p}.
\end{equation}
leading to a greater sensitivity to the large wave-modes
as one increases the loop order\footnote{But note that for the concordant model it converges at all order.}. 

Note also that the other diagrams (those coming from the $\Gamma-$ expansion) are generally better behaved. The reason is that
the property (\ref{IRasymp}) is then to be applied two times (on each side) gaining an overall factor $1/q^{2}$ for the convergence
in the UV domain (hence the $k^{4}$ contribution mentioned in the previous subsection). The loop contributions that are most sensitive to the UV domain are 
therefore those contributing to the propagator.

Unfortunately, at low redshift it implies that a regularization procedure in the UV domain should be used in order to make the contributions of these terms always 
realistic. This need is  marginal for the two-loop contributions at low redshift, it is essential if we ever want to include the three-loop diagrams. 

Let us mention possible solutions to this problem,
\begin{itemize}
\item  partial resummation of higher order terms could provide us with a self-consistent regularization 
scheme (in a similar way to the IR domain). A possible solution has been recently put forward in \shortciteN{2013arXiv1309.3308B} 
with the help of the Pad\'e ansatz. Furthermore other resummation schemes
mentioned in the introduction could lead to possible solutions as they exploit alternative series resummations;
\item a reformulation of PT with other field combinations (see next section) could lead to kernels that are less sensitive to the UV modes;
\item regularization could be associated from the development 
multi-flow regimes. In such cases none of the previous methods would work and one
should rely on other type of approaches. For instance it was advocated 
that the impact of the small scale physics could be captured with an effective theory 
approach where effective anisotropic pressure terms are introduced in the fluid \shortcite{2012JCAP...01..019P,2012JHEP...09..082C};
\item it is eventually always possible to rely on numerical results to actually measure the kernel
functions (although I am not aware of any attempt to do so in practice).
\end{itemize}

%%%%%%%%%%%%%%%%%%%%%%%%%%%%%%%%%%%%%%%%%%%%%%%%%%
\section{Alternative perturbation theory schemes}
\label{sec:otherPT}

It is to be noted that there is not a single way of doing perturbation theory. Indeed the choice of fields to represent the 
cosmic quantities is not unique. It is always possible to change into linearly related fields, for instance the potential 
instead of the density contrast, or the velocity potential instead of the velocity divergence, but it does not change
the structure of the perturbation series. A more dramatic change is to introduce nonlinear transforms of the field. A 
straightforward example is to replace the peculiar velocity field by the momentum field, $\vp(\vx,\eta)=\rho(\vx,\eta)\vu(\vx,\eta)$ 
(as exploited in a series of papers applied to the redshift space distortions starting with \citeNP{2011JCAP...11..039S}). 
It makes the continuity equation very simple, the divergence of $\vp$ is the time
derivative of the density to all orders but it makes the Euler equation a more cumbersome to manipulate. In particular the 
momentum field is no more potential to all orders.

An even more dramatic transformation is a change of the coordinate system itself. This is what the Lagrangian approach
does. This is a very popular, and useful, approach that we present in more detail in the following.

\subsection{Lagrangian PT}
Perturbation Theory calculations in Lagrangian coordinates deserve a special attention as it has been advocated as a very
efficient approach 
for a long time (see for instance \shortciteNP{1970Ap......6..164Z,1992MNRAS.254..729B,1995A&A...296..575B,1993A&A...267L..51B,1994MNRAS.267..811B,2008PhRvD..77f3530M}).

Here the fundamental quantity is the
displacement field, $\Psi(\vq)$, that relates the final position of each particle to its initial position (that we will denote $\vq$ in this section),
\begin{equation}
\vx=\vq+\Psi(\vq).
\end{equation}
The motion equation for a particle is simple. It simply comes from the expression of the acceleration. Taking the divergence of this equation
leads to the form,
\begin{equation}
{\partial_{x_{i}}}\left[
\psi_{i}''(\vq)+\left(\frac{3\Omega_{m}}{2f_{+}^{2}}-1\right)\psi_{i}'(\vq)
\right]=\frac{3\Omega_{m}}{2f_{+}^{2}}\delta(\vq),
\end{equation}
where a prime is the derivative with respect to $\eta$ and
$\delta(\vq)$ is the density contrast at the location occupied by the particles that were initially at position $\vq$.
Note that in full generality the source term of this equation should take into account all flows that have reached position $\vq+\Psi(\vq)$.
In the single flow approximation, there is precisely only one flow. The density at this location is given by the Jacobian of the transform from coordinates $\vq$
to $\vx$,
\begin{equation}
1+\delta(\vq)=\frac{1}{J(\vq)}=\frac{1}{\left\vert \left(\frac{\partial \vx}{\partial \vq}\right) \right\vert}.
\end{equation}
The motion equation can then be written
\begin{equation}
J(\vq){\partial_{x_{i}}}\left[
\psi_{i}''(\vq)+\left(\frac{3\Omega_{m}}{2f_{+}^{2}}-1\right)\psi_{i}'(\vq)
\right]=\frac{3\Omega_{m}}{2f_{+}^{2}}(J(\vq)-1).
\end{equation}
The Jacobian of the transform can easily be expressed in terms of the derivatives of the displacement field,
\begin{equation}
J(\vq)=\left\vert\det\left(\frac{\partial x_{i}}{\partial q_{j}}\right)\right\vert.
\end{equation}
Moreover in the single flow approximation the absolute value can be dropped, the sign of the determinant is changing for
each shell crossing. The determinant of the matrix $\partial x_{i}/\partial q_{j}$ is then given by
\begin{equation}
J(\vq)=\frac{1}{6}
\epsilon^{ijk}\epsilon^{i'j'k'}(\delta_{ii'}+\psi_{i,i'}(\vq))(\delta_{jj'}+\psi_{j,j'}(\vq))(\delta_{kk'}+\psi_{k,k'}(\vq))
\end{equation}
and appears to be a polynomial in the elements of $\psi_{i,j}$ of order 3.
Finally, the expression of $J(\vq){\partial_{x_{i}}}$ can also be expressed in terms of the displacement field. We first
note that
\begin{equation}
J(\vq){\partial_{x_{i}}}=
J(\vq)\frac{\partial q_{j}}{\partial x_{i}}{\partial_{q_{j}}}
\end{equation}
and $J(\vq)\frac{\partial q_{j}}{\partial x_{i}}$ is given by determinant of the co-matrices. More precisely we have,
\begin{equation}
J(\vq)\frac{\partial q_{i'}}{\partial x_{i}}=
\frac{1}{2}
\epsilon^{ijk}\epsilon^{i'j'k'}(\delta_{jj'}+\psi_{j,j'}(\vq))(\delta_{kk'}+\psi_{k,k'}(\vq))
\label{partialqx}
\end{equation}
which is a polynomial of order 2 in $\psi_{i,j}(\vq)$. The motion equation is therefore cubic in the displacement field
and reads,
\begin{eqnarray}
&&\epsilon^{ijk}\epsilon^{i'j'k'}(\delta_{jj'}+\psi_{j,j'}(\vq))(\delta_{kk'}+\psi_{k,k'}(\vq))
\left[
\psi_{i,i'}''(\vq)+\left(\frac{3\Omega_{m}}{2f_{+}^{2}}-1\right)\psi_{i,i'}'(\vq)
\right]=\nonumber\\
&&\frac{3\Omega_{m}}{f_{+}^{2}}
\left[\frac{1}{6}
\epsilon^{ijk}\epsilon^{i'j'k'}(\delta_{ii'}+\psi_{i,i'}(\vq))(\delta_{jj'}+\psi_{j,j'}(\vq))(\delta_{kk'}+\psi_{k,k'}(\vq))-1
\right].
\label{LagEqDiv}
\end{eqnarray}
It does not contain however all the required information: it is a scalar equation and as such it cannot allow
to reconstruct the full displacement field. One should also take into account the absence of curl modes in the velocity field
for the \emph{Eulerian} coordinates,
\begin{equation}
\nabla_{\vx}\times \Psi'(\vq)=0.
\label{LagNoCurl1}
\end{equation}
This constraint can easily be expressed using the form (\ref{partialqx}). This is the form which is usually given in
papers in the subject (see for instance \citeNP{2012JCAP...06..021R}). It is however possible to derive a simpler form which
is a re-expression of the Kelvin's circulation theorem applied to our situation where we know from the Eulerian analysis
that the curl term in the velocity field can be neglected \shortcite{2012JCAP...12..004R,2012arXiv1212.4333F}. It leads to\footnote{It is actually not
too difficult to derive this relation. The direct constraint, eqn~(\ref{LagNoCurl1}), leads to $(\delta_{kk'}+\psi_{k,k'})\epsilon^{i'j'k'}(\delta_{jj'}+\psi_{j,j'})\psi'_{j,i'}=0$ 
which is precisely satisfied if the relation (\ref{LagNoCurl2}) is valid provided the matrix $(\delta_{kk'}+\psi_{k,k'}) $ is regular.},
\begin{equation}
\epsilon^{ijk}\psi'_{l,i}(\delta_{lj}+\psi_{l,j})=0.
\label{LagNoCurl2}
\end{equation}
Armed with the relations (\ref{LagEqDiv}) and (\ref{LagNoCurl2}) one can then derive, order by order, the expression of the displacement 
field. The second order, third order and fourth order were subsequently derived in 
\shortciteN{1992ApJ...394L...5B}, \citeN{1994ApJ...427...51B}, \citeN{2012JCAP...06..021R}. The second order expression 
revealed particularly useful for setting the initial conditions of numerical
simulations as shown in \shortciteN{2006MNRAS.373..369C}

\subsection{The Zel'dovich approximation and higher order solutions}

The much celebrated Zel'dovich approximation in cosmology \cite{1970Ap......6..164Z} is nothing but the Lagrangian mapping when the displacement
field has been linearized. Note that in this case the divergence of the displacement field is solution of 
\begin{equation}
{\psi^{\zel}}_{i,i}''(\vq)+\left(\frac{3\Omega_{m}}{2f_{+}^{2}}-1\right){\psi^{\zel}}_{i,i}'(\vq)
-\frac{3\Omega_{m}}{2f_{+}^{2}}{\psi^{\zel}}_{i,i}=0
\end{equation}
which is precisely the equation followed by the density contrast at linear oder. Moreover it is easy to see
that at this order the displacement field is potential.

It is obviously possible to expand the solution to arbitrary orders. Note though that in general the
displacement is not necessarily potential. The third order displacement field in particular is not potential
anymore. 

Form a diagrammatic representation point of view that makes the development of perturbation theory calculations
much more intricate as potential and curl modes mix together. The diagrams are even more complicated 
since the equation of motion is third order in $\psi_{i,j}$ so that there are vertices with three incoming lines together
with vertices with only two incoming lines.

\subsection{From displacement fields to power spectra}

Let us finish this section with a final ingredient, the calculation of power spectra (in Eulerian space) out of
Lagrangian space calculations. The Eulerian Fourier modes are
\begin{equation}
\delta(\vk)=\int\frac{\dd^{3}\vx}{(2\pi)^{3/2}}\ e^{-\ii\vk.\vx}\ \delta(\vx)
\label{Lagdeltakdef1}
\end{equation}
or equivalently 
\begin{equation}
\delta(\vk)=\int\frac{\dd^{3}\vx}{(2\pi)^{3/2}}\ e^{-\ii\vk.\vx}\ \rho(\vx),
\label{Lagdeltakdef2}
\end{equation}
where the two expressions differ only by the $\vk=0$ mode value.
The expression of such a Fourier mode should then be expressed in terms of the Lagrangian variables.
Taking advantage of the change of variable $\vx\to\vq$ with $\dd^{3}\vq=\rho(\vq)\dd^{3}\vx$ one gets,
\begin{equation}
\delta(\vk)=\int\frac{\dd^{3}\vx}{(2\pi)^{3/2}}\ e^{-\ii\vk.(\vq+\Psi(\vq))}\ (1-J(\vq))
\end{equation}
in the first case and 
\begin{equation}
\delta(\vk)=\int\frac{\dd^{3}\vx}{(2\pi)^{3/2}}\ e^{-\ii\vk.(\vq+\Psi(\vq))}
\end{equation}
in the second.
The ensemble average of the product of two such wave modes can easily be done taking advantage of the statistical 
homogeneity of the universe. It leads to 
\begin{equation}
\langle\delta(\vk)\delta(\vk')\rangle
=
\Dirac(\vk+\vk')
\int\dd^{3}\vq\,
e^{-\ii\vk.\vq}
\langle
\exp[-\ii\vk.(\Psi(\vq)-\Psi(0))]
\rangle
\label{LagPk}
\end{equation}
(when one uses the second formulation for the Fourier modes). For the Zel'dovich approximation the displacement
field is Gaussian distributed, it is then possible to compute this expression using eqn~(\ref{momtocum}). 

In general though this is more cumbersome as the calculation of the ensemble average that appears in
eqn~(\ref{LagPk}) requires a full knowledge of the statistical properties of the displacement
field correlators, not only the two-point correlation function.

These complications make Lagrangian Perturbation Theory very difficult to implement and much less developed 
than the Eulerian approaches. The Zel'dovich approximation though is a fantastic toy
model to study the qualitative behavior of the cosmic web growth (see \shortciteNP{1996Natur.380..603B}) and has often been used to give insights in the
global statistical properties of the density fields. For instance,  in case of the Zel'dovich approximation this is made possible
because all the statistical properties of the deformation matrix, $\psi_{ij}$, in particular its eigenvalues are known~\shortcite{1970Ap......6..320D}.

%%%%%%%%%%%
\section{Other observables}
\label{sec:otherobs}

It is time to move to other types of observables  that can be built in continuous random fields. There are many of such observables and motivations for using one or 
the other are usually related to prejudices on the performances regarding the statistical errors or systematics. Let us mention
a few of popular ones,
\begin{itemize}
\item peak statistics that focus on most prominent features of the field, see the paper of \shortciteN{1986ApJ...304...15B} that set the stage for Gaussian fields and where the basics of such 
calculations are detailed;
\item topological invariant and Minkowski functionals introduced in \shortciteN{1988PASP..100.1307G} and in \shortciteN{1994A&A...288..697M} that aims at 
producing robust statistical indicators. How such observables are affected by weak deviations from
Gaussianity was investigated in particular by \shortciteN{1994ApJ...434L..43M}  with the use of the Edgeworth expression. We will
 present this tool in the following;
\item the study of extended topological invariants has been renewed with 
the later introduction of the notion of skeleton  \shortciteN{2000PhRvL..85.5515C} and \shortciteN{2012PhRvD..85b3011G};
\item counts-in-cells statistics and void probabilities with many studies starting with \citeN{1979MNRAS.186..145W} and a nice exhaustive study by \citeN{1989A&A...220....1B} we will exploit hereafter.
\end{itemize}
They all make intervene quantities that are beyond the two-point statistics. Whether they can be computed for the nonlinear density field
depends then strongly on our ability to compute the global statistical properties of the density field beyond the linear regime.
It turns out however that for most of those observables, calculations from first principles of
related quantities are often challenging and in practice can be done for only a limited range of parameters and for weak departure from 
Gaussian statistics. 

%\subsection{A chart of PT results}

\begin{table}
%\tableparts
{
\caption{This table indicates what has been computed from direct PT calculation as a function of the order in multi-point correlators (lines)
and the number of loops that have been taken into account in the calculations (columns).}
\label{ChartingPT}
}
{
\begin{tabular}{p{1.5cm}p{2cm}ccp{2cm}p{2cm}}
&tree&1-loop&2 loop&3-loop&p-loop
\vspace{.1cm}\\
\hline
\vspace{0.01cm}&&&&&\\
2-point stats &OK&OK&OK&partial integrations&partial resummation\\
3-point stats&OK&some terms&-&-&partial resummation\\
4-point stats&OK&-&-&-&-\\
N-point stats&OK in spec geometries&-&-&-&-\\
\hline
\end{tabular}
}
\end{table}

In Table \ref{ChartingPT}, we set the stage on what have be computed from the motion equations. In the previous sections, we focused
our interest in the first line corresponding to two-point correlators (either power spectra or two-point correlation functions.). The second
and third lines correspond to the computation of bispectra and trispectra and although known results there are less detailed, such calculations can be
addressed with the same tools.

In the next sections we would like to explore the first column. It corresponds to tree-order results obtained to correlators of arbitrary order.
These results are obtained however for specific geometries and with methods that are totally different from those presented previously. 
But before we enter the heart of those calculations, we first present generic tools that are useful in this context.

%\section{The Laplace transforms}
\subsection{The inverse Laplace transform}

The purpose of this section is to compute the inverse Laplace transform expression in the 
context of cosmological density field. 
It is interesting to do the calculation starting with a discrete version of the probability distribution function (PDF), 
i.e. the calculation of the
counts-in-cells probability, $P(N)$, that is the probability of having $N$ particles in one cell.
We assume here that the discrete field is a Poisson realization of a continuous field with a global averaged number density
of points given by $\Nb$, so that
\begin{equation}
P(N)=\int \dd\rho\ \Ppoisson(N\vert \rho\Nb)\  \mP(\rho)
\end{equation}
where $ \Ppoisson(N\vert \rho\Nb)$ is the Poisson probability, i.e. the probability of having $N$ particles in a cell
when the expected number density of particles is $\Nb\rho$,
\begin{equation}
 \Ppoisson(N\vert \rho\Nb)= \frac{(\Nb\rho)^{N}}{N!}\exp(-\Nb\rho).
\end{equation}
As a result one has
\begin{equation}
P(N)=\int \dd\rho\ \frac{(\Nb\rho)^{N}}{N!}\exp(-\Nb\rho)\ \mP(\rho).
\end{equation}
It is then easy to show that
\begin{equation}
\langle N\rangle=\Nb,\ \ \langle N(N-1)\rangle=\Nb^{2}\langle \rho^{2}\rangle, \dots\nonumber
\end{equation}
and more generally that,
\begin{equation}
\langle N\dots (N-p+1)\rangle=\Nb^{p}\langle \rho^{p}\rangle.
\end{equation}
The quantities $\langle N\dots (N-p+1)\rangle$ are called the factorial moments. They are directly proportional to 
the underlying density moments. 

Let us then define the counts-in-cells generating function,
\begin{equation}
\mgP(\mu)=\sum_{N}\mu^{N}\,P(N).
\end{equation}
One can note that 
\begin{equation}
\mgP(1)=1,\ \frac{\partial}{\partial\mu}\mgP(1)=\langle N\rangle, \  \frac{\partial^{p}}{\partial\mu^{p}}\mgP(1)=\Nb^{p}\langle\rho^{p}\rangle
\end{equation}
and it finally implies that
\begin{equation}
\mgP(\mu)=\mM(\Nb(\mu-1)),
\end{equation}
where $\mM(\lambda)$ is the local density moment generating function defined in eqn (\ref{fb:mMdef}).
Then we can use the residue theorem to relate the counts-in-cells statistics to the moment generating function,
\begin{equation}
P(N)=\oint\frac{\dd\mu}{2\pi \ii}\frac{1}{\mu^{(N+1)}}\mgP(\mu)=
\frac{-1}{\Nb}\oint\frac{\dd\mu}{2\pi\ii}e^{\left(-(N+1)\log\left[(1+\lambda)/\Nb\right]\right)}\mM(\lambda),
\end{equation}
where in the first integral the path line is a clockwise closed curve around the origin and in the second it is
counterclockwise closed curve around $-\Nb$.

The continuous $\mP(\rho)$ function is then obtained by taking the continuous limit $\Nb\to \infty$ while 
keeping $N/\Nb=\rho$ and $\lambda$ fixed,  which leads to,
\begin{equation}
\mP(\rho)\dd\rho=\dd\rho\int_{-\ii\infty}^{\ii\infty}\frac{\dd\lambda}{2\pi \ii}\exp(-\lambda\rho)\mM(\lambda),
\label{InvLapTrans}
\end{equation}
where the integration path can be deformed along the imaginary axis. There ought to be singular points along the real axis
in the positive domain otherwise the contour can be deformed away and the integral vanishes.
This is the inverse Laplace transform. It gives the formal expression of the density PDF~\footnote{A warning should be set here: 
there are cases where this inversion is not unique, that is, there exist families of distinct PDFs that exhibit the same
cumulants to all orders. This is the case for the Lognormal distribution but this is normally not the case for the cosmological density PDF. See a recent paper by 
\shortciteN{2012ApJ...750...28C} 
for explicit examples.} in terms of the moment generating function or equivalently 
of the cumulant generating function.

\subsection{The Edgeworth expansion}

The Edgeworth expansion has been developed in the context of small departure from a Gaussian statistics. It was originally developed
by Longuet-Higgins in the context of sea wave statistics~\cite{LonguetHiggins63}. It was later introduced in the cosmological context by \shortciteN{1995ApJ...442...39J} and \shortciteN{1995ApJ...443..479B}. There are different
approaches that can be used to obtain this result. Here I focus on the method used in \shortciteN{1995ApJ...443..479B} which is based on the inverse Laplace 
transform.

As we are aiming at describing the density PDF is a regime where the variance is small it is natural to rely on the scaling 
relation that have been derived previously. In particular in 
the previous section we have seen that the cumulant generating function could be written
\begin{equation}
\mC(\lambda)=\frac{1}{\sigma^{2}}\varphi(y)
\end{equation}
with $y=\lambda\sigma^{2}$ and where $\varphi(y)$ remains finite in the small variance limit. The Edgeworth expansion is then obtained by a simple
expansion of the moment generating function for terms that depart from a Gaussian distribution, i.e.
\begin{equation}
\exp\left(
\frac{1}{\sigma^{2}}\varphi(y)
\right)
\approx
\exp\left(\frac{1}{2\sigma^{2}}y^{2}\right)\left[1+\frac{1}{\sigma^{2}}S_{3}\frac{y^{3}}{3!}+\frac{1}{\sigma^{2}}S_{4}\frac{y^{4}}{4!}
+\frac{1}{\sigma^{4}}S_{3}^{2}\frac{y^{6}}{2(3!)^{2}}+\dots
\right].
\end{equation}
It is to be noted that in this expansion $y$ will be eventually of the order of $\sigma$ so that the terms appearing in the expansion
in the previous equation of the oder of $\sigma$, then of the order of $\sigma^{2}$, etc. The Edgeworh expansion is then obtained by
a simple integral over $y$ which is then easy to do as $y$ is Gaussian distributed. We eventually get
\begin{eqnarray}
\mP(\rho)=\frac{1}{(2\pi\sigma^{2})^{1/2}}\exp\left(-\frac{\delta^{2}}{2\sigma^{2}}\right)
&&\left[
1+\sigma\frac{S_{3}}{6}H_{3}\left(\frac{\delta}{\sigma}\right)+
\nonumber\right.\\
&&\left.+
\sigma^{2}\left(\frac{S_{4}}{24}H_{4}\left(\frac{\delta}{\sigma}\right)
+\frac{S_{3}^{2}}{72}
H_{4}\left(\frac{\delta}{\sigma}\right)
\right)
\right]
\end{eqnarray}
where $H_{n}$ are the Hermite polynomials. 

Such a construction can be extended to multiple variables (see for instance \shortciteNP{1995astro.ph..4029A}) and is therefore of
ubiquitous use when only a limited number of cumulants are known. 

%%%%%%%%%%%%%%%%%%%%%%%%%%%%%%%%%%%%%%
\section{Cumulants in spherical cells}
\label{sec:PDFs}

The aim of this section is to present explicitly the cumulant generating functions that can be explicitly derived from the single flow
Vlasov-Poisson system. They all concern variables that respect spherical symmetry: typically densities
in concentric cells.

\subsection{Direct calculation of low order cumulants}

Let us start with a very simple and illustrative example, the third order cumulant of the local density contrast filtered
at a given scale $R$. The shape of the filtering function is a priori arbitrary and we will denote it $\mW(k\,R)$ when
written in Fourier space. For functions that do not filter out the long-wave modes (such as Gaussian or top-hat  filters), $\mW(k\,R)$
is usually defined in such a way that $\mW(0)=1$. Then the local \emph{filtered}  density\footnote{Although the calculations are 
presented here for the density modes, it is possible to do it in the velocity divergence field as well. It would
follow exactly the same procedure.} field is,
\begin{equation}
\delta_{R}=\int\frac{\dd\vk^{3}}{(2\pi)^{3/2}}\delta_{R}(\vk),\ \ \ \delta_{R}(\vk)=\delta(\vk)\mW(k\,R)
\end{equation}
and this should obviously be true to all orders in perturbation theory,
\begin{eqnarray}
\delta_{R}^{(n)}=\int{\Pi}_{i}\left[\frac{\dd^{3}\vk_{i}}{(2\pi)^{3/2}}\delta^{\lin}(\vk_{i})\right]\ F_{n}(\vk_{1},\dots,\vk_{n})\,\mW\left(\left\vert \sum_{i} \vk_{i}\right\vert\,R
\right).
\end{eqnarray}
At linear order the only relevant quantity is the variance of the field, $\sigma_{R}$ given by,
\begin{equation}
\sigma_{R}^{2}=\langle\delta_{R}^{2}\rangle=\int\frac{\dd^{3}\vk}{(2\pi)^{3}}\,P^{\lin}(k)\,\mW^{2}(k\,R).
\end{equation}
Following the derivation of the scaling relation of Subsection \ref{fb:subsecscaling} higher order cumulants
will be given by a combination of factors taken at various orders.
The practical difficulty in these calculations comes from the fact that the angular dependences between the wave modes are
entangled in both the $F_{n}$ functions and the window function. For the third order cumulant the integral to compute is,
\begin{eqnarray}
\langle \delta_{R}^{3}\rangle_{c}&=&6\int\dd^{3}\vk_{1}\dd^{3}\vk_{2}\ F_{2}(\vk_{1},\vk_{2})
P^{\lin}(k_{1})P^{\lin}(k_{2})\,\nonumber\\
&&\times\mW(k_{1}\,R) \,\mW(k_{2}\,R)\,\mW(\vert\vk_{1}+\vk_{2}\vert R).
\label{S3general}
\end{eqnarray}
Such calculation has been computed first in \shortciteN{1993ApJ...412L...9J} for a Gaussian window function. It turns out however that the result takes a much 
simpler form
for a filter which appears at first sight more complicated, the real space top-hat window function whose Fourier shape is given by\footnote{This is nothing but the 
Fourier transform
of the characteristic function of a sphere of radius $R$.}
\begin{equation}
\mW(\vk\,R)=\sqrt{\frac{3\pi}{2}}\frac{J_{3/2}(k\,R)}{(k\,R)^{3/2}}
\end{equation}
in  3D where $J_{3/2}$ is the Bessel function of the first kind of index $3/2$.
%Following Fry 84, Goroff et al 85... 
The calculation\footnote{It is based on the exploitation of summation theorem enjoyed by the Bessel functions, relation 8.530 of \shortciteN{Gradshteyn}.} of 
(\ref{S3general}) makes indeed
intervene only the second moment and its variation with the smoothing scale so that~\shortcite{1994A&A...291..697B},
\begin{equation}
\frac{\langle \delta_{R}^{3}\rangle_{c}}{\langle\delta_{R}^{2}\rangle^{2}}=3 \nu_{2}+\frac{\dd\log \sigma^{2}_{R}}{\dd \log R}
\label{S3expression}
\end{equation}
where $\nu_{2}$ is directly related to $F_{2}$ as its angular average,
\begin{equation}
\nu_{2}=\int_{-1}^{1}\dd \mu\,F_{2}(\vk_{1},\vk_{2})
\end{equation}
($\mu$ is the $\cos$ of the angle between $\vk_{1}$ and $\vk_{2}$). For an Einstein-de Sitter universe we have $3\nu_{2}=34/7$.
Such relation between the spherical collapse dynamics and tree-order cumulant can actually be generalized to all 
orders. This is this connexion  that we will discuss in the rest of this section. First we need to explore a bit more the specificities
of the spherical collapse solutions.

%which is related to the spherical collapse dynamics as we will make it explicit in the next subsection.

\subsection{The spherical collapse}

\begin{figure}[ht] %  figure placement: here, top, bottom, or page
   \centering
 \includegraphics[width=8.5cm]{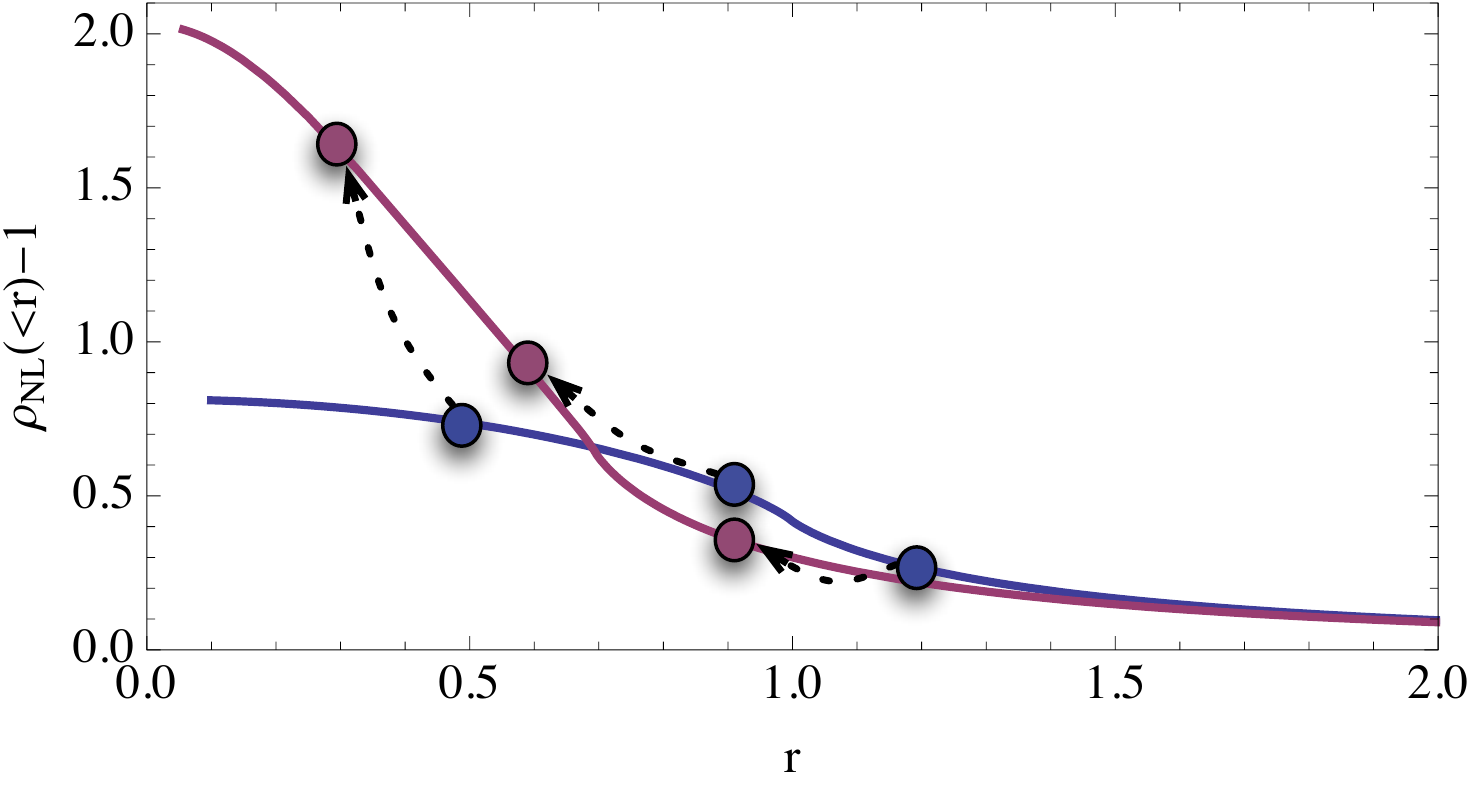} 
   \caption{Example of evolution of a density profile with the spherical collapse. 
In blue we give the linearly evolved profile (linear growing mode), in red its nonlinear evolution: given a density contrast within a radius in the
growing mode linear regime $\tau(<r)$ the subsequent shell size and density it encompasses are entirely determined by the spherical evolution -- examples of such evolutions are given by the blue and red circles -- which consequently determines the non-linear density profile
}
   \label{SphericalColl}
\end{figure}

The spherical collapse does not only give the time it takes for a spherically symmetric perturbation to collapse,
it also gives the explicit and exact solution of the nonlinear evolution
of the density field before shell crossing for a wide class of initial fields, those with initial spherical perturbations. Moreover, the Gauss theorem ensures that the 
radius evolution of a shell in such a geometry is entirely determined by the total mass it contains.  So let us consider a density contrast $\tau(<r)$
within the radius $r$. Let us call $R(\eta)$ the radius of that same shell during its nonlinear evolution and $\rho(<R,\eta)$ the total density
it contains. At an arbitrarily early time the amount of matter encompassed within such a radius is simply $4\pi/3 r^{3}\rhob(\eta_{0})$ and by matter conservation we 
have
\begin{equation}
\rho(<R,\eta)R^{3}(\eta)=\rhob(\eta^{0})r^{3}.
\end{equation}
The time evolution of $R$ can be solved in principle. It obeys the equation of motion 
\begin{equation}
\frac{\dd^{2}R}{\dd t^{2}}=-\frac{GM(<R)}{R^{2}}
\end{equation}
which is
nothing but the Friedman equation but for slightly different initial conditions. 
Matching the time variables in the two cases leads to an explicit form of the spherical collapse that relates the time dependent
nonlinear density to the initial linear one when the latter is taken in the linear growing mode\footnote{Setting the initial field in the growing mode is important otherwise one would need not only the initial density contrast but also the initial velocity gradient
to predict the subsequent density evolution.}. For an Einstein de Sitter background this equation is actually independent on
time once the initial density contrast is expressed in terms of its linear evolution. 
Its explicit form depends on whether the initial perturbation, evolved linearly, $\tau$ is negative or positive. In the former case, we have,
\begin{equation}
\rho(\tau)={9\over 2}{(\sinh\theta-\theta)^2\over
(\cosh\theta-1)^3},\ \ 
\tau=-{3\over 5}\left[{3\over
4}\left(\sinh\theta-\theta\right)\right]^{2/3}
\end{equation}
and in the latter case,
\begin{equation}
\rho(\tau)={9\over 2}{(\theta-\sin\theta)^2\over
(1-\cos\theta)^3},\ \ 
\tau={3\over 5}\left[{3\over
4}\left(\theta-\sin\theta\right)\right]^{2/3}.
\end{equation}
Another interesting peculiar case corresponds to the regime where the universe is almost empty ($\Omega_{m}\to 0$ with $\Omega_{\Lambda}=0$)
for which the spherical collapse solution takes a surprisingly simple form,
\begin{equation}
\rho(\tau)={1\over (1-2\tau/3)^{3/2}}.
\label{simplifiedzeta}
\end{equation}
In the following we denote $\zeta(\tau)$ the functional form that relates the linear density contrast to the nonlinear density.  
It can formally be expanded in,
\begin{equation}
\zeta(\tau)=\sum_{p}\nu_{p}\frac{\tau^{p}}{p!}.
\end{equation}
The a priori time dependent $\nu_{p}$ parameters encode all the spherical collapse dynamics.
And for the very same reason that the kernels $F_{n}$ and $G_{n}$ are almost independent on the background evolution, the function
$\rho$, expressed as a function of the initial linear density contrast, is very weakly dependent on the cosmological parameters.
The form (\ref{simplifiedzeta}) first proposed in \shortciteN{1992ApJ...392....1B}, is actually very accurate in practice\footnote{Although it predicts a critical value for 
the density 
contrast, 1.5, which is slightly below the value for an Einstein-de Sitter background, 1.69.}.

What it implies is that for any initial spherical profile, that can always be characterized 
by the function $\tau(<r,\eta_{0})$, the profile at time $\eta$ is given by
\begin{equation}
\rho(<R,\eta)=\zeta[e^{\eta-\eta_{0}}\,\tau(<r)],\ \ \hbox{with}\ \ \rho(<R,\eta)\,R^{3}=\rhob(\eta_{0})\,r^{3}.
\label{zetadef}
\end{equation}
Such a mapping is illustrated in Fig. \ref{SphericalColl}.

The explicit (or implicit) use of the spherical collapse solution is very common is cosmology and to a large extent to predict, at least roughly,
the number density of formed halos and their correlation properties. There are many developments about this idea in textbooks (see also
the review paper of \shortciteNP{2002PhR...372....1C}) but the purpose of these notes is not to cover this field.

The spherical collapse solution can also be related to the full ensemble of the density cumulants. 
In the following we will make explicit use of the fact that this mapping
provides an explicit non linear solution of the density field for
spherically symmetric initial conditions.

\subsection{The tree order cumulant generating function as a Legendre transform of the initial moments}

We are interested here in the leading order expression of $\varphi(\{\lambda_{i}\})$ for a finite number of concentric cells.
The central result is that can be obtained is that it is given by the Legendre transform of a function that can be built from the
initial moments of the density taken in concentric cells. We refer to 
 \citeN{2002A&A...382..412V} and \shortciteN{BernardeauPichonCodis2013} fo detailed demonstrations. In the following
 we simply explicit this result and present some of its applications.
 
So let us consider the joint cumulant distribution function of the cumulant of the densities in concentric cells,
\begin{equation}
\varphi(\{\lambda_{k}\})=\sum_{p_{i}=0}^{\infty}\ \langle \Pi_{i}\rho_{i}^{p_{i}}\rangle_{c}\frac{\Pi_{i}\lambda_{i}^{p_{i}}}{\Pi_{i}p_{i}!}\,,
\label{phidef}
\end{equation}
where the densities $\rho_{i}$ are obtained as the density within the radius $R_{i}$ et time $\eta$,
\begin{equation}
\rho_{i}(\eta)=\frac{1}{4\pi R_{i}^{3}/3}\int_{\vert\vx-\vx_{0}\vert<R_{i}}\dd^{3}\vx\ \rho(\eta,\vx).
\end{equation}

To each such density $\rho_{i}$ and radius $R_{i}$ one can associate an initial density contrast $\tau_{i}$
computed at the smoothing radius $r_{i}=R_{i}\rho_{i}^{1/3}$  that follows the spherical collapse mapping,
\begin{equation}
\rho_{i}(\eta)=\zeta(e^{\eta-\eta_{0}}\tau_{i}).
\end{equation}
Then what the Gauss theorem ensures is that  the cumulant generating at leading order
is entirely determined by the statistical properties of the variables $\tau_{i}$ at time $\eta_{0}$. 
So let us first define the cross-correlation of the initial density in cells of radii $r_{i}$
\begin{equation}
\Sigma_{ij}=\langle\tau\left(<R_{i}\rho_{i}^{1/3}\right)\tau\left(<R_{j}\rho_{j}^{1/3}\right)\rangle\,.
\end{equation}
One can then define the function $\Psi$ as,
\begin{equation}
\Psi(\{ \rho_{i}\})=\frac{1}{2}\sum_{ij}\Xi_{ij}\,\tau_{i}\tau_{j}\,,
\end{equation}
seen as a function of the variable $\rho_{i}$ where $\Xi_{ij}$ is the \emph{inverse} matrix of the cross-correlation
matrix $\Sigma_{ij}$.
The leading order expression of $\varphi$ is then given by the Legendre transform of the function $\Psi(\{\rho_{k}\})$,
\begin{equation}
\varphi(\{\lambda_{k}\})=\sum_{i}\lambda_{i}\rho_{i}-\Psi(\{ \rho_{k}\})\,,\label{phifromLeg}
\end{equation}
where the values of $\rho_{i}$ are determined by the stationary conditions,
\begin{eqnarray}
\lambda_{i}=\frac{\partial}{\partial \rho_{i}} \Psi(\{ \rho_{k}\}).\label{statCond3}
\end{eqnarray}
This is this general expression that we will exploit in the following. Note that this relation is true for $\eta=\eta_{0}$
as we then have $\rho_{i}(\eta_{0})=1+\tau_{i}$ and the fact that $\varphi(\{\lambda_{k}\})$ and $\Psi(\{\rho_{k}\})$
are Legendre transforms can easily be checked for a Gaussian field.

For $\eta>\eta_{0}$, the resulting cumulant generating function is no more quadratic. Higher order terms are induced due
to the nonlinear $\tau_{i}-\rho_{i}$ mapping.

Naturally, 
re-expanding the function $\varphi(\{\lambda_{k}\})$ with respect to $\lambda_{k}$ then gives back the individual cumulants. 
One can check for instance that the cubic term 
of the one-variable generating function following the prescription (\ref{phifromLeg}-\ref{statCond3}) gives back eqn (\ref{S3expression}).
This is not a very easy calculation can it can be implemented very efficiently in a formal calculator to any order
in the expansion. It gives the explicit expression of any such cumulant or joint cumulant computed at tree order (explicit expressions
have been given up to the tenth order cumulant!).

At this stage, what we have is a pure mathematical trick, an efficient way for obtaining full series of cumulant values
sparing the task of wave mode integrations, as in eqn (\ref{S3general}). This trick however works only at tree order.
Comparisons of such predictions with simulations 
have been made in  various papers and proved to be very accurate \shortcite{1995MNRAS.274.1049B,1994A&A...291..697B}.

It is however possible to go one step further by assuming that the generating function are given by eqns (\ref{phifromLeg}-\ref{statCond3}) 
which corresponds to the vanishing variance limit, is actually a good approximation of the actual cumulant generating functions
for \emph{finite} but small values of the variance. It is possible to exploit this construction to built the corresponding PDF or joint PDF
of the densities.

\section{Density PDFs and profiles with spherical cells}
\label{sec:profiles}

\subsection{Reconstructing the density PDFs}

\begin{figure}
\centering
\includegraphics[width=8cm]{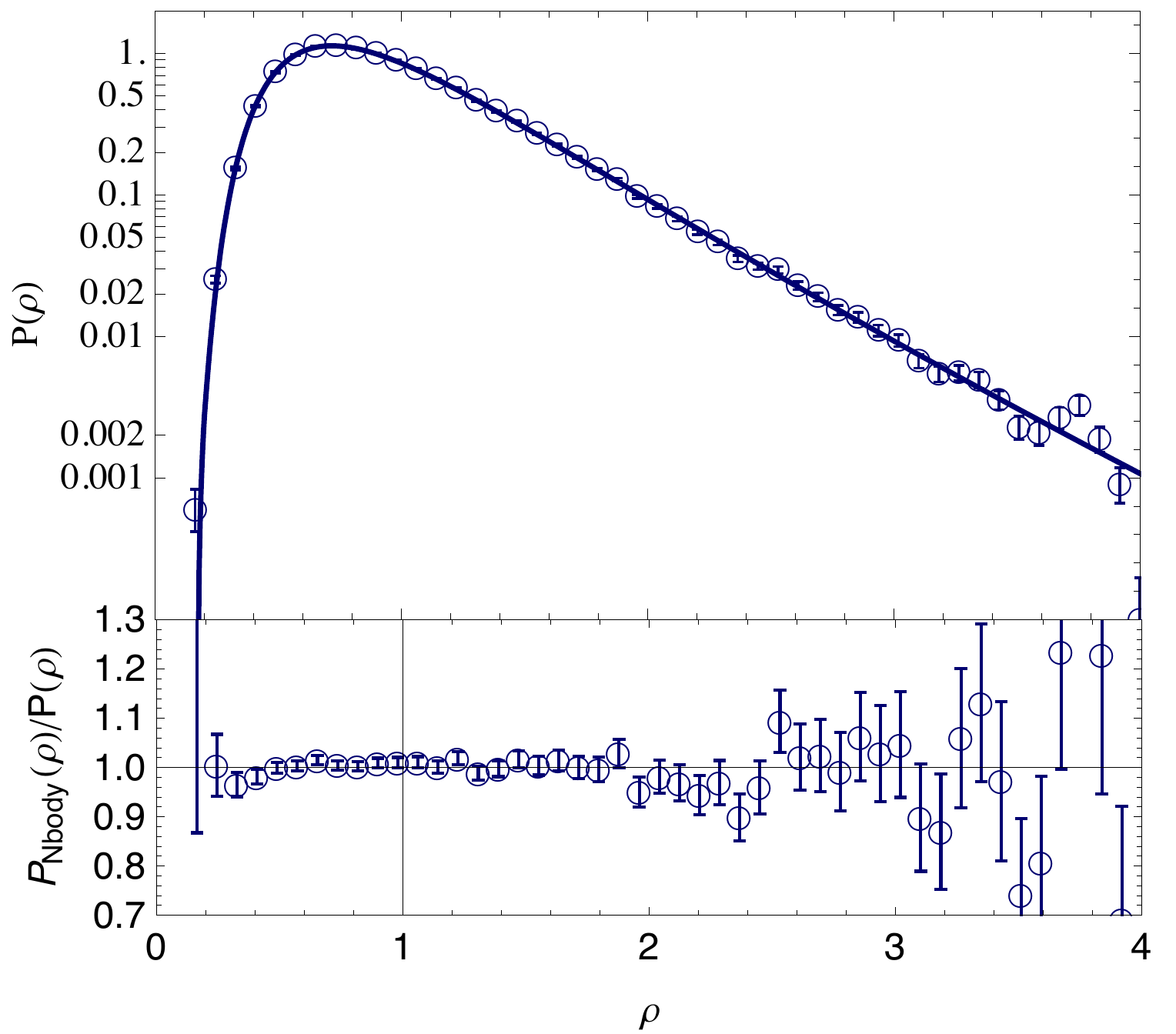} 
\caption{Comparison of the analytical prediction, blue solid line, with the numerical results. The bottom panel shows the residuals.}
\label{OnePointDensityPDF}
\end{figure}

The reconstruction of the density PDFs are based on the inverse Laplace transforms, eqn (\ref{InvLapTrans}), written
in term of the cumulant generating function,
\begin{equation}
\mP(\hrho)=\int_{-\ii\infty}^{\ii\infty}\frac{\dd\lambda}{2\pi \ii}\exp[-\lambda\hrho+\varphi(\lambda)].
\label{InvLapTrans2}
\end{equation}
In the following we will denote with a $\hat{\ }$ the random quantities for which we compute the statistical properties
(to avoid confusion with those that appear in the Legendre transform).
It is important to have in mind that the resulting one-point PDF depends strongly on the 
analytical structure of the cumulant generating functions (in this respect the key paper to explore in detail is \citeN{1989A&A...220....1B} 
which contains a description of all approximations and regimes one wishes to know about this
transformation). The latter turns out to be 
determined to a large extent by the structure of the stationary condition (\ref{statCond3}).
For the one-variable case,  it appears  that there is always a singular value on the real axis $\lambda_{c}>0$
beyond which there is no solution for $\lambda$ (this property appears on Fig. \ref{plotsMapPhiTheo} when $\varphi(\lambda,\mu)$ is restricted to $\mu=0$). It corresponds to a value $\rho=\rho_{c}$ at which location we 
have,
\begin{eqnarray}
0&=&\Psi''[\rho_{c}]\label{rhocdef}\\
\lambda_{c}&=&\Psi'[\rho_{c}]
\end{eqnarray}

Note that at $\rho=\rho_{c}$, $\Psi$ is regular\footnote{In particular this singular behavior in $\varphi(\lambda)$
we get is not related to any singularity in the spherical collapse.}.  The function $\varphi(\lambda)$ can be expanded near this point
and we typically get,
\begin{eqnarray}
\varphi(\lambda)&=&
\varphi _{c}+\left(\lambda -\lambda _{c}\right) \rho _{c}+\frac{2}{3} \sqrt{\frac{2}{\pi _3}}
   \left(\lambda -\lambda _{c}\right)^{3/2},   
   \label{sphiexp}
\end{eqnarray}
where $\pi_{i}={\partial^{i}\Psi}/{\partial\rho^{i}}$.  It is to be noted that the leading singular term scales like
 $\left(\lambda -\lambda _{c}\right)^{3/2}$. The coefficients $\pi_{i}$ are all related to the function $\Psi$ and are therefore
 model dependent~\footnote{Note that in practice $\pi_{3}$ is negative and that only the non regular parts appearing 
in the expansion of $\exp(\varphi(\lambda))$ will contribute.}

We can now re-construct the basic properties of the density PDF.

Let us focus to cases where the variance of the local density contrast is small.
Then the inverse transform can in turn be obtained from a steepest descent method. Formally
it leads to the conditions that should be met at the saddle point $\lambda_{s}$, 
\begin{eqnarray}
\frac{\partial}{\partial \lambda}\left[\lambda\hrho-\varphi(\lambda)\right]&=&0;\\
\frac{\partial^{2}}{\partial \lambda^{2}}\left[\lambda\hrho-\varphi(\lambda)\right]&<&0.
\end{eqnarray}
The first condition leads to $\rho(\lambda_{s})=\hrho$ and the second to $\lambda_{s}<\lambda_{c}$. It simply means that this approximation
can be used if $\hrho<\rho_{c}$. 
The resulting expression for the density PDF is easy to obtain and it leads to,
\begin{equation}
\mP(\hrho)=\frac{1}{\sqrt{2 \pi}}
\sqrt{\frac{\partial^{2}\Psi(\hrho)}{\partial\rho^{2}}}\exp\left[-\Psi(\hrho)\right].
\label{aPDFlowrho}
\end{equation}
It is valid as long as the  expression that appears in the square root is positive which amounts to say that
$\hrho<\rho_{c}$. 

When this latter condition is not satisfied, the singular behavior of $\varphi$ dominates the integral in the complex plane.
It  eventually leads to
\begin{equation}
\mP(\hrho)\approx
\frac{3\, \Im{(a_{\frac{3}{2}} )} }{4 \sqrt{\pi } \left(\hrho -\rho _{c}\right)^{5/2}}
\exp\left(\varphi _{c}- \lambda _{c}\hrho \right),
   \label{aPDFlargerho1}
\end{equation}
where $a_{j}$ are the coefficient in front of $\left(\lambda -\lambda _{c}\right)^{j}$ in Eq. (\ref{sphiexp}), e.g. 
$a_{3/2}={2}/{3} \sqrt{{2}/{\pi _3}}$ and $\Im(a_{3/2})$ is its imaginary part.
It leads to exponential cut-off at large $\hrho$ as $\exp(\lambda_{c}\hrho)$.  This property is actually robust and is preserved 
when one performs the inverse Laplace transform for finite values of the variance or even large value of the variance (see \citeNP{1989A&A...220....1B}).
It also gives a direct transcription of why $\varphi(\lambda)$ becomes singular: for values of $\lambda$ that are larger than $\lambda_{c}$,
$\int\dd\hrho \mP(\hrho) \exp(\lambda\hrho)$ is not converging.

None of these forms however are accurate for the full range of density values and in general one has to rely on numerical integrations
in the complex plane. Those can be performed accurately and quickly with a proper choice of the integration path in the
complex plane (as described for instance in \shortciteNP{1992ApJ...392....1B}). The resulting shape for the PDF is shown on Fig. \ref{OnePointDensityPDF} and 
compared to results of numerical simulation. In this comparison we have $\sigma_{R}=0.47$ and a power spectrum index close to $-1.47$. 
The predictions are shown to be accurate at percent level. They deteriorate when the variance gets comparable to unity.

Such calculations have been used for comparison with numerical simulation (as in \shortciteNP{1995MNRAS.274.1049B,1994A&A...291..697B} in the early papers and more recently in \cite{BernardeauPichonCodis2013}), 
for projected density PDFs such as in \shortciteN{2000A&A...356..771V} or more recently in  \shortciteN{2013MNRAS.429.1564M}.

%%%%%%%%%%%%%%%%%%%%%%%%%%%%%%%%%%%%%%%%%%%%%%%
%\section{The density profiles}

\subsection{Two-cell cumulant generating functions}

In this previous paragraph we applied the construction we found to the one-cell case. The whole procedure can actually
by applied to a whole set of concentric cells. With two concentric densities, one can define the local slope - the density difference within two close radii - or more generally the profile - how the density depends on the filtering radius once a constraint has bet set at a fixed one.
It is possible in particular to build marginal or constraint slope or profile PDF.

\begin{figure}[ht]
\centering
 \includegraphics[width=8cm]{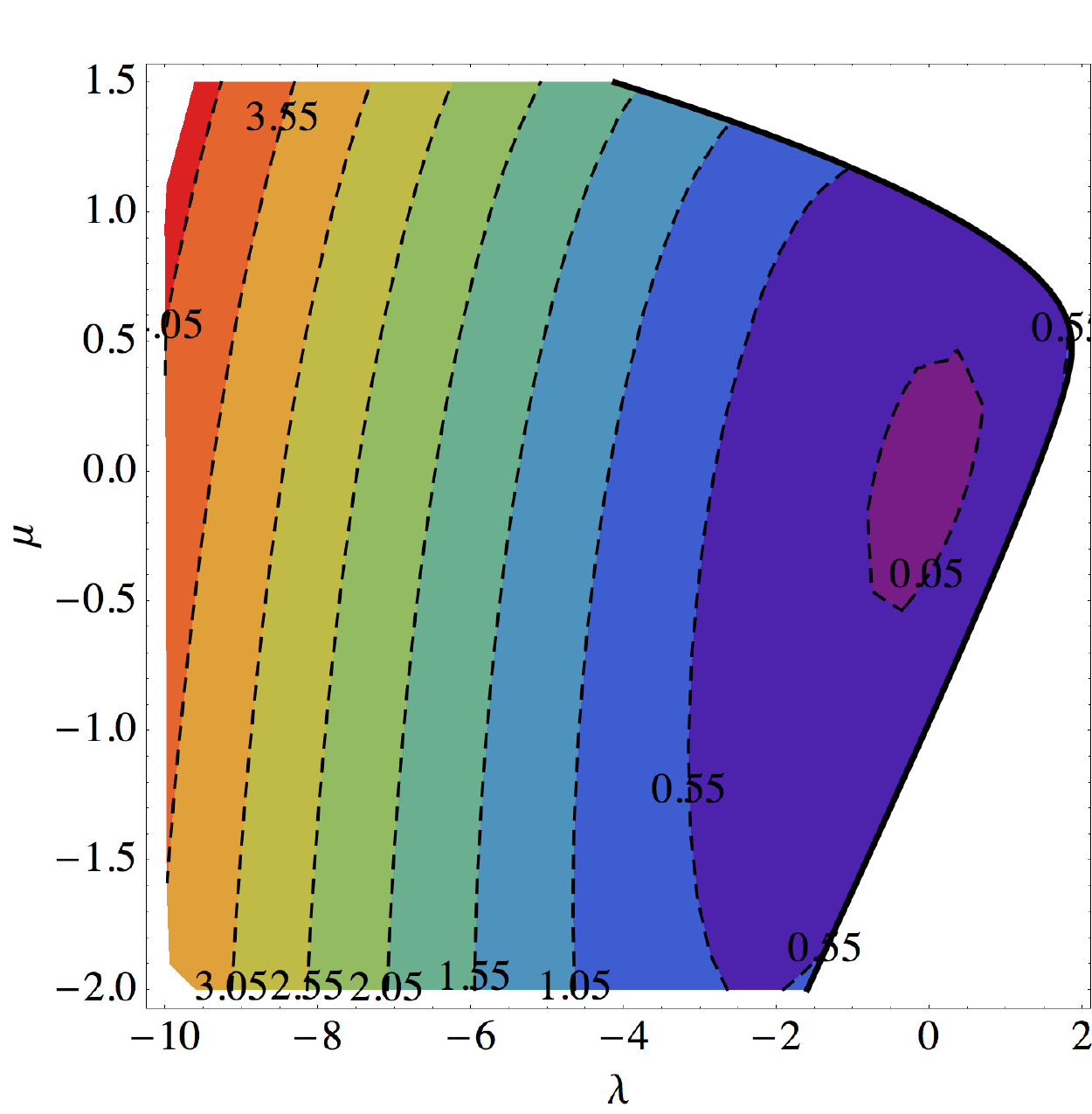}  
 \caption{Contour plot of $\varphi(\lambda,\mu)-\lambda$, left with a finite radius difference $\Dr/R=1/10$. The solid line shows the 
 critical region. On the axis $\mu=0$ one recovers the analytical structure of the one-point generating function which is defined in
 a semi-infinite interval $\lambda<\lambda_{c}$.
   \label{plotsMapPhiTheo}}
\end{figure}

All these constructions are based on the properties of the two-cell PDF or equivalently on the two-cell cumulant generating function.
So we now turn to the analytical properties of  $\varphi(\lambda_{1},\lambda_{2})$ for a finite radius difference. 
We represent this function, for a specific choice of radii but it does chance its expected properties, on Fig. \ref{plotsMapPhiTheo}.
And to make the resulting plot easier to read we rather use,
\begin{equation}
\lambda=\lambda_{1}+\lambda_{2},\ \ \ \ \mu=\frac{R_{2}-R_{1}}{R_{1}}\lambda_{1}.
\end{equation}
It is to be noted that, as for the 1D case, not all values of $\lambda_{1}$ and $\lambda_{2}$ (or equivalently $\lambda$ and $\mu$) are accessible. This is due to the 
fact that 
the $\rho_{i}$ -- $\lambda_{i}$ relation cannot always be inverted. This is signaled by the fact that the determinant of the transformation vanishes,
i.e., $\det\left[{\partial^{2}}\Psi(\{\rho_{i}\})/{\partial\rho_{i}\partial\rho_{j}}\right]=0$. This condition is met  for finite values of both
$\rho_{i}$ and $\lambda_{i}$. The resulting critical line is shown as a thick solid line on Figs. \ref{plotsMapPhiTheo}. Note that 
$\varphi(\lambda_{1},\lambda_{2})$ is also finite at this location. Within this line $\varphi$ is defined; beyond this line it is not.  

%The one-variable case (for the density in a given cell), correspond to the restriction of $\varphi(\lambda,\mu)$ to $\mu=0$. 

\subsection{The expected density slope and profile}

\begin{figure}[ht]
\centering
 \includegraphics[width=7cm]{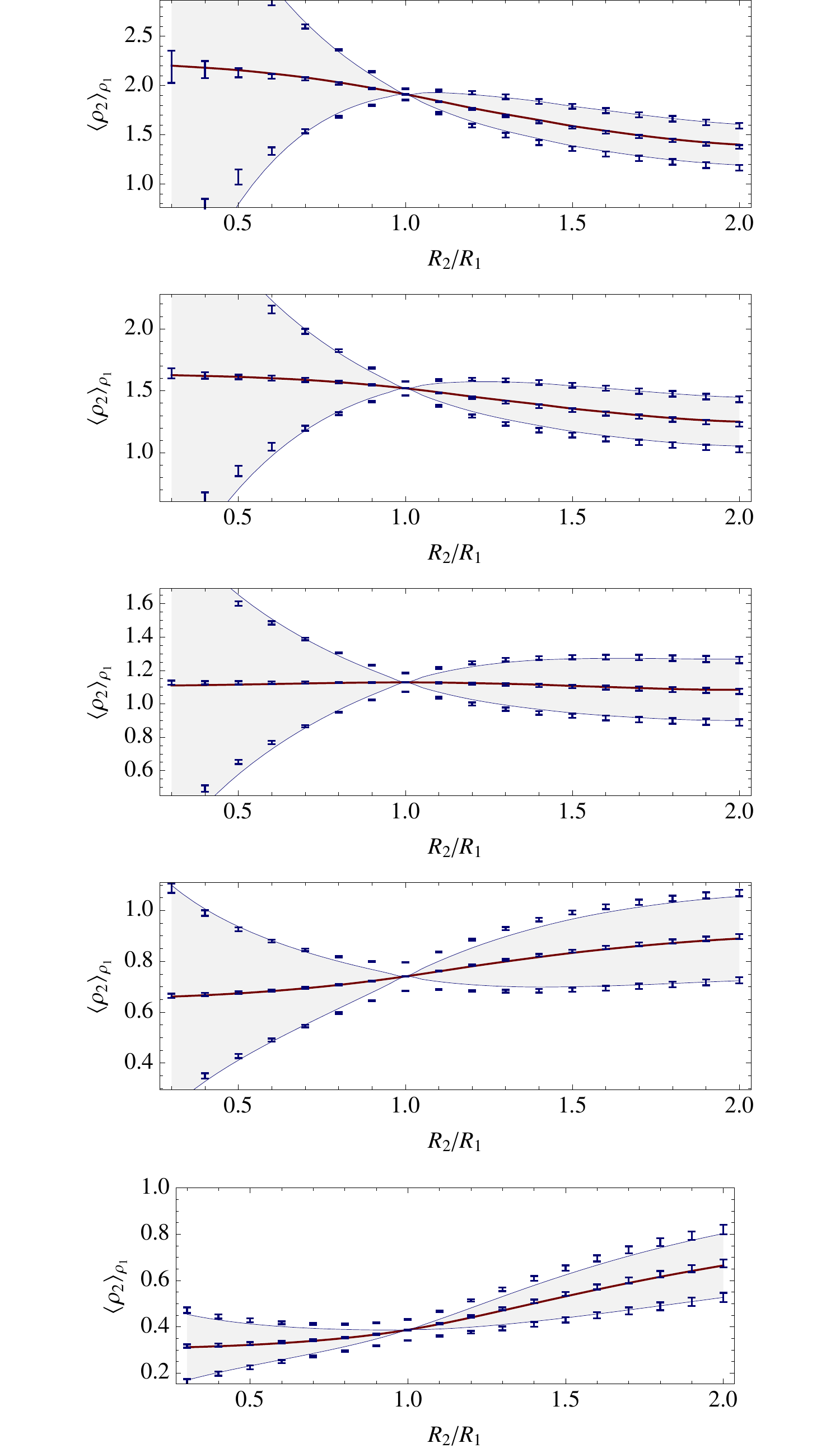} 
   \caption{The conditional profile as a function of $R_{2}$ and for different choices of $\hrho(R_{1})$ (to which $\hrho(R_{2})$ is equal to at $R_{2}=R_{1}$) as predicted by theory and as measured in simulations. The solid lines are the theoretical predictions and the points with error bars are the measurements for both the expected  value and its variance. The agreement is spectacular in particular for low density constraints. Taken from \protect \shortcite{BernardeauPichonCodis2013}.
    \label{ProfilesTheoNum}}
\end{figure}

As mentioned before, the general formalism allows in principle the computation of not only the one-point PDF but also of
multi-variable density PDFs such as $\mP(\hrho_{1},\hrho_{2})$. To illustrate one consequence of such
calculations, one can also make a change of variable and consider the joint PDF of the density and the density slope.
From the densities in two concentric cells, the latter can indeed be defined as\footnote{In the limit of a vanishing smoothing radius difference, $\hs$ will define the 
local density slope. In \protect\shortciteN{BernardeauPichonCodis2013} we explicitly show that
this limit is genuine and can be taken at the level of the generating function.},
\begin{equation}
\hs(R_{1},R_{2})=\frac{R_{1}}{R_{1}-R_{2}}\left[\hrho_{1}-\hrho_{2}\right].
\label{slopedef}
\end{equation}
One of the (many) quantities that can explicitly be computed is
the conditional slope that is the expected value of $\hs$ given the value of
$\hrho_{1}$,
\begin{equation}
\langle \hs\rangle_{\hrho_{1}}=-\frac{R}{\Dr}+\frac{R}{\Dr\, \mP(\hrho_{1})}\int\dd \hrho_{2} \ \hrho_{2} \ \mP(\hrho_{1},\hrho_{2}).
\end{equation}

Such a quantity can easily be expressed in terms of the joint cumulant generating function as,
\begin{equation}
\int\dd \hrho_{2} \ \hrho_{2} \ \mP(\hrho_{1},\hrho_{2})=\\
\int_{-\ii\infty+\epsilon}^{+\ii\infty+\epsilon}\frac{\dd\lambda_{1}}{2\pi \ii}
\left.\frac{\partial \varphi(\lambda_{1},\lambda_{2})}{\partial\lambda_{2}}\right\vert_{\lambda_{2}=0}
\exp(-\lambda_{1}\hrho_{1}+\varphi(\lambda_{1}))
\end{equation}
which can be obtained by explicit integration in the complex plane. Pursuing along the same line of calculations, 
the variance of $\hrho_{2}$ given $\hrho_{1}$  can be computed from the conditional value of $\hrho_{2}^{2}$. 
The latter is given by the second derivative of the moment generating function, therefore given by,
\begin{eqnarray}
\int\dd \hrho_{2} \ \hrho_{2}^{2} \ \mP(\hrho_{1},\hrho_{2})=
\int_{-\ii\infty+\epsilon}^{+\ii\infty+\epsilon}\frac{\dd\lambda_{1}}{2\pi \ii}
\left[\left.\frac{\partial^{2} \varphi(\lambda_{1},\lambda_{2})}{\partial\lambda_{2}^{2}}\right\vert_{\lambda_{2}=0}
\right.\nonumber\\
\left.+\left(\left.\frac{\partial \varphi(\lambda_{1},\lambda_{2})}{\partial\lambda_{2}}\right\vert_{\lambda_{2}=0}
\right)^{2}\right]
\exp(-\lambda_{1}\hrho_{1}+\varphi(\lambda_{1})).
\end{eqnarray}
Obviously such calculations can be pursued and can be used to predict the general statistical properties of the 
density profile as a function of the local density. Such predictions are shown on Fig. \ref{ProfilesTheoNum}, where the expected
value $\hrho_{2}(R_{2})$ are given as a function of $R_{2}$ -- we then build the constrained density profile -- for various values of $\hrho_{1}(R_{1})$, and compared the predictions with numerical simulations.
% in such a way that we have access to the density profile.
When the variance is small the predictions proved  to be very accurate, at percent level. The predictions are particularly appealing
in the low density region as the fluctuations about the mean profile are expected to be limited\footnote{The connexion with statistical indicators such as stacked voids \cite{2012ApJ...754..109L}
is however yet to be done!}.
With the advent of very large surveys, it opens the way to alternative ways of exploring
the statistical properties of the surveys.

\section{Conclusion and perspectives}
\label{sec:perspectives}

The Vlasov-Poisson system in its single flow approximation offers a nice playground for theoretical physicists. It describes the development  of gravitational 
instabilities in a pressureless medium before the first shell crossings and as such is an approximation
to the true motion equations, the Vlasov-Poisson system. It is nonetheless a well defined mathematical system on its own and
is the one on which perturbation theory calculations are all based. Progress has been made recently 
on the understanding of the interplay between dynamics, the structure of the cosmic field being expended to a given order
with respect to the initial fields, and statistics, in particular for Gaussian initial conditions.

One of the major efforts that have been undertaken is the computation of next-to-leading order and next-to-next-to-leading
order (the leading order being the linear theory) of the power cosmic field power spectra. In particular the mode coupling structure
of the corresponding terms has been analyzed in detail as presented in these notes, both in the infrared and the ultra-violet domains. 
The UV behavior, i.e. the importance of the impact of the small-scale modes on the large scale modes is still to be assessed and 
analyzed in detail as it represents potentially a major limitation to systematic PT calculations. Remedies to this limitation may be brought 
by EFT approaches that aim at regularizing the UV effects.

These efforts should be viewed as part of a grander work program that encompasses  the development of efficient 
analytical tools for predicting from first principles the behavior of the cosmic fields in the intermediate regime. It requires not only
the prediction for the density power spectrum but also the determination of the theoretical uncertainties related to those
predictions, the calculation of co-variances. 

Readers should also keep in mind that, contrary to what these notes suggest, these calculations should also be applicable not 
only to the density field
but also to quantities more closely related to actual observations. And if the application of those results to cosmic shear
for instance is a priori quite straightforward, its application to galaxy catalogue is more challenging as one should take into
account biasing, and the impact of redshift space distortions. The calculations should also be completed in order to accommodate
more realistic cosmological fluids such as massive neutrinos\footnote{See some attempts 
to incorporate the impact of massive neutrinos in perturbation theory calculations in \shortciteNP{2008PhRvL.100s1301S,2009PhRvD..80h3528S,2009JCAP...06..017L}.} (the existence of which we are sure of!). It should also be able
to handle modified gravity models\footnote{Some early analytical results have been obtained for instance in \shortciteNP{2009PhRvD..79l3512K,2012PhRvD..86f3512B,2013PhRvD..88b3527B,2013arXiv1309.6783T}.}.

One could also think of applying similar calculations to related or more elaborate observables such as
configuration space correlation functions and  higher order correlation functions. In these notes we also showed that alternative 
approaches could be used to explore this very system. As it has been shown, for some specific variables, it is possible
to compute the leading order generating function of  the moments and to build the corresponding PDFs or joint PDFs.
Similar construction may be doable for more elaborate quantities, alternative  methods or new techniques may still be uncovered 
that would help filling Table \ref{ChartingPT}.

\vspace{.5cm}
\leftline{\bf Acknowledgements}

I would like to thank to organizers of the school for giving me the opportunity to give this series of lecture and to write these notes. I hope they will be useful. I am 
particularly indebted to my recent collaborators Filippo Vernizzi, Rom\'an Scoccimarro, Mart\'{\i}n Crocce, Atsushi Taruya, Takahiro Nishimichi, Christophe Pichon 
and Sandrine Codis for a large fraction of the recent developments presented here.
This work is partially supported by the grant ANR-12-BS05-0002 of the French Agence Nationale de la Recherche.

\bibliographystyle{chicago}
\bibliography{LesHouches.bib}

\end{document}